\documentclass[%
 reprint,
 amsmath,amssymb,
 aps,
]{revtex4-1}

\usepackage{graphicx}
\usepackage{dcolumn}
\usepackage{bm}
\usepackage{amsfonts}
\usepackage{amssymb}

\begin{document}

\title{The Size of Local Bispectrum and Trispectrum in a Non-Minimal Inflation}
\author{Kosar Asadi} \homepage{k.asadi@stu.umz.ac.ir}
\author{Kourosh Nozari}
\homepage{knozari@umz.ac.ir} \affiliation{Department of Physics,
Faculty of Basic Sciences, University of Mazandaran, P. O. Box
47416-95447, Babolsar, Iran}
\date{\today}

\begin{abstract}
Focusing on the local type primordial non-Gaussianities,
we study the bispectrum and trispectrum during a non-minimal
slow-roll inflation. We use the so-called $\delta N$ formalism to
investigate the super-horizon evolution of the primordial
perturbations in this setup. Firstly we obtain the main equations
of the model and introduce the framework of the $\delta N$
formalism for this case. Then we give analytical expressions for
the nonlinear parameters describing the non-Gaussianity in the
slow-roll approximation. We analyze the bispectrum by its
non-linear parameter, $f_{NL}$. Furthermore, we calculate
$\tau_{NL}$ and $g_{NL}$ which are non-linear parameters
characterizing the amplitude of trispectrum. Finally, by adopting
a quadratic form for both the potential and non-minimal coupling
(NMC) function, we test our setup in the light of Planck2015 data
and constrain the model parameters space. Although the non-Gaussianity
parameters are so small in this setup, this model is
consistent with recent observation. We extend our analysis to see
the situation in the Einstein frame and compare the results in these two frames.

\begin{description}
\item[PACS numbers]98.80.Cq , 98.80.Es \item [Key Words]
Inflation, Non-Gaussianity, Bispectrum, Trispectrum, $\delta N$
formalism, Observational data.
\end{description}
\end{abstract}
\maketitle

\section{Introduction}

Cosmological inflation has become an important arena of cosmology
which can successfully address some shortcomings of the standard
model such as the horizon, flatness and relics problems
\citep{infa,infb,infc,infd,infe,inff}. Moreover, inflationary
cosmology provides a graceful mechanism which clarifies the
observed anisotropy of the CMB radiation and also explains the
origin of the almost scale invariant density perturbation leading
to the large scale structure formation in the universe
\citep{lssa,lssb,lssc,lssd,lsse}. To date, almost scale invariant
spectrum is confirmed and we even know that the violation of the
exact scale invariance is about 4$\%$ level \citep{si1}, which is
also consistent with the simplest slow-roll inflationary model.
Nevertheless, the non-linear dynamics of cosmological
perturbations causes non-Gaussianity of the temperature
fluctuations which is one of the most important achievements of
observational data \citep{NGa,NGb} and these observations may bring
us worth information about the dynamics of inflation. Since the
primordial non-Gaussianity contains a large amount of information
on the cosmological dynamics (which derives the initial
inflationary expansion of the universe), studying this feature of
the perturbation modes is indeed an important subject and any
inflationary scenario which can show the non-Gaussianity of the
primordial perturbation is somehow more favorable. For these
reasons, many authors studied the non-Gaussian feature of the
primordial perturbations so far. For a relatively extended list of
recent literature on non-Gaussianity, see
\citep{NGa,NGb,NGc,NGd,NGe,NGf,NGg,NGh,NGi,NGj,NGk,NGl,NGm,NGn,NGo,NGp,NGq,NGr,NGs,NGt,NGu,NGv,NGw,NGx,NGy,NGz,NG27,NG28,NG29,NG30,NG31,NG32,NG33,NG34,NG35,NG36,NG37,NG38,NG39,NG40,NG41}.

The non-Gaussianities which are generated in the primordial
universe can be specified by their shapes. For instance,
equilateral \citep{equ1} (which has a peak when all three wave
numbers are in similar sizes, $k_1=k_2=k_3$) or orthogonal
\citep{orth1} shapes indicate modifications of the kinetic
Lagrangian of the inflaton field. However, the local shape
non-Gaussianity \citep{NGa,NGb} (which has a peak in the squeezed
limit where two wave numbers are much larger than the third one,
$k_1=k_2\gg k_3$) illustrates large super-Hubble interactions.
Moreover, folded shapes \citep{equ1} are arisen from modified
initial conditions and intermediate shapes \citep{int1,int2} signal
the existence of interacting sectors with energy scale of the
Hubble order, respectively.

In order to specify an exact Gaussian distribution of
perturbations from a statistical approach, all we need is the
two-point correlation function (or its Fourier transform, the
power spectrum) \citep{pow1}. Hence the three-point correlation
function, or the Bispectrum which is its Fourier transform, gives
the lowest order statistics to be able to identify non-Gaussian
perturbations from Gaussian ones \citep{bis1,bis2}. Furthermore,
the higher order statistics of non-Gaussianity can be
characterized by the four-point correlation function or its
Fourier transform, the trispectrum \citep{tris1}. Most of the
research papers released so far have been focusing on constraining
the non-Gaussianity of primordial fluctuations by treating the
three-point correlation functions of the perturbations
\citep{NGa,3p1,3p2}. However, one can also constrain the four-point
correlation functions, by increasingly exact measurements (see
\citep{tris1,4p2,4p3}). As a comprehensive study of bispectrum and trispectrum one can be referred to \citep{Bar1,Bar2}. We focus on four-point correlation function
and trispectrum in this paper.

In this paper we consider an inflationary model with an ordinary
scalar field non-minimally coupled to gravity and we study both
bispectrum and trispectrum in the local type non-Gaussianities on
super-horizon scales. Moreover, we perform a transformation to Einstein frame to check the consistency conditions. We show that the results are the same as in Jordan frame, that is, we recover the small values of the non-Gaussianity parameters in this frame too. We note that as White et al. have shown, the non-Gaussianity of the curvature perturbation is identical in Jordan and Einstein frames for adiabatic perturbations in single-field inflation with a non-minimally coupled scalar field \citep{whi-13}.

In order to obtain the non-linear parameters
corresponding to bispectrum ($f_{NL}$) and trispectrum
($\tau_{NL}$ and $g_{NL}$) of the model under consideration, we
employ the $\delta N$ formalism which is based on the
\emph{separate universe} assumption
\citep{delN1,delN2,delN3,delN4,NGn,delN5}. This formalism provides
a powerful tool to analyze the evolution of the curvature
perturbation on scales larger than the horizon scale. As we know,
on super-horizon scales we can only use the evolution of
unperturbed separate universes and neglect spatial gradients. The
main result of this approach is that it allows the primordial
curvature perturbation to be related to the difference in the
number of e-folds between the perturbed universe ($N$) and the
homogeneous background one ($N_{0}$), which are calculated between
an initially flat hypersurface (with $t_{*}$ corresponding to the
time of horizon crossing) and a final uniform energy density
hypersurface (with $t_{e}$ referring to the end of inflation),
respectively. In other words, by applying this formalism one can
only use the field fluctuations at horizon exit and the
homogeneous field evolution thereafter in order to evolve
curvature perturbation $\zeta$ on super-horizon scales.

As we have mentioned, in testing the non-Gaussianity of
fluctuations in an inflationary scenario, important non-linear
parameters are presented such as $f_{NL}$, $\tau_{NL}$ and
$g_{NL}$ which can express the main properties of the cosmological
perturbations. Therefore, confrontation of the inflationary setup
with observation and constraining the model's parameters are
important tasks towards recognition of more natural scenarios. The
newest constraints on local non-Gaussianity to date are provided
by the analysis of data from the Planck2015 satellite
\citep{planck2015}. Planck2015 results in that the three non-linear
parameters $f_{NL}^{equil}$, $f_{NL}^{ortho}$ and
$f_{NL}^{local}$, which parameterize the overall amplitude of an
equilateral, orthonormal and local shapes of the bispectrum, are
constrained as
\begin{eqnarray}\label{1}
f_{NL}^{equil}=-4\pm43,\,\,\,\,\,f_{NL}^{ortho}=-26\pm21,\,\\\nonumber
\,\,\,\,f_{NL}^{local}=0.8\pm5.0\,,\hspace{2cm}
\end{eqnarray}
at the $\%68$ confidence level. This represents a substantial step
forward relative to the Planck2013 \citep{planck2013} with error
bars shrinking by $43\%$ equilateral, $46\%$ orthogonal and $14\%$
local shape. Also, the Planck team have performed an analysis of
trispectrum shapes in the local case and beyond, and obtained that
the amplitude of primordial trispectrum in the local model is
constrained to be
\begin{eqnarray}\label{2}
g^{local}_{NL}=(-9.0\pm 7.7)\times 10^4\,,
\end{eqnarray}
at the $68\%$ CL. These results significantly improved the earlier
best constraints on the trispectrum from WMAP
\citep{wmap1,wmap2,wmap3,wmap4,wmap5,wmap6,wmap7} and large-scale
structure \cite{lss6,lss7,lss8}. Furthermore, the authors of
\citep{Fen15} have reported constraint on $g^{local}_{NL}$ from
Planck2013 data as $g^{local}_{NL}=(-13\pm 18)\times 10^4$.
Although the more recent central amplitude agrees well with this
result, however the statistical error is smaller by a factor of
$2.3$\,. This improvement is partly due to the lower noise level
in Planck2015 data and partly due to usage of a more optimal
estimator.

This paper is organized as follows: After introducing the model
and reviewing the basic equations in Sec. II, we study the local
type bispectrum and trispectrum of the curvature perturbations in
Sec. III. To this end, we use the $\delta N$ formalism and work on
super-horizon scales. Then in Sec. IV, by adopting a quadratic
form for the potential and non-minimal coupling function, we
obtain the non-linear parameters associated to both first and
second order non-Gaussianities of the model focusing on the local
shapes ($f_{NL}^{local}$, $\tau_{NL}^{local}$ and
$g_{NL}^{local}$). After analyzing the evolution of the mentioned
non-linear parameters in Sec. V, we test the model in the light of
Planck data (Planck2013 for $g^{local}_{NL}$ versus
$\tau^{local}_{NL}$ and Planck2015 for $g^{local}_{NL}$ versus
$f^{local}_{NL}$) and we obtain constrains on the parameters space
of the model. Furthermore, we study the non-minimal inflation in Einstein frame in section VI. Then in sections VII and VIII we present a detailed calculation of the non-linear parameters associated to bispectrum and trispectrum in Einstein frame and we study the model numerically in this frame. Finally, we give our summary and conclusions in Sec. IX.

\section{The Model}

We consider a model of cosmological inflation driven by a scalar
field which is non-minimally coupled to the Ricci scalar and is
described by the following action
\begin{eqnarray}\label{A1}
{\cal{S}}=\hspace{8cm}\\\nonumber\int d^{4}x
\sqrt{-g}\left[\frac{1}{2\kappa^{2}}R+f(\phi)R-\frac{1}{2}
\partial_{\mu}\phi\partial^{\mu}\phi-V(\phi)\right],\hspace{0.5cm}
\end{eqnarray}
where $R$ is the 4-dimensional Ricci scalar, $\phi$ is an ordinary
scalar field as inflaton and $V(\phi)$ is its potential. $f(\phi)$
shows an explicit non-minimal coupling of the scalar field with
the Ricci scalar. Considering a spatially flat
Friedmann-Robertson-Walker space-time and varying the action
(\ref{A1}) with respect to the scalar field, results the following
equation of motion
\begin{eqnarray}\label{A2}
\ddot{\phi}+3H\dot{\phi}+V_{,\phi}=f_{,\phi}R\,.
\end{eqnarray}
Furthermore, variation of the action (\ref{A1}) with respect to
the metric leads to the Friedmann equation as
\begin{eqnarray}\label{A3}
H^2=\frac{{\kappa}^2}{3(1+2\kappa^2
f)}\left[\frac{1}{2}\dot{\phi}^2-6H\dot{\phi}f_{,\phi}+V(\phi)\right]\,.
\end{eqnarray}
Now, we apply the slow-roll approximation to these main equations
in which we assume $\ddot{\phi}\ll |3H\dot{\phi}|$ and
$\dot{\phi}^2\ll V(\phi)$. The equation of motion of the
non-minimally coupled scalar field and the Friedmann equation in
the slow-roll limit, are given respectively as
\begin{eqnarray} \label{A4}
3H\dot{\phi}=f_{,\phi}R-V_{,\phi}\,,
\end{eqnarray}
and
\begin{eqnarray}\label{A5}
H^2=\frac{{\kappa}^2}{3(1+2\kappa^2 f)}\left[V-2f_{,\phi}^2
R+2f_{,\phi}V_{,\phi}\right]\,.
\end{eqnarray}
Moreover, the number of e-folds during inflationary era is defined
as
\begin{eqnarray}\label{A6}
N=\int_{t_{*}}^{t_{e}} H dt\,,
\end{eqnarray}
and takes the following expression in our setup
\begin{eqnarray}\label{A7}
N(\phi)=\int_{\phi_{*}}^{\phi_{e}}\frac{\kappa^2\left
(V-2f_{,\phi}^2R+2f_{,\phi}V_{,\phi}\right)}{(1+2\kappa^2
f)(f_{,\phi}R-V_{,\phi})} d\phi\,,
\end{eqnarray}
where $\phi_{*}$ refers to the value of the scalar field when the
universe scale observed today crosses the Hubble horizon during
inflation and $\phi_{e}$ denotes the value of $\phi$ when the
universe scale exits the inflationary phase.

\section{local Non-Gaussianity Using $\delta N$ formalism}

The amplitude of quantum fluctuation of the scalar field $\phi$
(inflaton) during slow-roll inflation is given by
\begin{eqnarray}\label{B1}
\delta\phi=\frac{H}{2\pi}\,.
\end{eqnarray}
We note that, on scales as large as the Horizon radius, since the
amplitude of the field perturbation depends on the chosen gauge,
its meaning is not so clear. However, choosing the flat slicing
gauge, the field's perturbation equation becomes very simple and
is look alike to the case without gravitational perturbation. This
is because choosing this gauge the trace of the spatial curvature
remains unperturbed \citep{Muk92}. Thus the amplitude of field
perturbation can be realized as that in the flat slicing gauge and
can be interpreted as the dimensionless curvature perturbation on
a uniform energy density hypersurface. The primordial curvature
perturbation on uniform density spatial hypersurfaces is denoted
by $\zeta$ \citep{Tan10}. Transformation law is determined by the
following expression
\begin{eqnarray}\label{B2}
\zeta=H\delta t\,,
\end{eqnarray}
with $\delta t$ being the shift in time coordinate for this
transformation. Applying the time derivative of the background
field, $\dot{\phi}$, we have $\delta t=\frac{\delta
\phi}{\dot{\phi}}$ and using (\ref{B1}), equation (\ref{B2}) can
be rewritten as
\begin{eqnarray}\label{B3}
\zeta= \frac{H^2}{2\pi\dot{\phi}}\,.
\end{eqnarray}
Since the evolution path of the universe is unique, the curvature
perturbation does not evolve on super-horizon scales and hence it
is useful to write the perturbation amplitude in terms of the
dimensionless curvature perturbation, $\zeta$. Stability of
$\zeta$ makes the analysis of the density perturbation during
inflation with a single field model easier.

It is important to note that the statistical properties of $\zeta$
are constrained by observations. Moreover, these properties are
commonly measured in terms of $\zeta$'s power spectrum,
bispectrum, and trispectrum. As mentioned previously, various
types of non-Gaussianities are proposed, but in this paper we
focus on the so-called \emph{local-type non-Gaussianity} which is
specified by the existence of a one-to-one local map between the
physical curvature perturbation, $\zeta(x)$, and the variable
which follows the Gaussian statistics, $\zeta_G(x)$, at respective
spatial points. In other words, the curvature perturbation can be
expanded as
\begin{eqnarray}\label{B4}
\zeta(x)=\zeta_{G}(x)+\frac{3}{5}f_{NL}\zeta^2_G(x)+
\frac{9}{25}g_{NL}\zeta^3_G(x)+...\,,\hspace{0.8 cm}
\end{eqnarray}
and the two-point correlation function can be determined by the
power spectrum ${P}_{\zeta}$ as
\begin{eqnarray}\label{B5}
\langle \zeta_{\textbf{k}_1}\zeta_{\textbf{k}_2} \rangle\equiv
(2\pi)^3 \delta^3(\textbf{k}_1+\textbf{k}_2){P}_{\zeta}(k_1)\,.
\end{eqnarray}
The three point correlation function can be defined by the
following expression
\begin{eqnarray}\label{B6}
\langle
\zeta_{\textbf{k}_1}\zeta_{\textbf{k}_2}\zeta_{\textbf{k}_3}
\rangle\equiv(2\pi)^3\delta^3(\textbf{k}_1+\textbf{k}_2+\textbf{k}_3)
{\cal{B}}_{\zeta}(k_1,k_2,k_3)\,,\hspace{0.8 cm}
\end{eqnarray}
where ${\cal{B}}_{\zeta}$ is the bispectrum given by
\begin{eqnarray}\label{B7}
{\cal{B}}_{\zeta}(k_1,k_2,k_3)=\hspace{5.5cm}\\\nonumber
\frac{6}{5}f_{NL} [{P}_{\zeta}(k_1){P}_{\zeta}(k_2)+
{P}_{\zeta}(k_2){P}_{\zeta}(k_3)+
{P}_{\zeta}(k_3){P}_{\zeta}(k_1)]\,,
\end{eqnarray}
with $f_{NL}$ being the non-linear parameter \citep{NGa}. We
emphasize that, although other types of non-Gaussianities can be
generated, but focusing on super-horizon scales only leads to the
local-type non-Gaussianity which is the aim of this study. To
proceed further, we use the $\delta N$ formalism which is based on
the separate universe assumption \citep{delN2,delN3,delN4,delN6}.
This formalism provides a powerful tool to evaluate the evolution
of the curvature perturbation on super-horizon scales. Since on
scales larger than the horizon spatial gradients can be neglected,
each spatial point can be considered to evolve as a separate
Friedmann-Robertson-Walker universe. To put it simply,
super-horizon dynamics is locally characterized by the FRW
universe. We choose an initially flat hypersurface at $t=t_{*}$
(when the observational scales crossed the cosmological horizon)
and a later uniform energy density hypersurface at $t=t_{e}$
(which corresponds to the time of the end of inflation). In fact
we evolve the space-time until reaching the final surface at
$t=t_e$ from the initial surface at $t=t_*$. After this
consideration, the number of e-folds can be written as a function
of the initial and final time, $t_{*}$ and $t_{e}$, on the
relevant hypersurfaces in the following form
\begin{eqnarray}\label{B8}
N(t_e,t_*,x)=\int _{t_*}^{t_e} dt H(t,x)\,,
\end{eqnarray}
and the primordial curvature perturbation on the final
hypersurface can be expressed as
\begin{eqnarray}\label{B9}
\zeta(t_e,x)\equiv\delta N(t_e,x)=N(t_e,t_*,x)-N_0(t_e,t_*)\,,
\end{eqnarray}
which is the heart of the $\delta N$ formalism and
\begin{eqnarray}\label{B10}
N_0(t_e,t_*)=\int _{t_*}^{t_e} dt H_0(t,x)\,.
\end{eqnarray}
Since in an expanding universe each horizon patch is causally
disconnected from the others, its evolution is defined as a local
process.

By considering $t_*$ to be the time of horizon exit ($kc_s = a
H$), one can write the curvature perturbation on super-horizon
scales in terms of partial derivatives of $N$ with respect to the
initial field values on the flat slicing (unperturbed scalar field
values at horizon exit). In fact, during the slow-roll inflation,
evolution of the universe is assumed to be specified by one scalar
field or more. Choosing the flat slicing gauge and considering
perturbations, one can expand each scalar field to a homogeneous
background and local perturbation,
$\phi_i(t_*,x)=\phi_i(t_*)+\delta\phi_i(t_*,x)$. Therefore $\zeta$
can be expanded in powers of $\delta\phi$
\begin{eqnarray}\label{B11}
\zeta(t_e,x)=\sum_{i}N_{,i}\delta\phi^i+\hspace{4cm}\\\nonumber
\sum_{ij}\frac{1}{2}N_{,ij}\delta\phi^i\delta\phi^j
+\sum_{ijk}\frac{1}{6}N_{,ijk}\delta\phi^i\delta\phi^j\delta\phi^k+...
\end{eqnarray}
where $N$ is the number of e-folds from $t_*$ to $t_e$ and
$N_{,i}=\frac{\partial{N}}{\partial{\phi_i}}$ and
$\delta{\phi}_i$'s are the field fluctuations on the initial flat
hypersurface shortly after horizon exit, $t_*$. We note that this
is while the initial and final hypersurfaces are remained
constant. Furthermore, we emphasize that the inflaton field
velocities are functions of the field position,
$\dot{\phi}(t_*,x)$. Subsequently, even when the evolution runs
away from slow-roll, the number of e-folds between the initial
hypersurface and the final one is a function of the initial field
values on initial flat hypersurface, $N(t_e,\phi(t_*,x))$.

Now employing the $\delta N$ expansion and using the two-point
function (\ref{B5}), one finds the dimensionless power spectrum as
\citep{delN2}
\begin{eqnarray}\label{B12}
{\cal{P}}_{\zeta}(k_1)=N_{,i}^2{\cal{P}}_{*}\,,
\end{eqnarray}
where ${\cal{P}}_{\zeta}=\frac{k^3}{2\pi^2}P_{\zeta}$ and
${\cal{P}}_{*}=\left(\frac{H_{*}}{2\pi}\right)^2$. Now we proceed
to the lowest order non-Gaussianity which is the three-point
correlation function, or its Fourier transform, the bispectrum,
${\cal{B}}_{\zeta}$. Using the $\delta N$ formalism, equation
(\ref{B6}) can be written in terms of the derivatives of $N$ as
\citep{NGn}
\begin{eqnarray}\label{B13}
\langle
\zeta_{\textbf{k}_1}\zeta_{\textbf{k}_2}\zeta_{\textbf{k}_3}
\rangle=\sum_{ijk}N_{,i}N_{,j}N_{,k}
\langle\delta\phi_{{\textbf{k}}_1}^i \delta\phi_{{\textbf{k}}_2}^j
\delta\phi_{{\textbf{k}}_3}^k\rangle+\hspace{1.2cm}\\\nonumber
\left(\frac{1}{2}\sum_{ijkl}N_{,i}N_{,j}N_{,kl}\langle
{\delta\phi_{{\textbf{k}}_1}^i
\delta\phi_{{\textbf{k}}_2}^j(\delta\phi^k \star
\delta\phi^l)_{\textbf{k}_3}}\rangle+ 2\,
\textrm{perms.}\right),
\end{eqnarray}
where $\star$ denotes a convolution and ?perms? denotes cyclic
permutations over the momenta. This expansion of the three-point
function, (\ref{B13}), demonstrates two distinct contributions.
Its first term contributes to non-Gaussianity in the primordial
curvature perturbation through the inherent non-Gaussianity of
$\delta{\phi}^i$, which is generated by quantum field interactions
on sub-horizon scales. Although for Gaussian perturbations, this
term vanishes identically, however, $\zeta$ can still be
non-Gaussian through the second term. This second term is due to
the non-linear behavior in curvature perturbation on super-horizon
scales and is mostly referred to as \emph{local-type
non-Gaussianity} (one can refer to \citep{Bab04} for a discussion
of the shape dependence of the bispectrum). Therefore, we mainly
focus on this contribution which is toward our aim in this paper.

Now let us neglect the connected part of the four-point function
and apply Wick?s theorem which helps us to rewrite the four-point
correlation functions in terms of two-point functions. Then the
second contribution in (\ref{B13}) can be written as
\citep{NGn,NGu}
\begin{eqnarray}\label{B14}
\frac{1}{2}\sum_{ijkl}N_{,i}N_{,j}N_{,kl}\langle
{\delta\phi_{{\textbf{k}}_1}^i
\delta\phi_{{\textbf{k}}_2}^j(\delta\phi^k \star
\delta\phi^l)_{\textbf{k}_3}}\rangle+ 2\,\,
\textrm{perms.}\hspace{0.7cm}\\\nonumber
=(2\pi)^{3}4\pi^{4}{\cal{P}}_{*}^2\frac{\sum_{i}{k_i}^3}
{\Pi_{i}{k_i}^3}\sum_{ij}N_{,i}N_{,j}N_{,ij}
\delta^3(\textbf{k}_1+\textbf{k}_2+\textbf{k}_3)\,.
\end{eqnarray}
Adopting the notation of \citep{Ver06}, the bispectrum can be
written as follows
\begin{eqnarray}\label{B15}
{\cal{B}}_{\zeta}(k_1,k_2,k_3)=4\pi^4{\cal{P}}_{\zeta}^2
\frac{\sum_{i}{k_i}^3}{\Pi_{i}{k_i}^3}
\left(\frac{6}{5}f_{NL}(k_1,k_2,k_3)\right)\,,\hspace{0.5cm}
\end{eqnarray}
where
\begin{eqnarray}\label{B16}
f_{NL}(k_1,k_2,k_3)=f_{NL}^{(3)}(k_1,k_2,k_3)+f_{NL}^{(4)}\,.
\end{eqnarray}
Here $f^{(3)}_{NL}$ is the momentum dependent parameter which
accounts for the sub-horizon contribution. However, since we are
primarily interested in local non-Gaussianity which is confirmed
on super-horizon scales, we skip over this term. Whilst
$f^{(4)}_{NL}$ is momentum independent parameter and accounts for
the super-horizon contribution (one can refer to
\citep{Ken15,Ken16} for more discussions).

One of our aim in this paper is to calculate $f^{(4)}_{NL}$ for
the case with a non-minimally coupled scalar field. This quantity
is generally described by the following expression
\begin{eqnarray}\label{B17}
\frac{6}{5}f^{(4)}_{NL}=\frac{\sum_{ij}N_{,i}N_{,j}N_{,ij}}{\left(\sum_k
N_{,k}^2\right)^2}\,.
\end{eqnarray}
Now we describe the leading order contributions to the primordial
trispectrum in a perturbative expansion for the local shape of
non-Gaussianity using the $\delta N$ formalism. To this end, we
derive the four-point correlation function of the field
fluctuations which is given by
\begin{eqnarray}\label{B18}
\langle
\zeta_{\textbf{k}_1}\zeta_{\textbf{k}_2}\zeta_{\textbf{k}_3}
\zeta_{\textbf{k}_4}\rangle\equiv \hspace{4.5cm} \\\nonumber
(2\pi)^3\delta^3
(\textbf{k}_1+\textbf{k}_2+\textbf{k}_3+\textbf{k}_4)
{\cal{T}}_{\zeta}(k_1,k_2,k_3,k_4)\,,
\end{eqnarray}
where
\begin{eqnarray}\label{B19}
{\cal{T}}_{\zeta}(k_1,k_2,k_3,k_4)=\hspace{5cm}\\\nonumber
\tau_{NL}[{P}_{\zeta}(k_{13}){P}_{\zeta}
(k_3){P}_{\zeta}(k_4)+(11\, \textrm{perms.})]+\\\nonumber
\frac{54}{25} g_{NL} [{P}_{\zeta}(k_2){P}_{\zeta}(k_3)
{P}_{\zeta}(k_4)+(3\, \textrm{perms.})]\,\,.
\end{eqnarray}

Now, as performed previously for the bispectrum, using the
expression for the curvature perturbation (\ref{B11}) one can
obtain these new observational parameters, $\tau_{NL}$ and
$g_{NL}$, in terms of the derivatives of $N$. The trispectrum of
the primordial curvature perturbation (\ref{B19}) at the leading
order can be written as
\begin{eqnarray}\label{B20}
{\cal{T}}_{\zeta}(k_1,k_2,k_3,k_4)=\hspace{5cm}\\\nonumber
\sum_{ijk}N_{,ij}N_{,ik}N_{,j}N_{,k} [{P}(k_{13})
{P}(k_3){P}(k_4)+11\,\, \textrm{perms.}]+\\\nonumber
\sum_{ijk}N_{,ijk}N_{,i}N_{,j}N_{,k}[{P}(k_{2}) {P}(k_3){P}(k_4)+
3\,\,\textrm{perms.}]\,.\hspace{0.25cm}
\end{eqnarray}
By comparing (\ref{B19}) and (\ref{B20}) and recalling (\ref{B12})
we have
\begin{eqnarray}\label{B21}
\tau_{NL}^{local}=\frac{\sum_{ijk}N_{,ij}N_{,ik}N_{,j}
N_{,k}}{\sum_{l}(N_{,l}N_{,l})^3}\,,
\end{eqnarray}
\begin{eqnarray}\label{B22}
g_{NL}^{local}=\frac{25}{54}\frac{\sum_{ijk}N_{,ijk}N_{,i}
N_{,j}N_{,k}}{\sum_{l}(N_{,l}N_{,l})^3}\,.
\end{eqnarray}

We note that in order to check the local type non-Gaussianity in
the trispectrum, terms with different momentum dependence are
factored out and only two momentum independent non-linearity
parameters, $\tau_{NL}^{local}$ and $g_{NL}^{local}$, are defined.
This can give the facility that observational data may be able to
identify between the two parameters \citep{tris1}. Up to this point
we introduced the $\delta N$ formalism in order to calculate the
non-linear parameters $f_{NL}^{local}$, $\tau_{NL}^{local}$ and
$g_{NL}^{local}$. In the next section we present the results for
the case with a non-minimally coupled scalar field.

\section{Primordial Non-Gaussianity in a Non-Minimal Inflation}

Up to this point, by considering an inflationary scenario with an
inflaton field non-minimally coupled to gravity, we obtained the
basic equations. Then we have presented the general form of the
non-linear parameters in the local type non-Gaussianities. Now we
are in the position to calculate the local type non-Gaussianities
of the primordial fluctuations in our non-minimal setup.

Using (\ref{B11}) the primordial curvature perturbation takes the
following form in our model
\begin{eqnarray}\label{c1}
\zeta(t_e,x)=N_{,\phi}\delta\phi+\frac{1}{2}N_{,\phi\phi}(\delta\phi)^2
+\frac{1}{6}N_{,\phi\phi\phi}(\delta\phi)^3\,.\hspace{0.8cm}
\end{eqnarray}
As mentioned previously, in order to study local type
non-Gaussianity it is enough to focus on super-horizon scales and
in this region $f_{NL}^{local}$ is defined by (\ref{B17}) which
takes the following form in our model
\begin{eqnarray}\label{c2}
\frac{6}{5}f_{NL}^{local}=\frac{N_{,\phi\phi}}{(N_{,\phi})^2}\,.
\end{eqnarray}
It is clear that since there is only one field and one
$\delta\phi_i$, we have rewritten $N_{,i}$ as $N_{,\phi}$ and
$N_{,ii}$ as $N_{\phi\phi}$ for simplicity. According to the
equation (\ref{A7}), we obtain the following relation for the
first derivative of the number of e-folds with respect to the
inflaton field
\begin{eqnarray}\label{c3}
N_{,\phi}=-\kappa^2\frac{\left(V-2f_{,\phi}^2R+2f_{,\phi}
V_{,\phi}\right)}{(1+2\kappa^2f)(V_{,\phi}-f_{,\phi}R)}\,.
\end{eqnarray}
Next, the second derivative of $N$ gives
\begin{eqnarray}\label{c4}
N_{,\phi\phi}=\kappa^2\left(\frac{2\kappa^2f_{,\phi}
(V-2f_{,\phi}^2R+2f_{,\phi}V_{,\phi})}{(1+2\kappa^2
f)^2(V_{,\phi}-f_{,\phi}R)}\right)-\hspace{1cm}\\\nonumber
\kappa^2\left(\frac{V_{,\phi}-4f_{,\phi}f_{,\phi\phi}R+
2f_{,\phi\phi}V_{,\phi}+2f_{,\phi}V_{,\phi\phi}}{(1+2\kappa^2
f)(V_{,\phi}-f_{,\phi}R)}\right)+\\\nonumber
\kappa^2\left(\frac{(V-2f_{,\phi}^2R+2f_{,\phi}V_{,\phi})
(V_{,\phi\phi}-f_{,\phi\phi}R)}{(1+2\kappa^2
f)(V_{,\phi}-f_{,\phi}R)^2}\right)\,.
\end{eqnarray}
Applying (\ref{c3}) and (\ref{c4}) in the main equation for
$f_{NL}^{local}$, that is, (\ref{c2}), we obtain our first
non-linear parameter in the local configuration as
\begin{eqnarray}\label{c5}
\frac{6}{5}f_{NL}^{local}=\frac{(1+2\kappa^2f)(V_{,\phi}-f_{,\phi}R)}
{\kappa^2(V-2f_{,\phi}^2R+2f_{,\phi}V_{,\phi})}
\times\hspace{1.5cm}\\\nonumber\Bigg[\frac{4f_{,\phi}
f_{,\phi\phi}R-V_{,\phi}-2f_{,\phi\phi}V_{,\phi}-2f_{,\phi}V_{,\phi\phi}}
{V-2f_{,\phi}^2R+
2f_{,\phi}V_{,\phi}}+\\\nonumber\frac{2\kappa^2f_{,\phi}}{1+2\kappa^2
f}+\frac{V_{,\phi\phi}-f_{,\phi\phi}R}{V_{,\phi}-f_{,\phi}R}\Bigg]\,.
\end{eqnarray}
Then we calculate the trispectrum of the model. According to
equations (\ref{B21}) we have
\begin{eqnarray}\label{tau1}
\tau_{NL}^{local}=\frac{(N_{,\phi\phi})^2}{(N_{,\phi})^4}\,,
\end{eqnarray}
which leads to the following relation in our model
\begin{eqnarray}\label{tau1}
\tau_{NL}^{local}=\frac{(1+2\kappa^2f)^2(V_{,\phi}-f_{,\phi}R)^2}{\kappa^4(V-2f_{,\phi}^2R+2f_{,\phi}
V_{,\phi})^4}\Bigg[4f_{,\phi}f_{,\phi\phi}R-\hspace{0.7cm}\\\nonumber
V_{,\phi}-2f_{,\phi\phi}V_{,\phi}-2f_{,\phi}
V_{,\phi\phi}+(V-2f_{,\phi}^2R+2f_{,\phi}
V_{,\phi})\\\nonumber\times\left(\frac{2\kappa^2f_{,\phi}}{1+2\kappa^2f}+
\frac{V_{,\phi\phi}-f_{,\phi\phi}R}{V_{,\phi}-f_{,\phi}R}\right)\Bigg]^2\,.
\end{eqnarray}
Also equation (\ref{B22}) results the following form for
$g_{NL}^{local}$
\begin{eqnarray}\label{c6}
\frac{54}{25}g_{NL}^{local}=\frac{N_{,\phi\phi\phi}}{(N_{,\phi})^3}\,.
\end{eqnarray}
Thus, in order to obtain $g_{NL}^{local}$, we should first obtain
the third derivative of $N$ which takes the following expression

\begin{widetext}
\begin{eqnarray}\label{c7}
N_{,\phi\phi\phi}=\Bigg[V_{,\phi\phi}-4f_{,\phi\phi}^2R-4f_{,\phi}
f_{,\phi\phi\phi}R+2f_{,\phi\phi\phi}V_{,\phi}+4f_{,\phi\phi}
V_{,\phi\phi}+2f_{,\phi}V_{,\phi\phi\phi}-\left
(\frac{2\kappa^2f_{,\phi}}{1+2\kappa^2f}+\frac{V_{,\phi\phi}-
f_{,\phi\phi}R}{V_{,\phi}-f_{,\phi}R}\right)
\times\hspace{0.8cm}\\\nonumber\left(V_{,\phi}-4f_{,\phi}f_{,\phi\phi}R+
2f_{,\phi\phi}V_{,\phi}+ 2f_{,\phi}V_{,\phi\phi}\right) -\left
(\frac{2\kappa^2f_{,\phi\phi}}{1+2\kappa^2
f}-\frac{4\kappa^4f_{,\phi}^2}{(1+2\kappa^2
f)^2}+\frac{V_{,\phi\phi\phi}-f_{,\phi\phi\phi}R}{V_{,\phi}-
f_{,\phi}R}-\frac{(V_{,\phi\phi}-f_{,\phi\phi}R)^2}
{(V_{,\phi}-f_{,\phi}R)^2}\right)\times\\\nonumber
\left(V-2f_{,\phi}^2R+2f_{,\phi}V_{,\phi}\right)\Bigg]
\left[\frac{-\kappa^2}{(1+2\kappa^2f)(V_{,\phi}-f_{,\phi}R)}\right]
+\left[\frac{\kappa^2}{(1+2\kappa^2f)
(V_{,\phi}-f_{,\phi}R)}\left(\frac{2\kappa^2f_{,\phi}}
{1+2\kappa^2f}+\frac{V_{,\phi\phi}-f_{,\phi\phi}R}
{V_{,\phi}-f_{,\phi}R}\right)\right]\times\\\nonumber
\left[V_{,\phi}-4f_{,\phi}f_{,\phi\phi}R+2f_{,\phi\phi}V_{,\phi}+
2f_{,\phi}V_{,\phi\phi}-\left(\frac{2\kappa^2f_{,\phi}}
{1+2\kappa^2f}+\frac{V_{,\phi\phi}-f_{,\phi\phi}R}{V_{,\phi}-
f_{,\phi}R}\right)(V-2f_{,\phi}^2R+2f_{,\phi}V_{,\phi})\right]\,.\,
\end{eqnarray}
\end{widetext}
Finally, using (\ref{c6}), the trispectrum non-linear parameter in
a non-minimal inflation reads
\begin{widetext}
\begin{eqnarray}\label{c8}
\frac{54}{25}g_{NL}^{local}=\Bigg[V_{,\phi\phi}-4f_{,\phi\phi}^2R-4f_{,\phi}f_{,\phi\phi\phi}R
+2f_{,\phi\phi\phi}V_{,\phi}+4f_{,\phi\phi}V_{,\phi\phi}+2f_{,\phi}V_{,\phi\phi\phi}-
\hspace{3cm}\\\nonumber\left(V_{,\phi}-4f_{,\phi}f_{,\phi\phi}R+2f_{,\phi\phi}V_{,\phi}+
2f_{,\phi}V_{,\phi\phi}\right)\left(\frac{2\kappa^2f_{,\phi}}{1+2\kappa^2f}+\frac{V_{,\phi\phi}
-f_{,\phi\phi}R}{V_{,\phi}-f_{,\phi}R}\right)-\left(V-2f_{,\phi}^2R+2f_{,\phi}
V_{,\phi}\right)\times\\\nonumber\left(\frac{2\kappa^2f_{,\phi\phi}}{1+2\kappa^2
f}-\frac{4\kappa^4f_{,\phi}^2}{(1+2\kappa^2f)^2}+\frac{V_{,\phi\phi\phi}
-f_{,\phi\phi\phi}R}{V_{,\phi}-f_{,\phi}R}-\frac{(V_{,\phi\phi}-f_{,\phi\phi}R)^2}
{(V_{,\phi}-f_{,\phi}R)^2}\right)\Bigg]
\left[\frac{(1+2\kappa^2f)^2(V_{,\phi}-f_{,\phi}R)^2}{\kappa^4(V-2f_{,\phi}^2R+
2f_{,\phi}V_{,\phi})^3}\right]+\\\nonumber\left[V_{,\phi}-4f_{,\phi}f_{,\phi\phi}R+2f_{,\phi\phi}
V_{,\phi}+2f_{,\phi}V_{,\phi\phi}-\left(\frac{2\kappa^2f_{,\phi}}
{1+2\kappa^2f}+\frac{V_{,\phi\phi}-f_{,\phi\phi}R}{V_{,\phi}-f_{,\phi}R}\right)
(V-2f_{,\phi}^2R+2f_{,\phi}V_{,\phi})\right] \times\\\nonumber
\left(\frac{2\kappa^2f_{,\phi}}{1+2\kappa^2f}+\frac{V_{,\phi\phi}-
f_{,\phi\phi}R}{V_{,\phi}-f_{,\phi}R}\right)\left
[-\frac{(1+2\kappa^2f)^2(V_{,\phi}-f_{,\phi}R)^2}
{\kappa^4(V-2f_{,\phi}^2R+2f_{,\phi}V_{,\phi})^3}\right]\,.
\end{eqnarray}
\end{widetext}

Up to here we have calculated the main equations of the setup
toward analyzing both bispectrum and trispectrum of the local-type
non-Gaussianity. Now we are in the position to proceed further by
specifying the form of the potential, $V(\phi)$, and the
non-minimal coupling function, $f(\phi)$. In this regard we
consider a quadratic form for these functions of the scalar field
as, $V(\phi)=\frac{1}{2}m_{\phi}^{2}\phi^2$ and
$f(\phi)=\frac{1}{2}\xi\phi^2$. Adopting these functions in main
equations of section (II) leads to the following number of e-folds
in our scenario
\begin{eqnarray}\label{c9}
N(\phi)=-\int_{\phi}^{\phi_{e}}\frac{\kappa^2
(\frac{1}{2}m_{\phi}^2\phi^2-2\xi^2\phi^2R+2\xi\phi^2m_{\phi}^2)}
{(m_{\phi}^2\phi-\xi\phi
R)(1+\kappa^2\xi\phi^2)}d\phi\,,\hspace{0.6cm}
\end{eqnarray}
which gives
\begin{eqnarray}\label{c10}
N(\phi)=\left(\frac{m_{\phi}^2-4\xi^2 R+4\xi m_{\phi}^2}{4\xi(\xi
R
-m_{\phi}^2)}\right)\ln\left(\frac{1+\kappa^2\xi\phi_{e}^2}{1+\kappa^2\xi\phi^2}\right)\,,\hspace{0.5cm}
\end{eqnarray}
Through Eq. (\ref{c10}) we can obtain the value of the field
during the slow-roll inflation in terms of the number of e-folds
\begin{eqnarray}\label{c11}
\phi(N)=\frac{\left(-1+(1+\kappa^2\xi\phi_{e}^2)e^{\frac{4N\left(-m_{\phi}^2+\xi
R\right)\xi}{4\xi^2
R-4m_{\phi}^2\xi-m_{\phi}^2}}\right)^{\frac{1}{2}}}{\xi^{\frac{1}{2}}\kappa}\,.\hspace{0.5cm}
\end{eqnarray}
Moreover, adopting the mentioned forms of $V(\phi)$ and $f(\phi)$,
equation (\ref{c5}) results the following expression for the local
non-linear parameter, $f_{NL}^{local}$
\begin{eqnarray}\label{c12}
\frac{6}{5}f_{NL}^{local}=\Bigg[\frac{m_{\phi}^2\phi-4\xi^2\phi R+4\xi
m_{\phi}^2 \phi}{\frac{1}{2}m_{\phi}^2\phi^2-2\xi^2\phi^2
 R+2\xi\phi^2m_{\phi}^2}+\hspace{1.5cm}\\\nonumber\frac{-m_{\phi}^2+\xi R}
 {m_{\phi}^2\phi-\xi\phi R}-\frac{2\kappa^2\xi\phi}{
1+\kappa^2\xi\phi^2}\Bigg]\big(1+\kappa^2\xi\phi^2\big)
\big(m_{\phi}^2\phi-\xi\phi R\big)\\\nonumber
\times\left[\kappa^2\left(\frac{1}{2}m_{\phi}^2\phi^2-
2\xi^2\phi^2 R+2\xi\phi^2m_{\phi}^2\right)\right]^{-1}\,.
\end{eqnarray}
Furthermore, the local non-linear parameters associated to the
trispectrum of the model (\ref{tau1}) and (\ref{c8}), finally are
given by
\begin{eqnarray}\label{tau2}
\tau_{NL}^{local}=\Bigg(\frac{m_{\phi}^2\phi-4\xi^2\phi R+4\xi
m_{\phi}^2\phi}{\frac{1}{2}m_{\phi}^2\phi^2-2\xi^2\phi^2
R+2\xi\phi^2 m_{\phi}^2}+\hspace{2cm}\\\nonumber
\frac{-m_{\phi}^2+\xi R}{m_{\phi}^2\phi-\xi\phi R}-\frac{2
\kappa^2\xi\phi}{1+\kappa^2\xi\phi^2}\Bigg)^2
(m_{\phi}^2\phi-\xi\phi R)^2\times\hspace{0.7cm}
\\\nonumber(1+\kappa^2\xi\phi^2)^2\left[\kappa^2\left(
\frac{1}{2}m_{\phi}^2\phi^2-2\xi^2\phi^2 R+2\xi\phi^2
m^2\right)\right]^{-2}\,,
\end{eqnarray}
and
\begin{widetext}
\begin{eqnarray}\label{c13}
\frac{54}{25}g_{NL}^{local}= \Bigg[m_{\phi}^2-4\xi^2 R+4\xi
m_{\phi}^2-\left(m_{\phi}^2\phi-4\xi^2\phi R+4\xi
m_{\phi}^2\phi\right)
\left(\frac{2\kappa^2\xi\phi}{1+\kappa^2\xi\phi^2}+\frac{m_{\phi}^2-\xi
R}{m_{\phi}^2\phi-\xi\phi R}\right)-\hspace{1.5cm}\\\nonumber
\left(\frac{1}{2}m_{\phi}^2\phi^2-2\xi^2\phi^2 R+2\xi\phi^2
m_{\phi}^2\right)\left(\frac{2\kappa^2\xi}{1+\kappa^2\xi\phi^2}-
\frac{4\kappa^4\xi^2\phi^2}{(1+k^2\xi\phi^2)^2}-\frac{(m_{\phi}^2-\xi
R)^2}{(m_{\phi}^2\phi-\xi\phi R)^2}\right)\Bigg]
\big(1+\kappa^2\xi\phi^2\big)^2\big(m_{\phi}^2\phi-\xi\phi
R\big)^2\\\nonumber\times\left[\kappa^4\left(\frac{1}{2}
m_{\phi}^2\phi^2-2\xi^2\phi^2
R+2\xi\phi^2m_{\phi}^2\right)^2\right]^{-1}-\kappa^8
\left(\frac{1}{2}m_{\phi}^2\phi^2-2\xi^2\phi^2
R+2\xi\phi^2m_{\phi}^2\right)^3\left(\frac{2\kappa^2\xi\phi}{
1+\kappa^2\xi\phi^2}+\frac{m_{\phi}^2-\xi
R}{m_{\phi}^2\phi-\xi\phi R}\right)\\\nonumber
\times\left[m_{\phi}^2\phi-4\xi^2\phi R+4\xi m_{\phi}^2\phi-
\left(\frac{2\kappa^2\xi\phi}{1+\kappa^2\xi\phi^2}+\frac{m_{\phi}^2-\xi
R}{m_{\phi}^2\phi-\xi\phi R}\right)\left
(\frac{1}{2}m_{\phi}^2\phi^2-2\xi^2\phi^2 R+2\xi\phi^2
m_{\phi}^2\right)\right]\\\nonumber \times\Bigg[(1+\kappa^2\xi
\phi^2)(m_{\phi}^2\phi-\xi\phi R)\Bigg]^{-4}\,.
\end{eqnarray}
\end{widetext}

Now one can easily substitute the field value (\ref{c11}) in
(\ref{c12}), (\ref{tau2}) and (\ref{c13}) to obtain the intended
non-linear parameters as a function of $N$ (which imply the level
of non-Gaussianity in both bispectrum and trispectrum to be
depended on the non-minimal coupling strength and the number of
e-folds). In what follows we perform some numerical analysis on
the model's parameters space and draw the relevant figures for
$f_{NL}^{local}$, $\tau_{NL}^{local}$ and ${g_{NL}^{local}}$ to
see the behaviors of these quantities intuitively. Furthermore, we
eventually test our results in the light of Planck observational
data.

\section{Testing the evolution of the non-linear parameters
in confrontation with observation}

In previous section we obtained the non-linear parameters
corresponding to both the first and higher order of
non-Gaussianity. Eq. (\ref{c12}) shows how the first level of
non-Gaussianity, $f_{NL}^{local}$, depends on the inflaton field
and non-minimal coupling parameter. Since the scalar field $\phi$
is a function of the number of e-folds (through equation
(\ref{c11})), we can depict the evolution of $f_{NL}^{local}$ in
$\xi$ and $N$ space. In this regard, we should obtain the value of
the inflaton at the end of inflation, $\phi_{end}$.

\begin{figure*}[ht!]
\begin{center}
\scalebox{0.55}[0.55]{\includegraphics{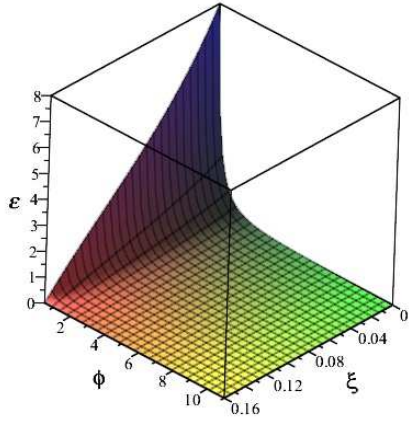}}
\caption{The behavior of $\epsilon$ versus
$\phi$ and $\xi$.}
\label{fig:epsilon}
\end{center}
\end{figure*}

\begin{figure*}[htp]
\flushleft\leftskip0em{
\includegraphics[width=.46\textwidth,origin=c,angle=0]{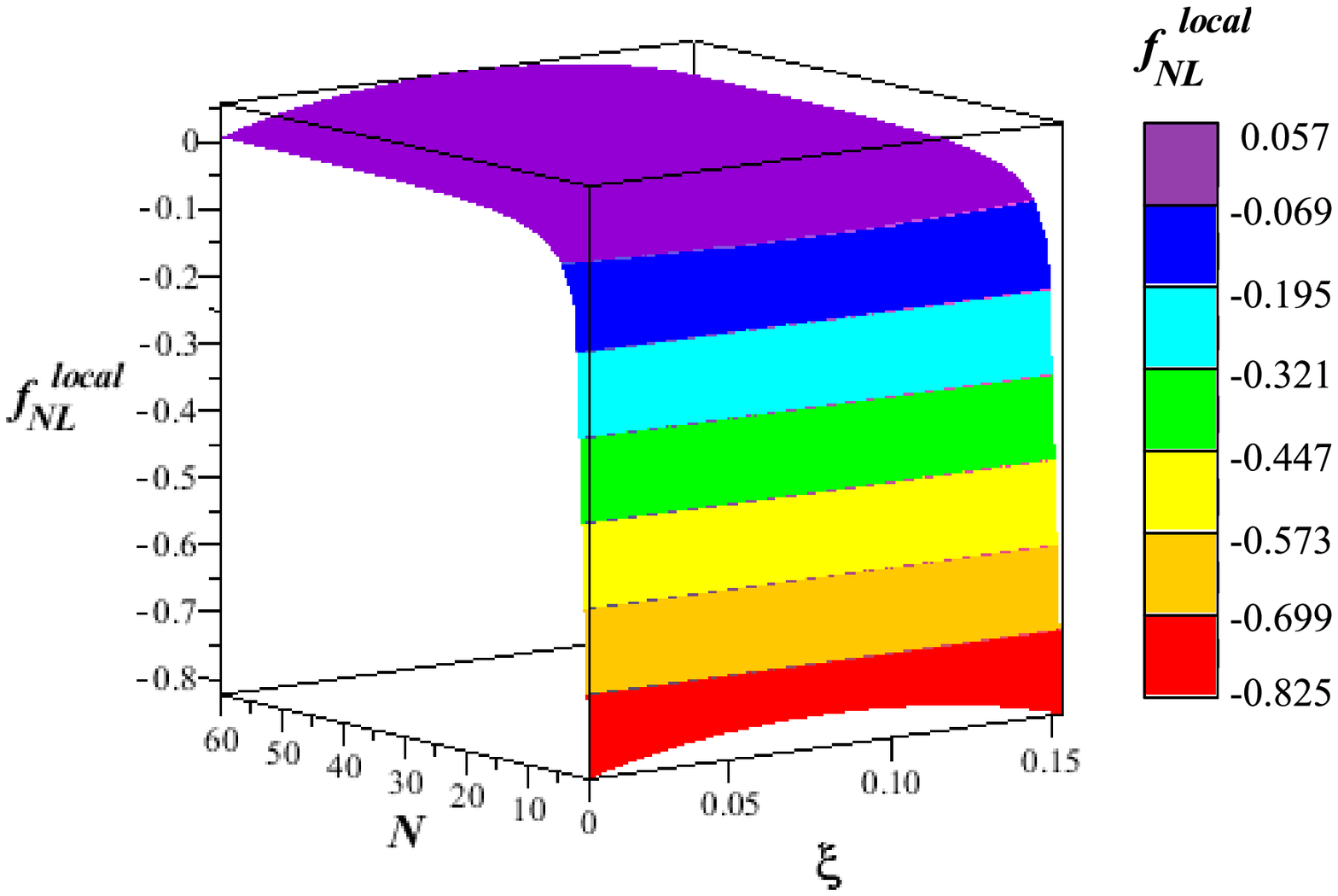}
\hspace{1cm}
\includegraphics[width=.42\textwidth,origin=c,angle=0]{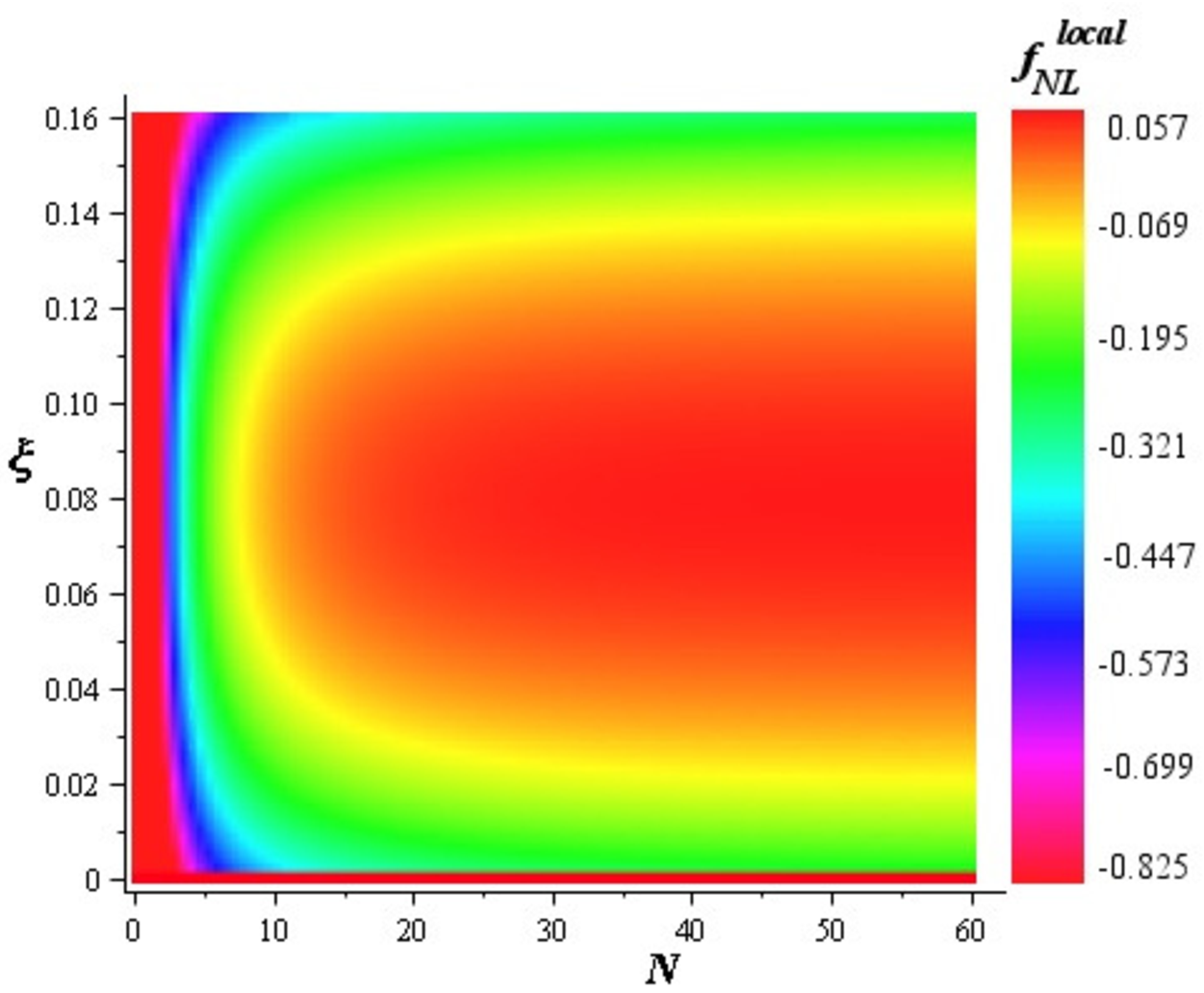}}
\caption{\label{fig:1} The non-linear parameter $f_{NL}^{local}$
as a function of the non-minimal coupling parameter $\xi$ and the
number of e-folds $N$.}
\end{figure*}

\begin{figure*}[htp]
\flushleft\leftskip0em{
\includegraphics[width=.45\textwidth,origin=c,angle=0]{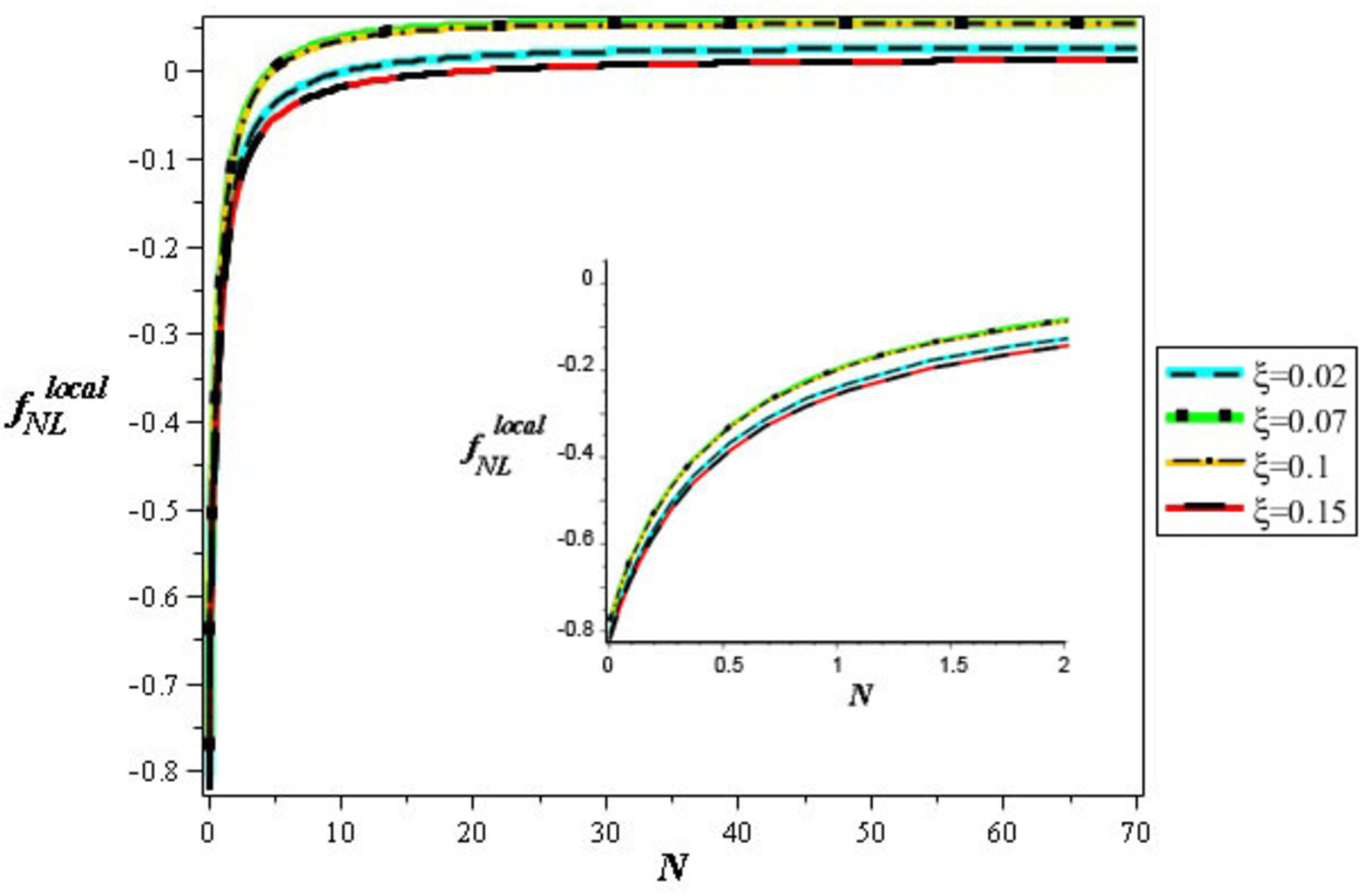}
\hspace{1cm}
\includegraphics[width=.40\textwidth,origin=c,angle=0]{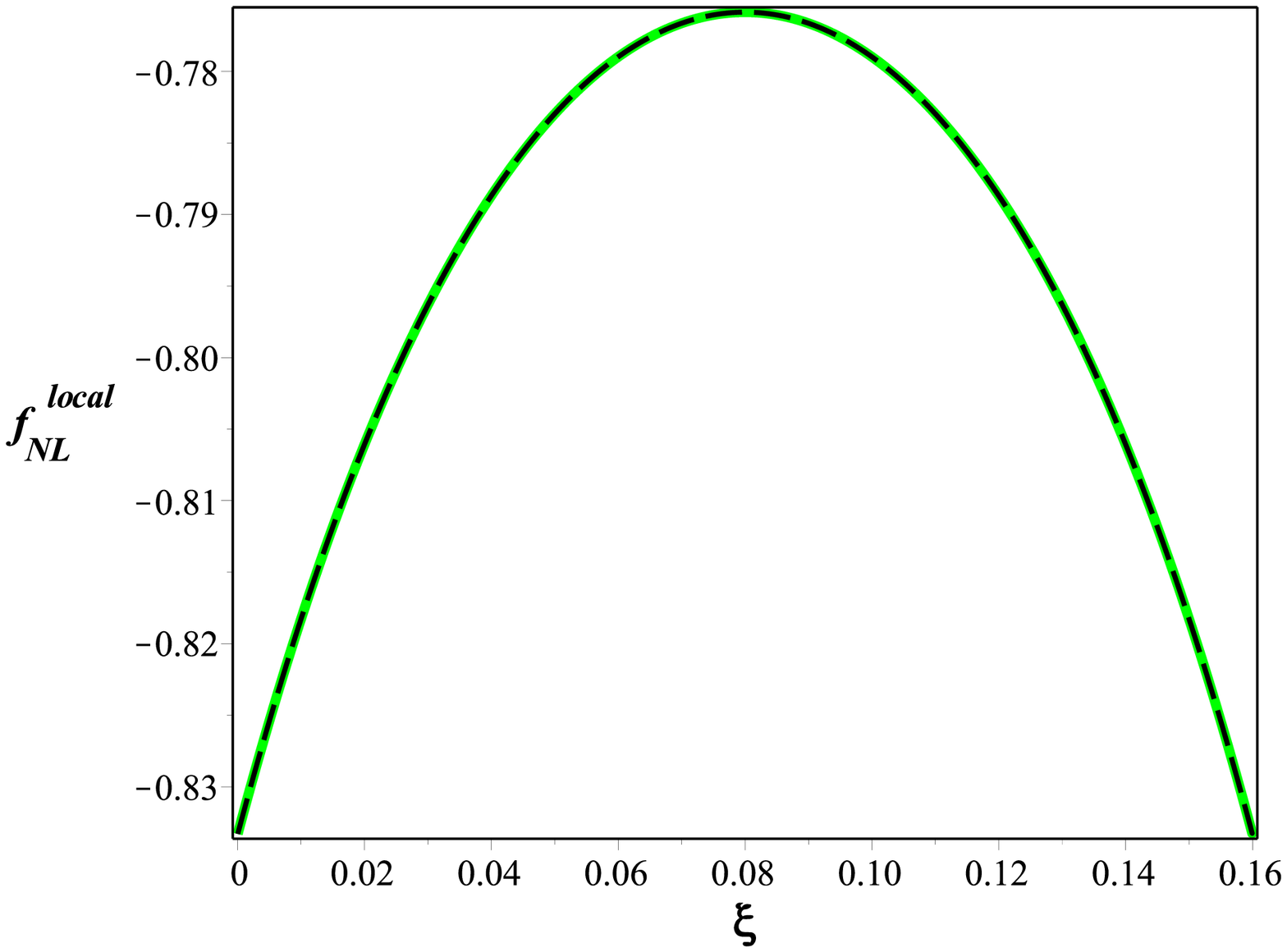}}
\caption{\label{fig:2} Left Panel: The non-linear parameter
$f_{NL}^{local}$ for the local type non-Gaussianity versus the
number of e-folds for various values of the non-minimal coupling
parameter. Right Panel: The non-linear parameter $f_{NL}^{local}$
for the local type non-Gaussianity as a function of the
non-minimal coupling parameter about the time that cosmological
scales exit the Hubble horizon.}
\end{figure*}

We note that during the inflationary era, the evolution of the
Hubble parameter is so slow, so that in this era the conditions
$\epsilon\ll 1$ and $\eta\ll 1$ are satisfied. As one of these two
slow-roll parameters reaches unity, the inflation phase
terminates. Thus by studying the evolution of these parameters,
one can find the value of the inflaton field for which the
inflation ends ($\epsilon$ or $\eta$ reach unity). Using the
definition of the slow-roll parameter,
$\epsilon=-\frac{\dot{H}}{H^2}$ we obtain the following expression

\begin{eqnarray}
\epsilon=\frac{f_{,\phi}(f_{,\phi}R-V_{,\phi})}{(V-2f_{,\phi}^2R+2f_
{,\phi}V_{,\phi})}+\left(\frac{1+2\kappa^2
f}{2\kappa^2}\right)\times\hspace{1cm}\\\nonumber\frac
{(V_{,\phi}-4f_{,\phi}f_{\phi\phi}R+2f_{,\phi\phi}
V_{,\phi}+2f_{,\phi}V_{,\phi\phi})(V_{,\phi}-f_{,\phi}R)}
{(V-2f_{,\phi}^2R+2f_{,\phi}V_{,\phi})^2}\,,
\end{eqnarray}
and also $\eta=-\frac{\ddot{H}}{\dot{H}{H}}$ gives
\begin{eqnarray}
\eta=2\epsilon-\frac{\dot{\epsilon}}{H\epsilon}\,.
\end{eqnarray}
Fig.~\ref{fig:epsilon} shows the behavior of $\epsilon$ versus
$\phi$ and $\xi$. This figure confirms that the value of the field
to have $\epsilon=1$, that is, $\phi_{e}$, is directly dependant
on the non-minimal coupling parameter, $\xi$. This relationship is
given by the following expression
\begin{eqnarray}
\phi_{e}=\pm\frac{\sqrt{2(\xi
R-m_{\phi}^2)(4\xi^2R-m_{\phi}^2-4\xi
m_{\phi}^2)}}{\kappa(4\xi^2R-m_{\phi}^2-4\xi m_{\phi}^2)}\,.
\end{eqnarray}
In fact for smaller values of $\xi$, inflation ends sooner (that
is, it occurs in larger values of $\phi$ for large field
inflation). We continue our analysis by adopting $\kappa=1$
and $m=0.4$. Using these values, we obtain
$\phi_{e}\simeq\pm\frac{0.7\sqrt{25\xi-4}}{25\xi^2-1-4\xi}$. To
have positive $\phi_{e}$ for $0\leq\xi\leq \frac{1}{6}$ (we
restrict ourselves to this range of $\xi$), we choose the minus
sign for this relation. After obtaining the field value at the end
of inflation, now we are in the position to study the
non-Gaussianity of the model at hand through its non-linear
parameters. Fig.~\ref{fig:1} demonstrates the relation between
$f_{NL}^{local}$ and parameters $\xi$ and $N$ intuitively. This
figure shows that for any value of $\xi$, the absolute value of
$f_{NL}^{local}$ increases during the inflationary era. This
property is more clarified in the left panel of Fig.~\ref{fig:2},
which shows the evolution of $f_{NL}^{local}$ in the number of
e-folds space for various values of $\xi$. We note that in our
numerical study the values of $m_{\phi}$ and $\phi_{e}$ are chosen
so that the inflation phase terminates gracefully about the end of
inflation (which is equivalent to $N=0$ in our setup). For
instance, we have set $m_{\phi}=0.4$ which results in
$\phi_{e}\simeq-\frac{0.7\sqrt{25\xi-4}}{25\xi^2-1-4\xi}$. The
right panel of Fig.~\ref{fig:2} also shows this evolution in the
$\xi$ space around the end of inflation. As an important result we
can see from the left panel of Fig.~\ref{fig:2} that the first
order non-Gaussian effect is almost constant and weak at the
initial stage of the inflationary era and grows gradually to
values of order unity at the time of horizon crossing.

\begin{figure*}[htp]
\flushleft\leftskip0em{
\includegraphics[width=.46\textwidth,origin=c,angle=0]{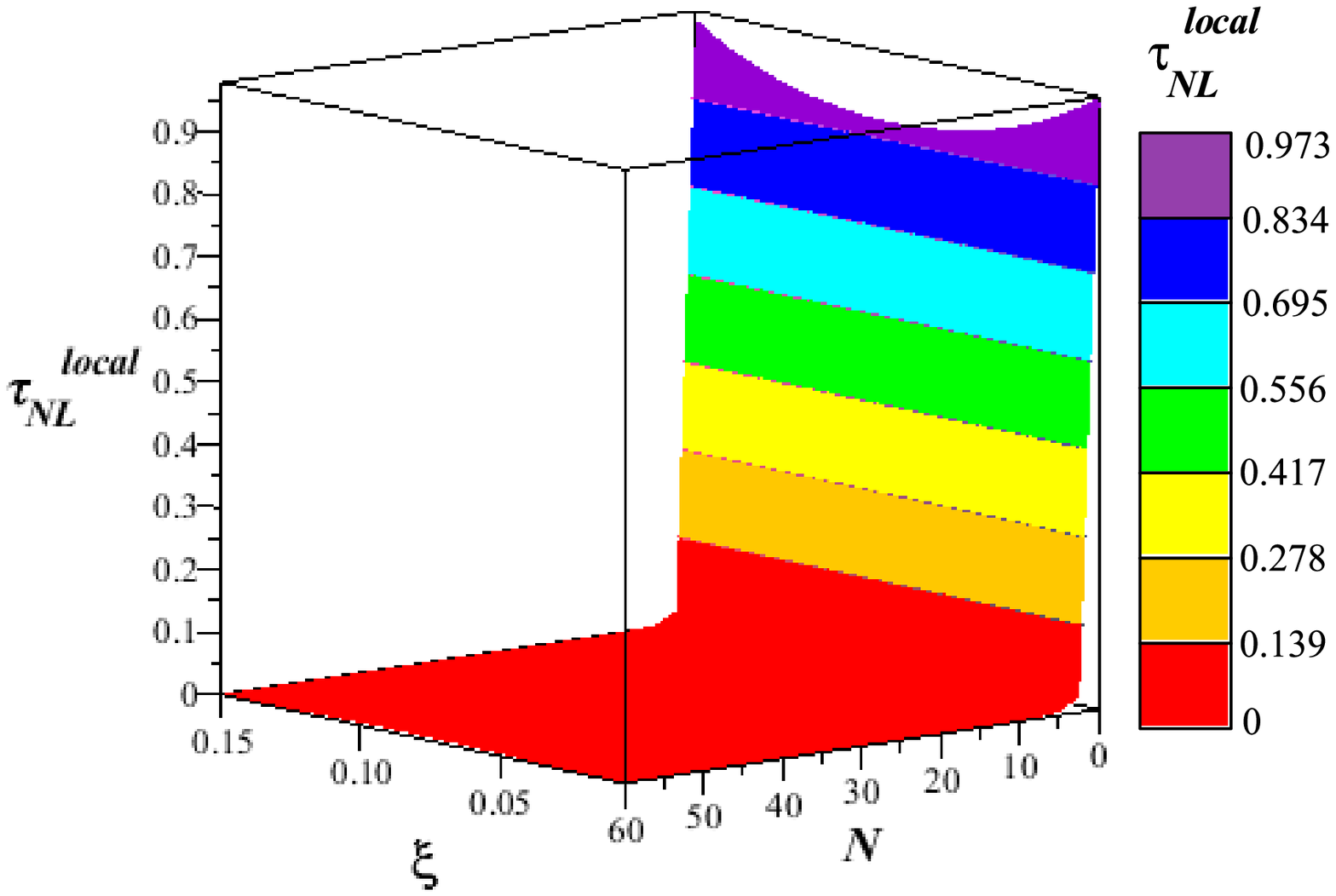}
\hspace{1cm}
\includegraphics[width=.39\textwidth,origin=c,angle=0]{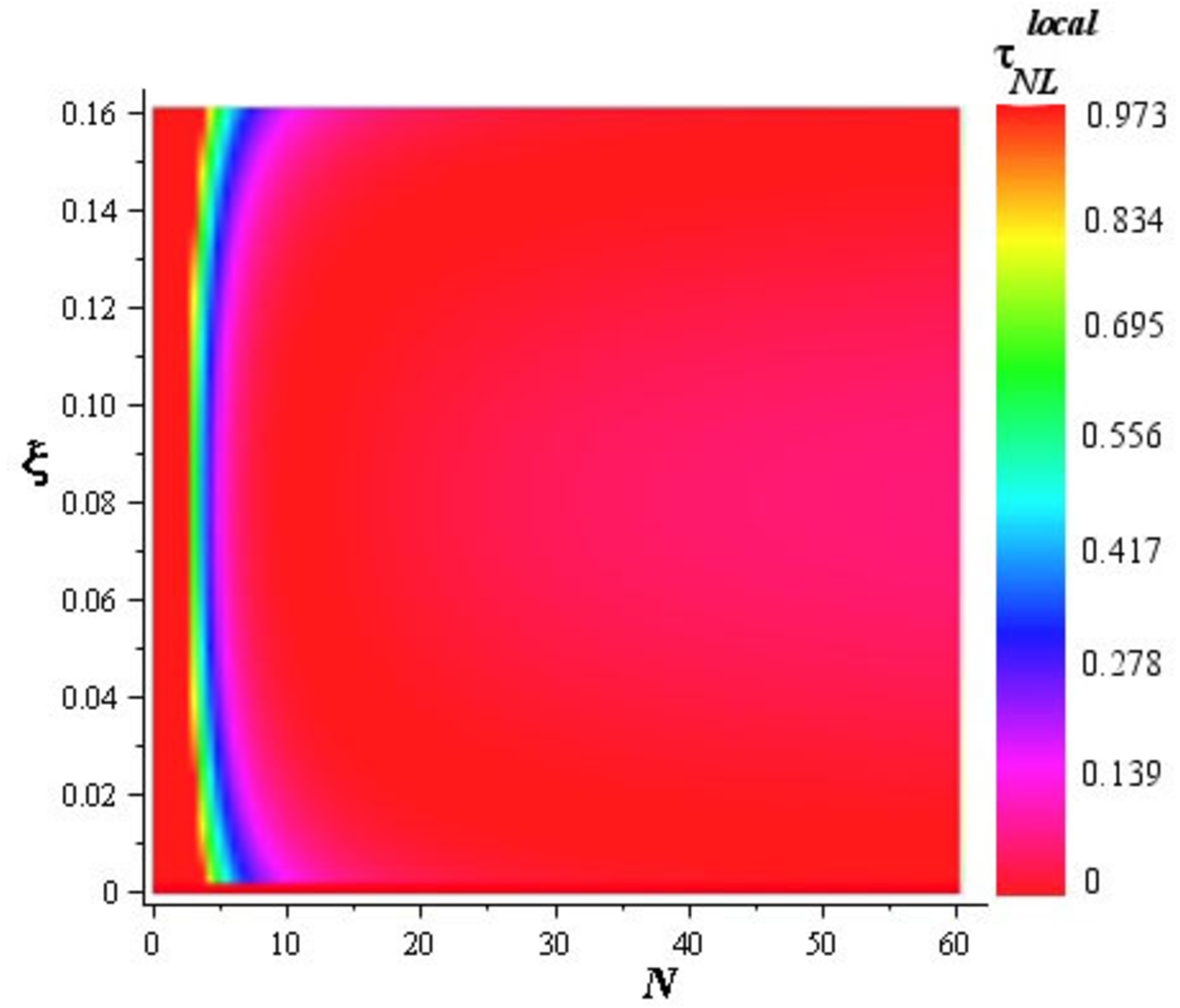}}
\caption{\label{fig:3} The non-linear parameter
$\tau_{NL}^{local}$ as a function of the non-minimal coupling
parameter $\xi$ and the number of e-folds $N$. }
\end{figure*}

\begin{figure*}[htp]
\flushleft\leftskip0em{
\includegraphics[width=.42\textwidth,origin=c,angle=0]{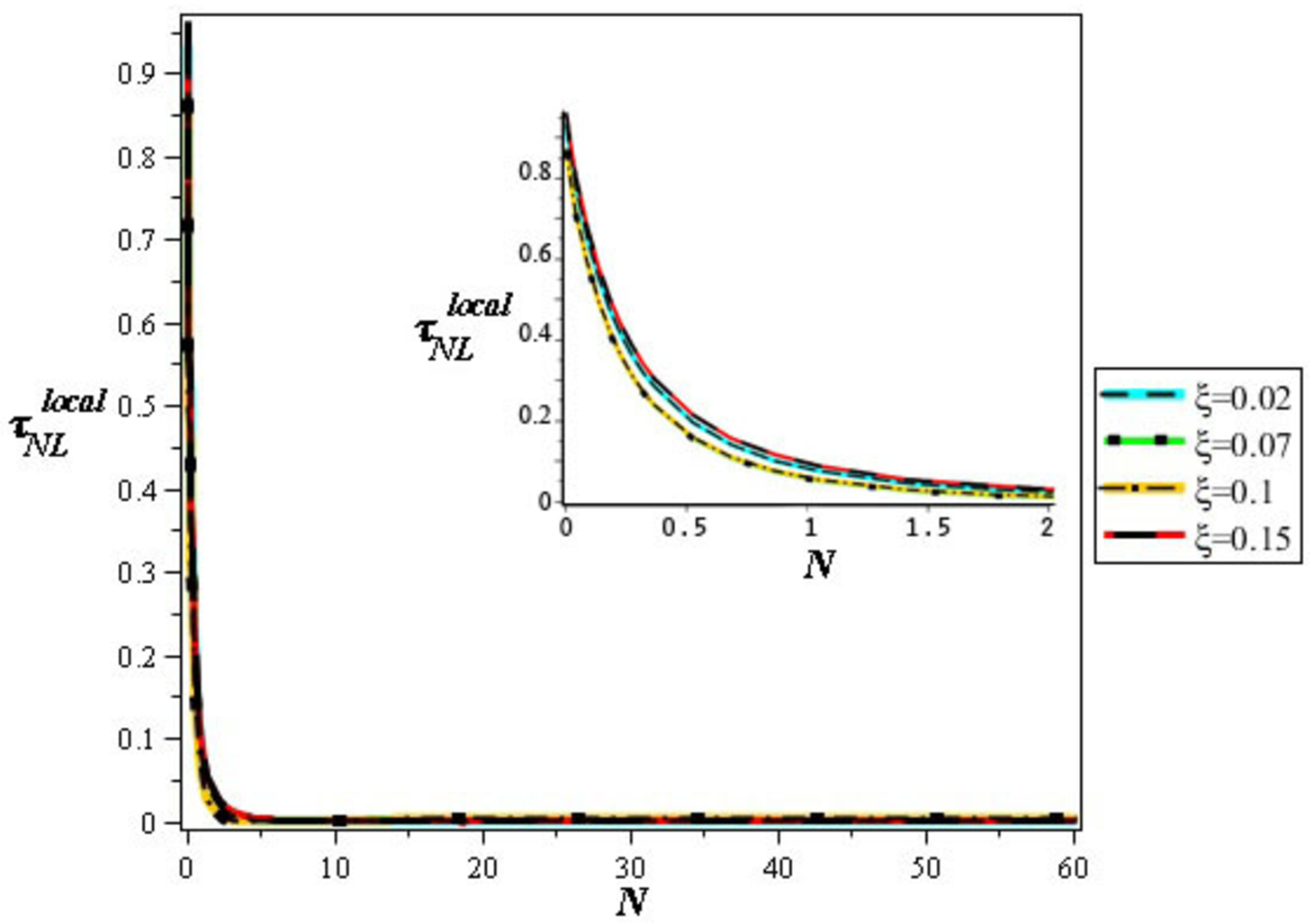}
\hspace{1.5cm}
\includegraphics[width=.37\textwidth,origin=c,angle=0]{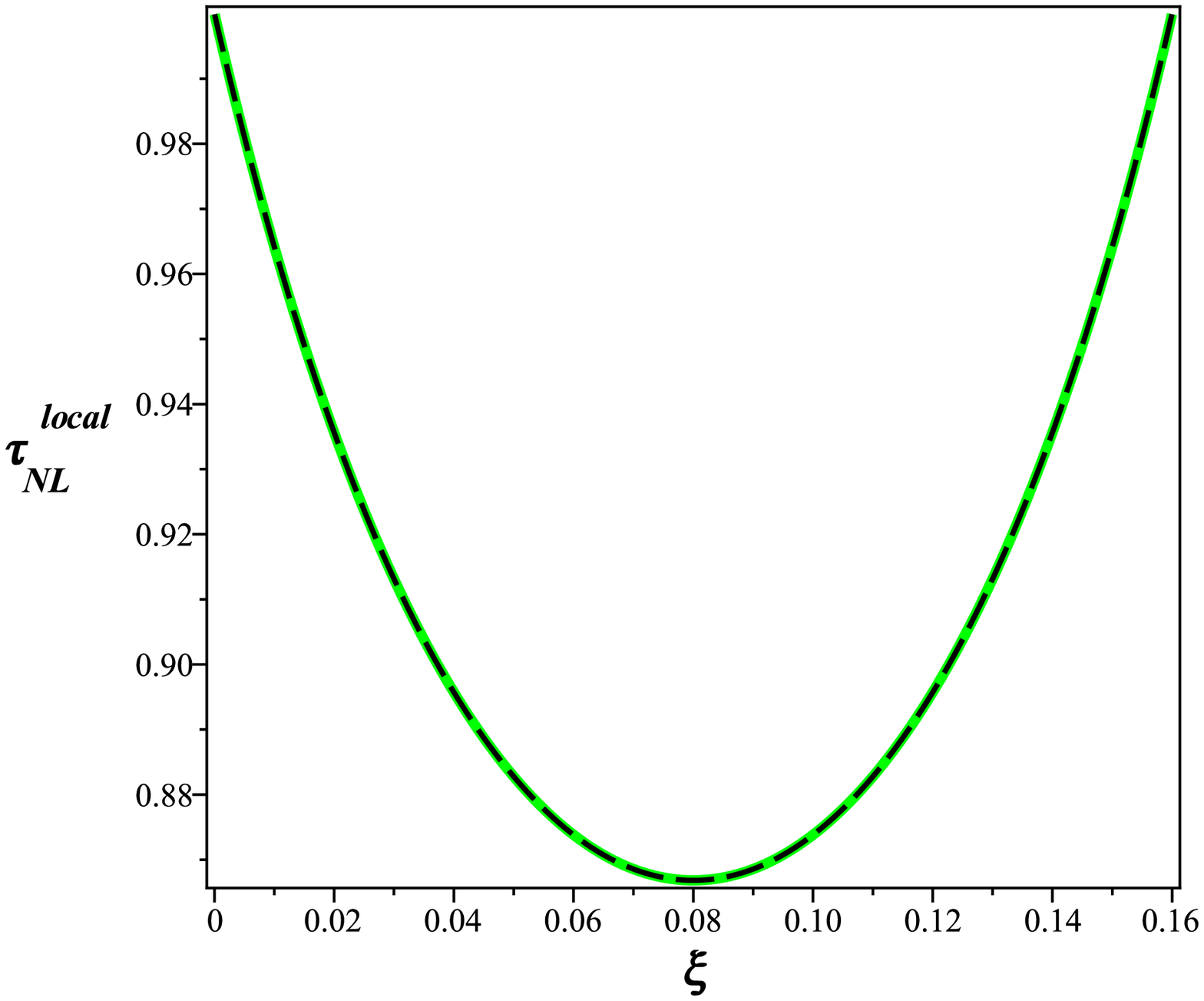}}
\caption{\label{fig:4} Left Panel: The non-linear parameter
$\tau_{NL}^{local}$ for the local type non-Gaussianity versus the
number of e-folds for various values of the non-minimal coupling
parameter. Right Panel: The non-linear parameter
$\tau_{NL}^{local}$ for the local type non-Gaussianity as a
function of the non-minimal coupling parameter about the time that
cosmological scales exit the Hubble horizon.}
\end{figure*}

\begin{figure*}[htb]
\flushleft\leftskip0em{
\includegraphics[width=.42\textwidth,origin=c,angle=0]{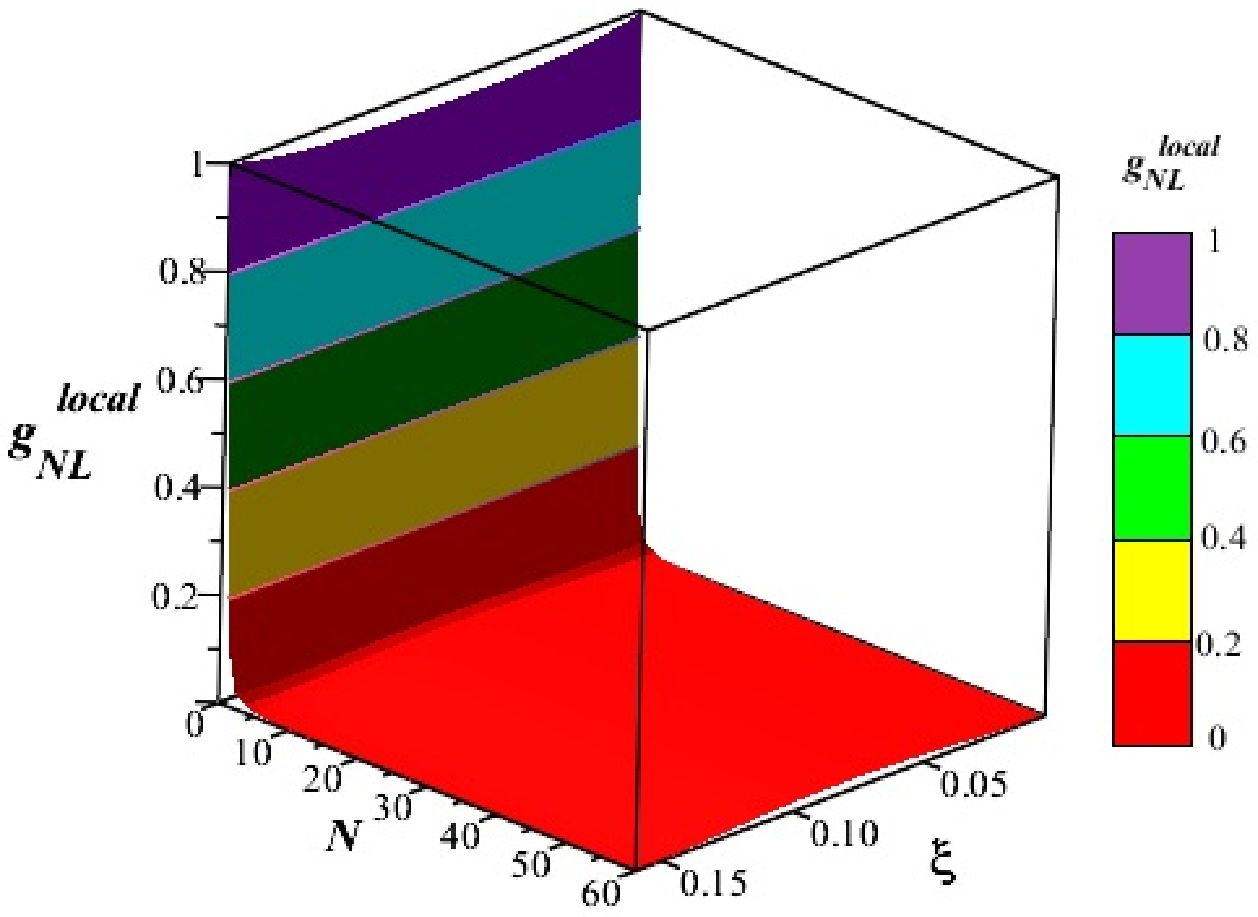}
\hspace{1cm}
\includegraphics[width=.44\textwidth,origin=c,angle=0]{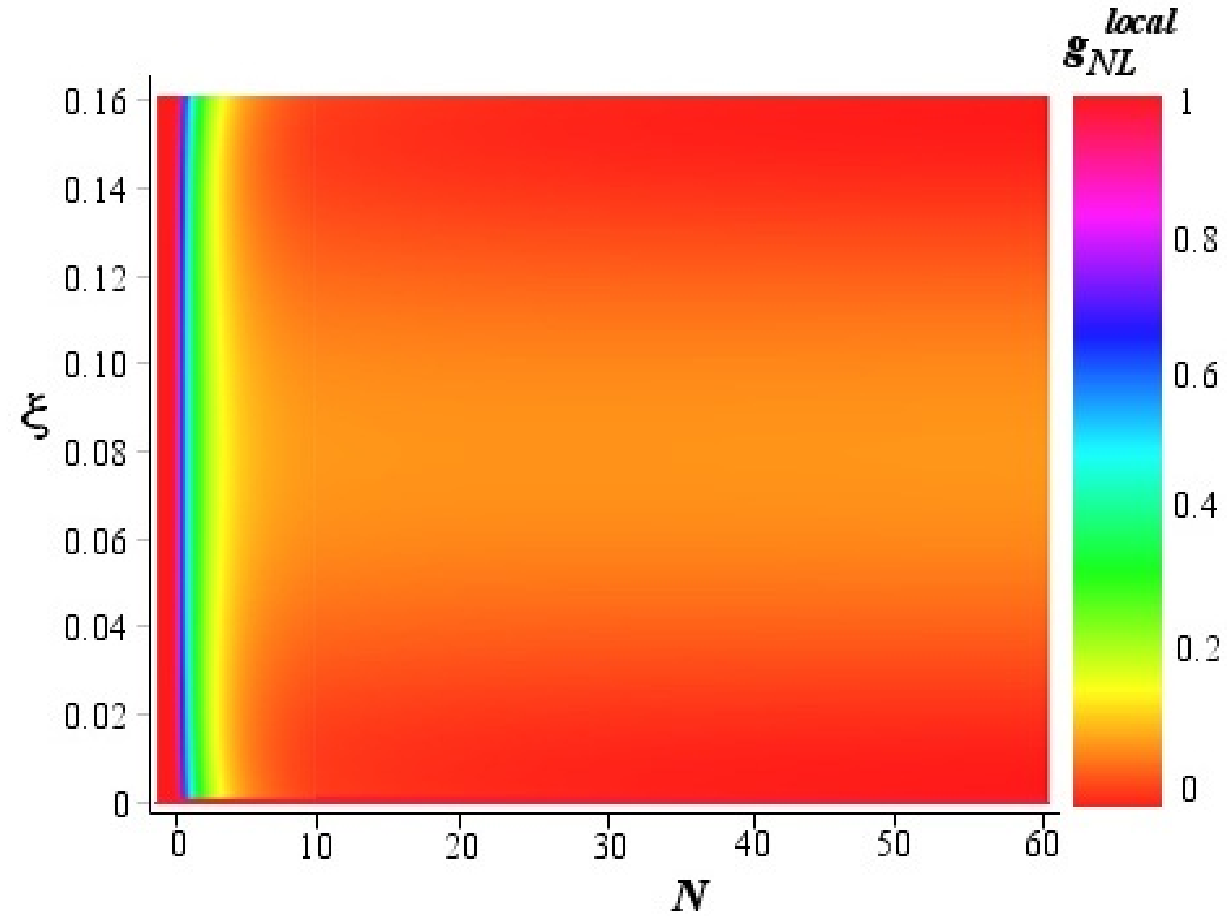}}
\caption{\label{fig:5} The non-linear parameter $g_{NL}^{local}$
as a function of the non-minimal coupling parameter $\xi$ and the
number of e-folds $N$.}
\end{figure*}

\begin{figure*}
\flushleft\leftskip0em{
\includegraphics[width=.44\textwidth,origin=c,angle=0]{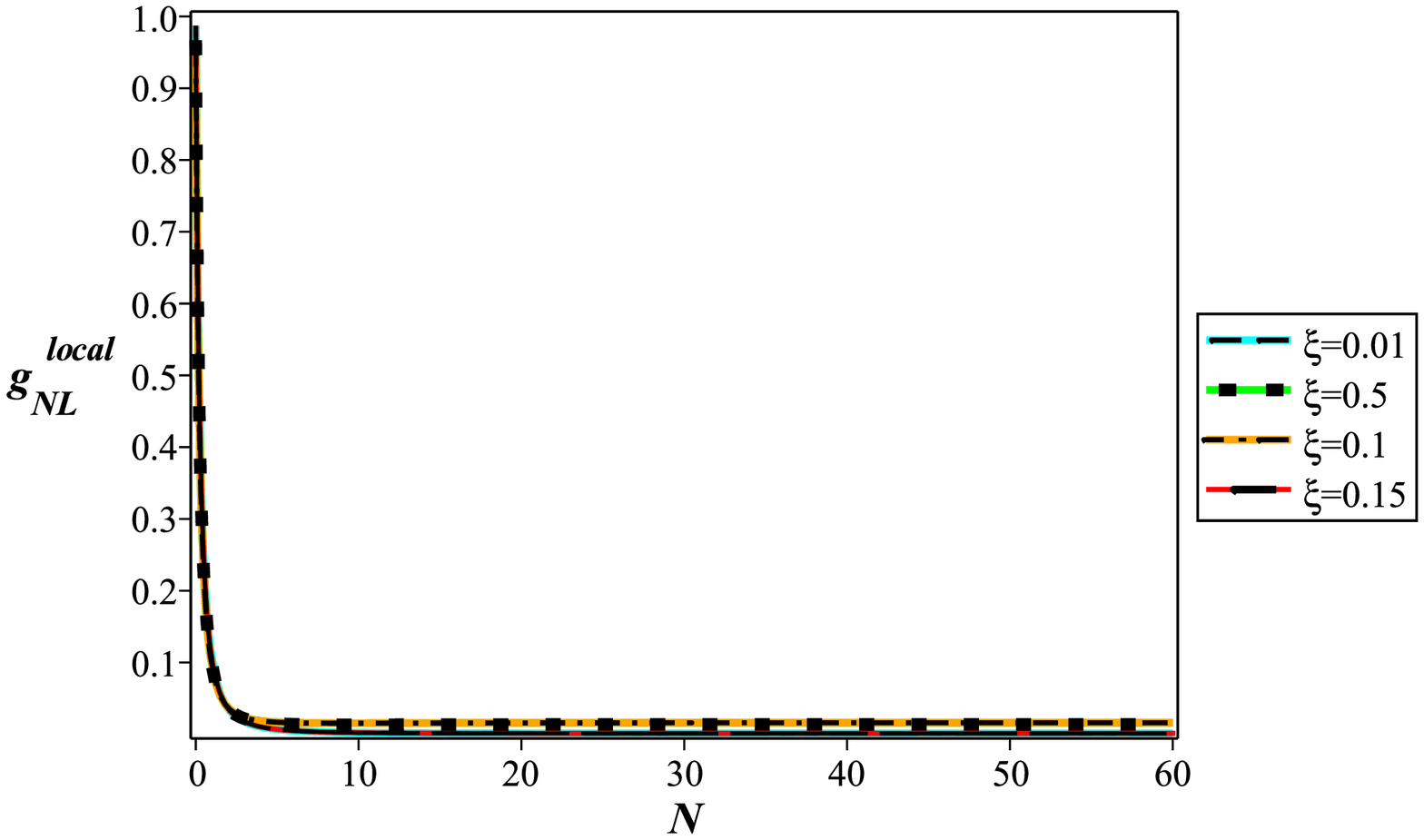}
\hspace{1cm}
\includegraphics[width=.39\textwidth,origin=c,angle=0]{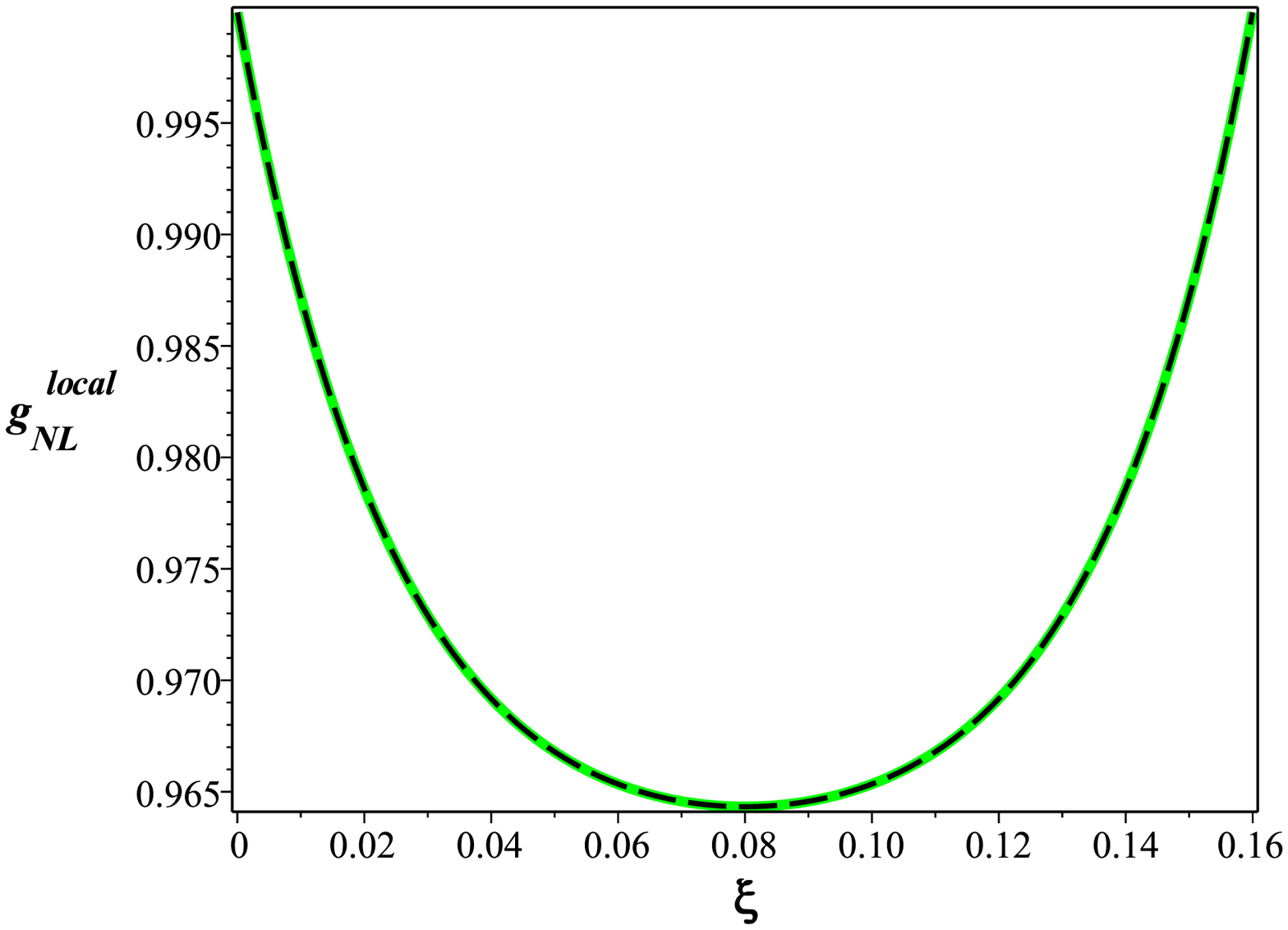}}
\caption{\label{fig:7} Left Panel: The non-linear parameter
$g_{NL}^{local}$ for the local type non-Gaussianity versus the
number of e-folds for various values of the non-minimal coupling
parameter. Right panel: The non-linear parameter $g_{NL}^{local}$
for the local type non-Gaussianity as a function of the
non-minimal coupling parameter about the time that cosmological
scales exit the Hubble horizon.}
\end{figure*}

\begin{figure*}
\flushleft\leftskip0em{
\includegraphics[width=.43\textwidth,origin=c,angle=0]{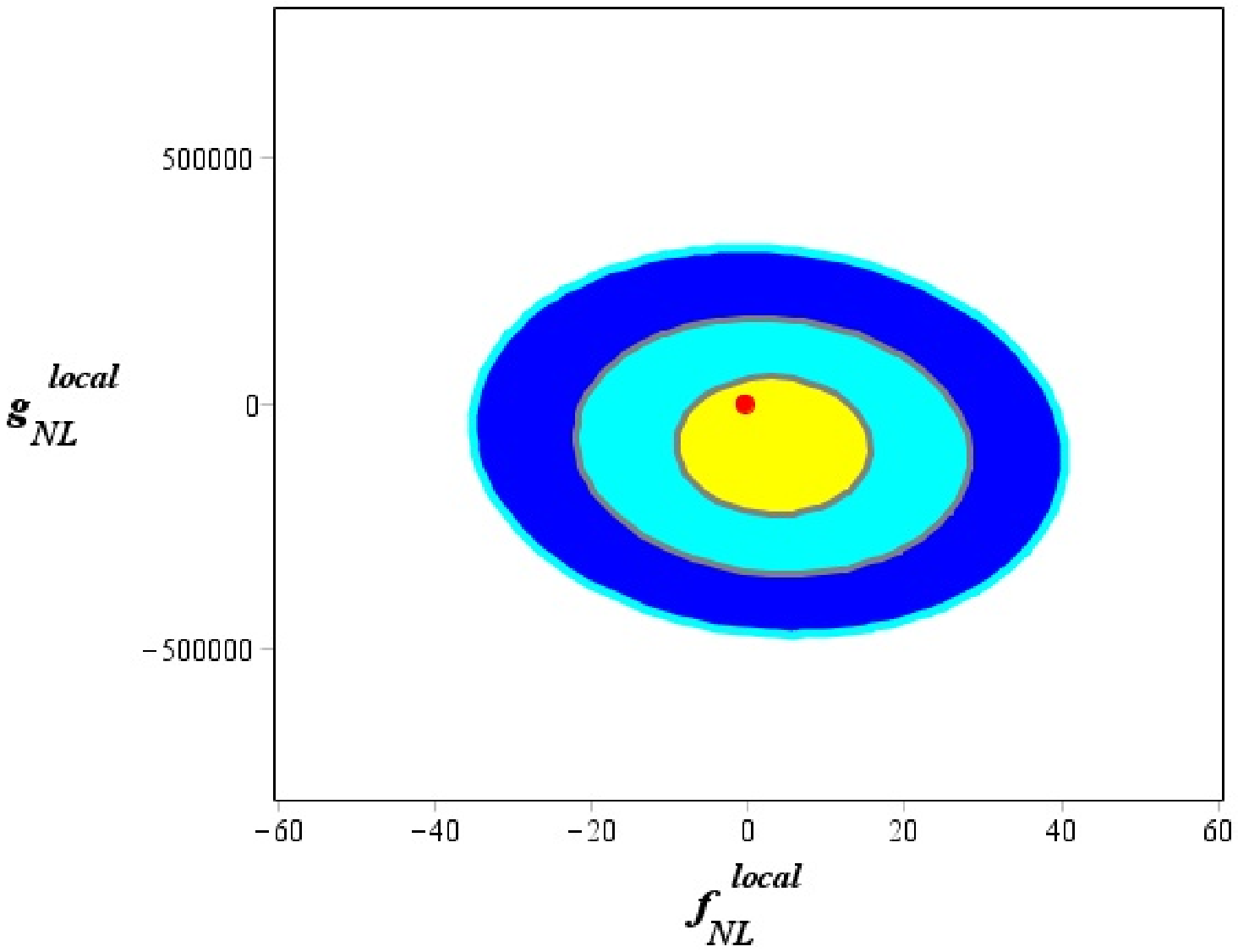}
\hspace{1cm}
\includegraphics[width=.42\textwidth,origin=c,angle=0]{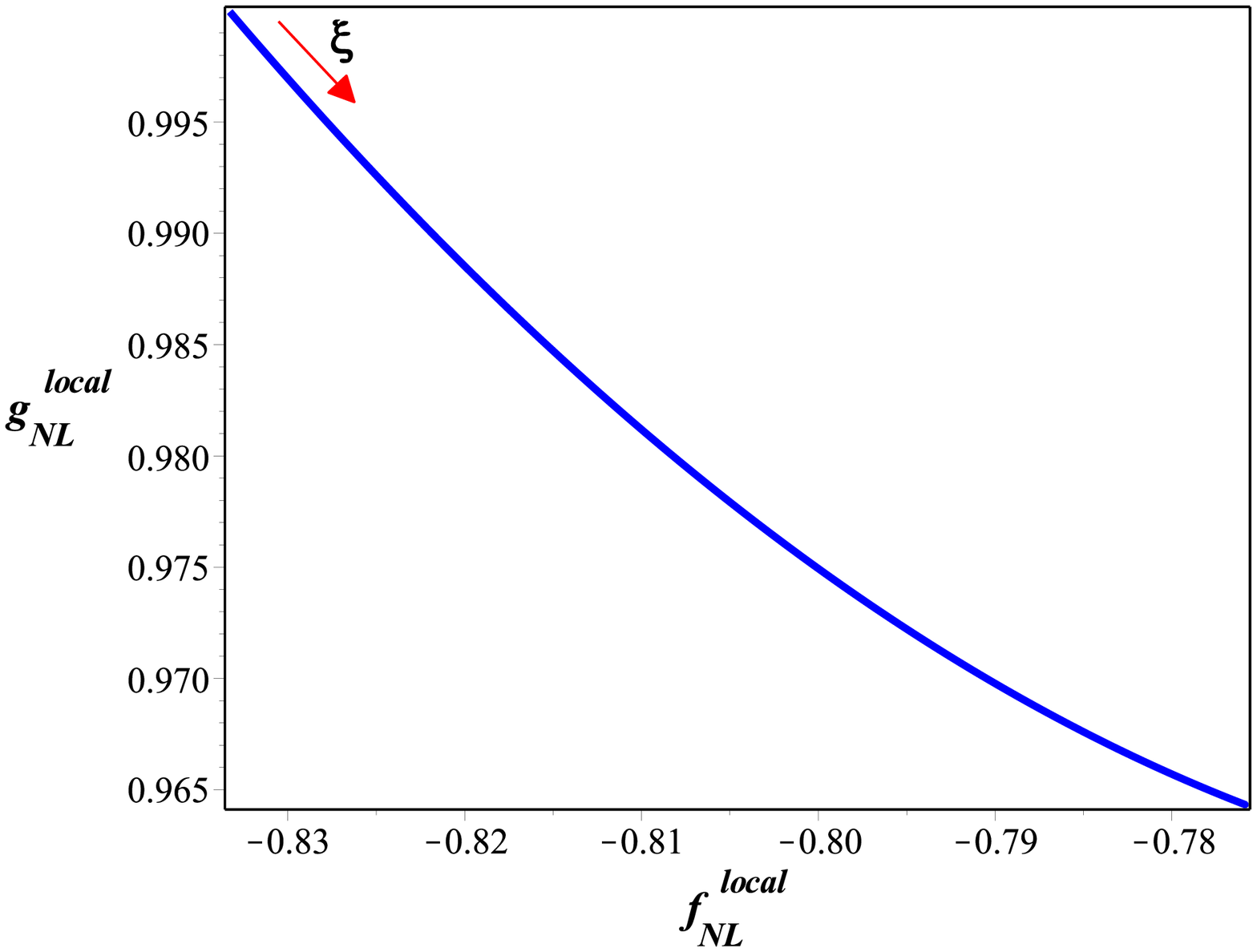}}
\caption{\label{fig:8} The amplitude of $g_{NL}^{local}$ versus
$f_{NL}^{local}$ for local configuration of non-Gaussianity in the
background of Planck2015 data. These figures are plotted about the
end of inflation for a quadratic potential and the non-minimal
coupling function as $f(\phi)\sim\xi\phi^2$. We note that the red
spot in this figure shows the position of our result in the background of the observational data.
The right panel highlights more details of $g_{NL}^{local}$ versus
$f_{NL}^{local}$ in terms of $\xi$. }
\end{figure*}

\begin{figure*}
\flushleft\leftskip0em{
\includegraphics[width=.46\textwidth,origin=c,angle=0]{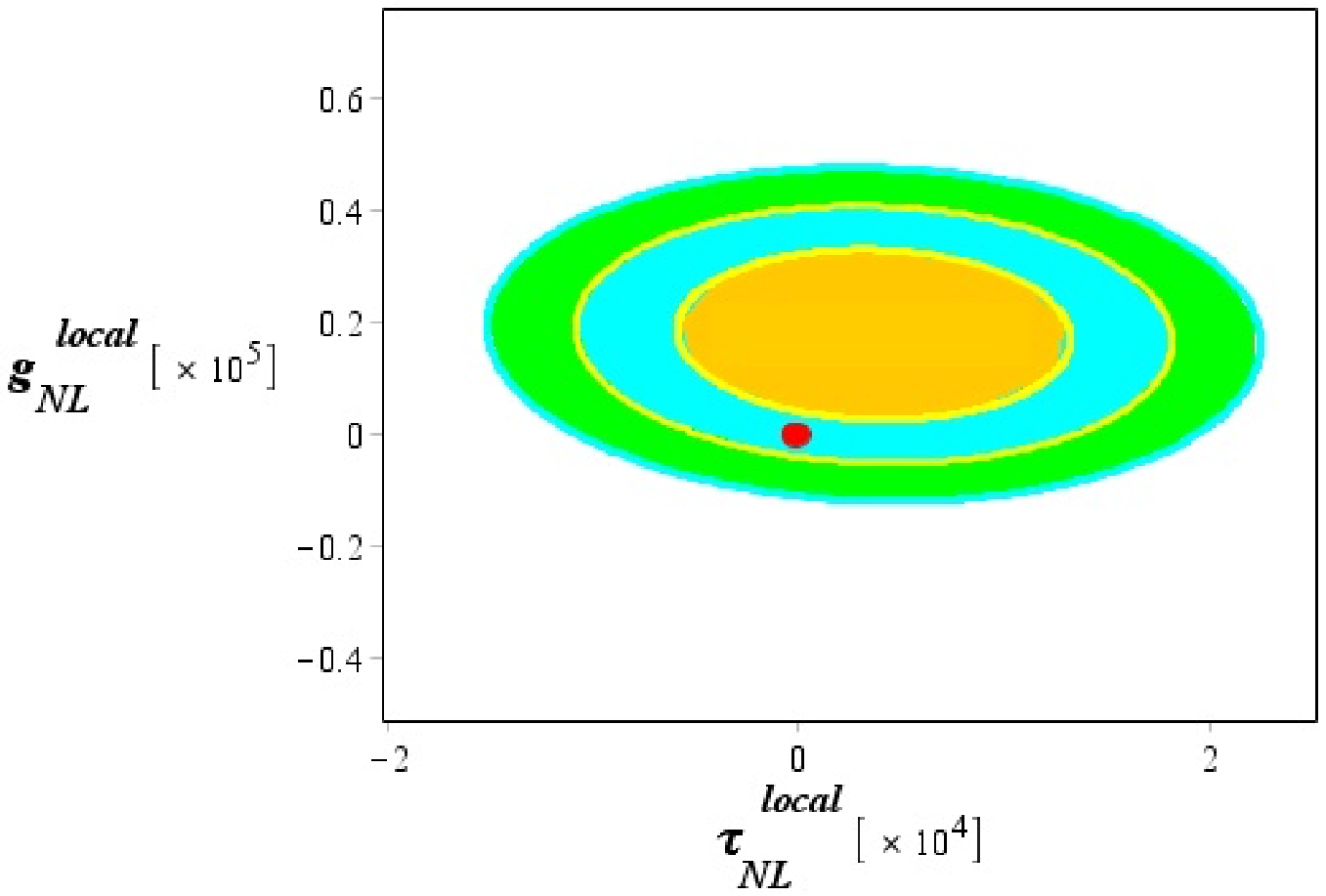}
\hspace{1cm}
\includegraphics[width=.37\textwidth,origin=c,angle=0]{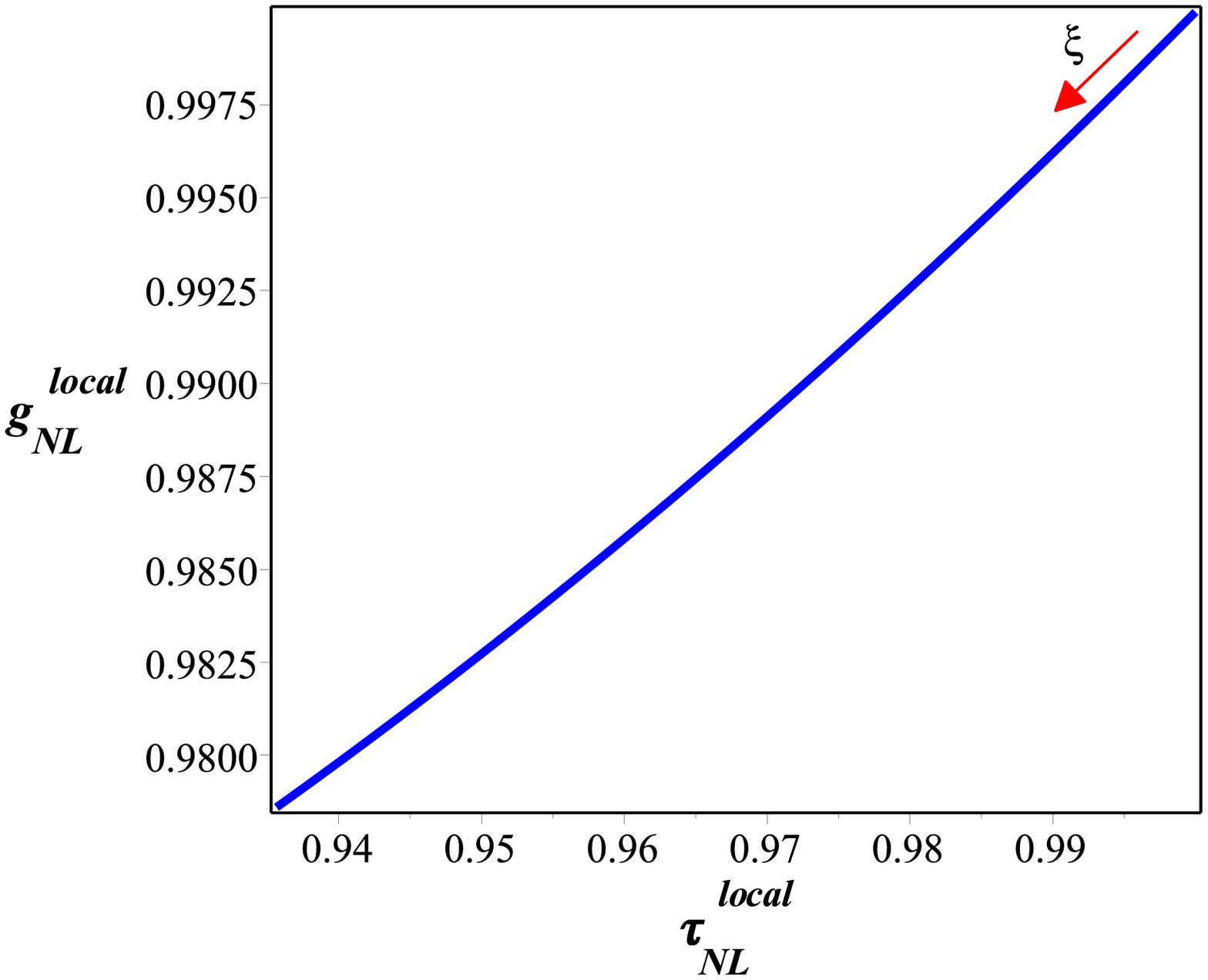}}
\caption{\label{fig:9} The amplitude of $g_{NL}^{local}$ versus
$\tau_{NL}^{local}$ for local configuration of non-Gaussianity in
the background of Planck2013 data. These figures are plotted about
the end of inflation for a quadratic potential and the non-minimal
coupling function as $f(\phi)\sim\xi\phi^2$. We note that the red
spot in this figure shows the position of our result in the background of the observational data.
 The right panel highlights more details of $g_{NL}^{local}$ versus
$\tau_{NL}^{local}$ in terms of $\xi$.}
\end{figure*}

Now we proceed further by investigating evolution of the higher
order non-linear parameters of local shapes, $\tau_{NL}^{local}$
and $g_{NL}^{local}$. Eqs. (\ref{tau2}) and (\ref{c13}) show how
the second and third levels of non-Gaussianity depend on the
inflaton field and the non-minimal coupling parameter. Again we
can depict the behavior of $\tau_{NL}^{local}$ and
$g_{NL}^{local}$ in $\xi$ and $N$ space by using Eq. (\ref{c11}).
These behaviors are depicted in Figs.~\ref{fig:3} and
~\ref{fig:5}. Comparing Figs.~\ref{fig:1} and ~\ref{fig:3} shows
that the first and second order of non-Gaussianinties behave in
the same way as the relation $\tau_{NL}=(\frac{6}{5}f_{NL})^2$
shows. Also it can be seen from Fig.~\ref{fig:5} that
$g_{NL}^{local}$ grows rapidly towards the end of inflation but
stays small even in this situation. Furthermore, the behavior of the
trispectrum non-linear parameters during expansion of the universe
are depicted in the left panels of Figs. ~\ref{fig:4} and
~\ref{fig:7}. In order to see the influence of the non-minimal
coupling parameter on the level of trispectrum, the right panels
of Figs. ~\ref{fig:4} and~\ref{fig:7} have been plotted about the
end of inflation.

After illustrating the evolution of bispectrum and trispectrum
separately, now we are in the position to constrain the model with
recent observational data. In this regard, we explore the behavior
of the $f_{NL}$ versus $g_{NL}$ with the local configuration in
the background of the Planck2015 data to see the viability of this
theoretical model in confrontation with the recent observations.
This behavior is shown in Fig.~\ref{fig:8} which confirms that
although our non-Gaussianity result in non-minimal inflation is
somehow small, however it is well inside the
confidence region allowed by the Planck observations. We also note that this
figure is depicted about the end of inflation and for
$0<\xi\leq\frac{1}{6}$. Confronting these evolutions and the
Planck2015 data, we find the values of the non-minimal coupling
parameter for which the model is consistent. As it is clear from
the Fig.~\ref{fig:8}, our model is consistent with observation for
all chosen values of $\xi$.

Moreover, as mentioned previously, with the trispectrum or the
4-point correlation function of the CMB anisotropies \citep{Hu01},
we are able to evaluate the second and third order non-Gaussian
parameters $\tau_{NL}$ and $g_{NL}$. Using the correlations
between (square temperature)-(square temperature) and cubic
(temperature)-(temperature) anisotropies, the authors in
\citep{Fen15} have reported measurements of the kurtosis power
spectra of the CMB temperature as mapped by Planck2013. In
combination with noise simulations, they have found the best joint
estimates to be $\tau_{NL}^{local}=(0.3\pm 0.9)\times 10^4$ and
$g_{NL}^{local}=(-1.2\pm 2.8)\times 10^5$. They have also obtained
$g_{NL}^{local}=(-1.3\pm 1.8) \times 10^5$ if
$\tau_{NL}^{local}=0$. Their analysis shows that
$\tau_{NL}^{local}$ and $g_{NL}^{local}$ are consistent with zero
for all the combinations. Using the results of \citep{Fen15}, we
can study the trispectrum parameters of non-Gaussianity and
constrain the model with observation. Fig.~\ref{fig:9} shows the
amplitude of $\tau_{NL}^{local}$ versus $g_{NL}^{local}$ for local
configuration of non-Gaussianity for a quadratic potential and the
non-minimal coupling function as $f(\phi)\sim\xi\phi^2$ in the
background of Planck2013 data. This figure confirms that although
the non-Gaussianity in the non-minimal inflation is
almost small, however it is well inside the region allowed by the
Planck observations. Confronting these behaviors and the
Planck2013 data, we find that the model is consistent with observation for all values of $\xi$.

\section{Analysis in Einstein Frame}

Now we can move to the Einstein frame with a conformal transformation as
\citep{ein-1,ein-2,ein-3}
\begin{eqnarray}\label{ein-g}
\hat{g}_{\mu\nu}=\Omega^2g_{\mu\nu}\,,
\end{eqnarray}
where the parameter $\Omega$ is given by the following expression
\begin{eqnarray}\label{ein-omega}
\Omega^2=(1+2\kappa^2f(\phi))\,.
\end{eqnarray}
In Einstein frame the new scalar field is redefined as follows
\begin{eqnarray}\label{ein-phi}
\frac{d\hat{\phi}}{d\phi}=\left(\frac{(1+2\kappa^2f)+6\kappa^2f_{,\phi}^2}
{(1+2\kappa^2f)^2}\right)^\frac{1}{2}\,.
\end{eqnarray}
We further define a new scalar potential $\hat{V}$ by
\begin{eqnarray}\label{ein-V}
\hat{V}(\hat{\phi})=\Omega^{-4}V(\phi)\,,
\end{eqnarray}
which gives
\begin{eqnarray}\label{ein-V2}
\hat{V}(\hat{\phi})=\frac{V(\phi)}{(1+2\kappa^2f)^2}\,,
\end{eqnarray}
Furthermore, the action (\ref{A1}) in this frame becomes as
\begin{eqnarray}\label{ein-S}
S_{E}=\int
d^4x\sqrt{-\hat{g}}\left[\frac{1}{2\kappa^2}\hat{R}-\frac{1}{2}
\partial_{\mu}\hat{\phi}\partial^{\mu}\hat{\phi}-\hat{V}(\hat{\phi})\right]\,.\hspace{0.8cm}
\end{eqnarray}
The dynamics of the model is given by the Friedmann equation which
takes the following form in Einstein frame
\begin{eqnarray}\label{ein-H2}
\hat{H}^2=\frac{\kappa^2}{3}\left[\frac{1}{2}\left(\frac{d\hat{\phi}}{d\hat{t}}\right)^2+\hat{V}(\hat{\phi})\right]\,,
\end{eqnarray}
where we defined
$\hat{a}(\hat{t})=\Big(1+2\kappa^2f(\phi)\Big)^{\frac{1}{2}}a(t)$
and $\hat{t}=\Big(1+2\kappa^2f(\phi)\Big)^{\frac{1}{2}}t$. The equation
of motion of the scalar field in Einstein frame is given by
\begin{eqnarray}\label{ein-motion}
\ddot{\hat{\phi}}+3\hat{H}\dot{\hat{\phi}}+\hat{V}_{,\hat{\phi}}=0\,,
\end{eqnarray}
where a dot marks a derivative with respect to $\hat{t}$. With slow-roll approximations as
$\Big(\frac{d\hat{\phi}}{d\hat{t}}\Big)^2\ll\hat{V}(\hat{\phi})$
and $\Big(\frac{d^2\hat{\phi}}{d\hat{t}^2}\Big)\ll|3\hat{H}\frac{d\hat{\phi}}{d\hat{t}}|$, we find
\begin{eqnarray}\label{ein-H2-SR}
\hat{H}^2=\frac{\kappa^2}{3}\hat{V}(\hat{\phi})\,,
\end{eqnarray}
and
\begin{eqnarray}\label{ein-motion-SR}
3\hat{H}\frac{d\hat{\phi}}{d\hat{t}}+\frac{d\hat{V}}{d\hat{\phi}}=0\,.
\end{eqnarray}
The number of e-folds with a non-minimally coupled scalar field in
Einstein frame is given by
\begin{eqnarray}\label{ein-N}
\hat{N}=-\kappa^2\int_{\hat{\phi}_{*}}^{\hat{\phi}_{e}}
\frac{\hat{V}}{\hat{V}_{,\hat{\phi}}}d\hat{\phi}\,,
\end{eqnarray}
which can be rewritten as
\begin{eqnarray}\label{ein-N-2}
\hat{N}=-\kappa^2\int
\frac{\hat{V}}{(\frac{d\hat{V}}{d\phi})}\left(\frac{d\hat{\phi}}{d\phi}\right)^2d\phi\,,
\end{eqnarray}
where we have used
$\frac{d\hat{V}}{d\hat{\phi}}=\frac{d\hat{V}}{d{\phi}}\frac{d{\phi}}{d\hat{\phi}}$.
Finally the number of e-folds in Einstein frame becomes
\begin{eqnarray}\label{ein-N-3}
\hat{N}=\hspace{7cm}\\\nonumber
-\kappa^2\int_{\phi_{*}}^{\phi_{e}}
\frac{V\left((1+2\kappa^2f)+6\kappa^2f_{,\phi}^2\right)}{(1+2\kappa^2f)
\left[V_{,\phi}(1+2\kappa^2f)-4\kappa^2Vf_{,\phi}\right]}d\phi\,.
\end{eqnarray}
Now we study bispectrum and trispectrum of perturbations during a non-minimal slow-roll
inflation in Einstein frame.

\section{Primordial non-Gaussianity in Einstein frame}

The general form of the non-linear parameters
in the local type non-Gaussianities have been presented in section
3. Now we calculate the local type
non-Gaussianities of the primordial fluctuations in Einstein frame. To study
local non-Gaussianities, it is enough to focus on super-horizon
scales and in this region the non-linear parameters regarding to
the bispectrum and trispectrum, that is, $f_{NL}^{local}$,
$\tau_{NL}^{local}$ and $g_{NL}^{local}$, are defined by Eqs.
(\ref{c2}), (\ref{tau1}) and (\ref{c6}). Thus, in order to obtain
these non-Gaussian parameters, we should obtain the first, second
and third derivatives of the number of e-folds with respect to the
scalar field (see Appendix \textbf{A}). The results for the
non-linear parameters in Einstein frame are given in Appendix \textbf{B}. After
calculating the main equations of the model to analyze both
bispectrum and trispectrum of the local-type non-Gaussianities in
Einstein frame, now we are in the position to proceed further by
specifying the form of the potential. The potential in the
Einstein frame is related to the potential in
Jordan frame through Eq. (\ref{ein-V2}). Similar to section 4, we
continue our treatment by considering the quadratic form of the functions $V$
and $f$ as $V(\phi)=\frac{1}{2}m_{\phi}^2\phi^2$ and
$f(\phi)=\frac{1}{2}\xi\phi^2$. So, the potential in Einstein frame is given by
\begin{eqnarray}\label{ein-V3}
\hat{V}(\hat{\phi})=\frac{m_{\phi}^2\phi^2}{2(1+\kappa^2\xi\phi^2)^2}\,.
\end{eqnarray}
With these choices, the number of e-folds in Einstein frame is as follows
\begin{eqnarray}\label{ein-N(phi)}
\hat{N}(\phi)=\hspace{7cm}\\\nonumber
\frac{3\xi+1}{4\xi}\ln{\Bigg[\frac{\kappa^2\xi\phi_e^2-1}{\kappa^2\xi\phi^2-1}\Bigg]}
+\frac{3}{4}\ln{\Bigg[\frac{\kappa^2\xi\phi_e^2+1}{\kappa^2\xi\phi^2+1}\Bigg]}\,.
\end{eqnarray}
The value of the field during the slow-roll
inflation in Einstein frame in terms of the number of e-folds is given by
\begin{eqnarray}\label{ein-phi(N)}
\phi(\hat{N})\simeq\frac{\Bigg(e^{\frac{4\hat{N}\xi}{3\xi+1}}\xi\Big(e^{\frac{4\hat{N}\xi}{3\xi+1}}+
\kappa^2\xi\phi_{e}^2-1\Big)\Bigg)^\frac{1}{2}}{e^{\frac{4\hat{N}\xi}{3\xi+1}}\xi\kappa}\,.\hspace{0.5cm}
\end{eqnarray}
Finally, adopting the mentioned forms of the potential and the
non-minimal coupling function results in the following expressions
for the local non-linear parameters associated to both bispectrum
and trispectrum of the model in Einstein frame

\begin{widetext}
\begin{eqnarray}\label{ein-fNL}
\frac{6}{5}\hat{f}_{NL}^{local}=\frac{\left( {m_{\phi}}^{2}\phi\, \left(
1+{\kappa}^{2}\xi\,{\phi}^{2 } \right)
-2\,{\kappa}^{2}{m_{\phi}}^{2}{\phi}^{3}\xi \right) ^{2}}{{\kappa}^{
4}{m_{\phi}}^{4}{\phi}^{4}}\Bigg( -{\frac {4{\kappa}^{2}{m_{\phi}}^{2}\phi
\left( 1+{\kappa}^{2}
\xi\,{\phi}^{2}+6\,{\kappa}^{2}{\xi}^{2}{\phi}^{2} \right)}{
\left( 1 +{\kappa}^{2}\xi{\phi}^{2} \right)  \left( {m_{\phi}}^{2}\phi
\left( 1+{ \kappa}^{2}\xi{\phi}^{2} \right)
-2\,{\kappa}^{2}{m_{\phi}}^{2}{\phi}^{3} \xi
\right)}}-\hspace{0.5cm}\\\nonumber {\frac
{2{\kappa}^{2}{m_{\phi}}^{2}{\phi}^{2} \left( 2\,{
\kappa}^{2}\xi\,\phi+12{\kappa}^{2}{\xi}^{2}\phi \right) }{\left(
1 +{\kappa}^{2}\xi\,{\phi}^{2} \right)  \left( {m_{\phi}}^{2}\phi\,
\left( 1+{ \kappa}^{2}\xi\,{\phi}^{2} \right)
-2{\kappa}^{2}{m_{\phi}}^{2}{\phi}^{3} \xi \right) }}+{\frac
{4{\kappa}^{4}{m_{\phi}}^{2}{\phi}^{3} \left( 1+{\kappa}
^{2}\xi{\phi}^{2}+6\,{\kappa}^{2}{\xi}^{2}{\phi}^{2} \right) \xi}{
 \left( 1+{\kappa}^{2}\xi\,{\phi}^{2} \right) ^{2} \left( {m_{\phi}}^{2}\phi
\, \left( 1+{\kappa}^{2}\xi\,{\phi}^{2} \right)
-2\,{\kappa}^{2}{m_{\phi}}^{2 }{\phi}^{3}\xi \right) }}+\\\nonumber
{\frac {2{\kappa}^{2}{m_{\phi}}^{2}{\phi}^{2}
 \left( 1+{\kappa}^{2}\xi\,{\phi}^{2}+6\,{\kappa}^{2}{\xi}^{2}{\phi}^{
2} \right)  \left( {m_{\phi}}^{2} \left( 1+{\kappa}^{2}\xi\,{\phi}^{2}
 \right) -4\,{m_{\phi}}^{2}{\phi}^{2}{\kappa}^{2}\xi \right) }{ \left( 1+{
\kappa}^{2}\xi\,{\phi}^{2} \right)  \left( {m_{\phi}}^{2}\phi\left( 1+{
\kappa}^{2}\xi\,{\phi}^{2} \right)
-2\,{\kappa}^{2}{m_{\phi}}^{2}{\phi}^{3} \xi \right) ^{2}}} \Bigg)
\frac{\left( 1+{\kappa}^{2}\xi\,{\phi}^{2}
 \right) ^{2}}{\left(1+{\kappa}^{2}\xi\,{\phi}^{2}+6\,{\kappa} ^{2}{\xi}^{2}{\phi}^{2}
\right) ^{2}} \,,
\end{eqnarray}
\begin{eqnarray}\label{ein-TNL}
\hat{\tau}_{NL}^{local}=\frac{16\left( {m_{\phi}}^{2}\phi\left(
1+{\kappa}^{2}\xi\,{\phi}^{2 } \right)
-2\,{\kappa}^{2}{m_{\phi}}^{2}{\phi}^{3}\xi
\right)^{4}}{{\kappa}^{8}{m_{\phi}}^{8}{\phi}^{8}}\Bigg(-{\frac{{\kappa}^{2}{m_{\phi}}^{2}\phi\left(
1+{\kappa}^{2} \xi{\phi}^{2}+6\,{\kappa}^{2}{\xi}^{2}{\phi}^{2}
\right)}{\left(1 +{\kappa}^{2}\xi{\phi}^{2}\right)\left(
{m_{\phi}}^{2}\phi\left(1+{\kappa}^{2}\xi{\phi}^{2}\right)
-2{\kappa}^{2}{m_{\phi}}^{2}{\phi}^{3}\xi
\right)}}-\hspace{0.5cm}\\\nonumber
{\frac{{\kappa}^{2}{m_{\phi}}^{2}{\phi}^{2} \left(2\,{
\kappa}^{2}\xi\,\phi+12\,{\kappa}^{2}{\xi}^{2}\phi \right)}{
2\left( 1 +{\kappa}^{2}\xi\,{\phi}^{2} \right)\left( {m_{\phi}}^{2}\phi\,
\left( 1+{ \kappa}^{2}\xi\,{\phi}^{2} \right)
-2\,{\kappa}^{2}{m_{\phi}}^{2}{\phi}^{3} \xi \right) }}+{\frac
{{\kappa}^{4}{m_{\phi}}^{2}{\phi}^{3} \left( 1+{\kappa}
^{2}\xi\,{\phi}^{2}+6\,{\kappa}^{2}{\xi}^{2}{\phi}^{2} \right)
\xi}{\left( 1+{\kappa}^{2}\xi\,{\phi}^{2} \right) ^{2} \left(
{m_{\phi}}^{2}\phi \, \left( 1+{\kappa}^{2}\xi\,{\phi}^{2} \right)
-2\,{\kappa}^{2}{m_{\phi}}^{2 }{\phi}^{3}\xi \right) }}+\\\nonumber
{\frac {{\kappa}^{2}{m_{\phi}}^{2}{\phi}^{2}
 \left( 1+{\kappa}^{2}\xi\,{\phi}^{2}+6\,{\kappa}^{2}{\xi}^{2}{\phi}^{
2} \right)  \left( {m_{\phi}}^{2} \left( 1+{\kappa}^{2}\xi\,{\phi}^{2}
 \right) -4\,{m_{\phi}}^{2}{\phi}^{2}{\kappa}^{2}\xi \right) }{2\left( 1+{
\kappa}^{2}\xi\,{\phi}^{2} \right)  \left( {m_{\phi}}^{2}\phi\, \left(
1+{ \kappa}^{2}\xi\,{\phi}^{2} \right)
-2\,{\kappa}^{2}{m_{\phi}}^{2}{\phi}^{3} \xi \right) ^{2}}} \Bigg)^{2}
\frac{\left( 1+{\kappa}^{2}\xi\,{\phi}^{2}
 \right)^{4}}{\left(
1+{\kappa}^{2}\xi\,{\phi}^{2}+6\,{\kappa} ^{2}{\xi}^{2}{\phi}^{2}
\right)^{4}} \,,
\end{eqnarray}
and
\begin{eqnarray}\label{ein-gNL}
\frac{54}{25}\hat{g}_{NL}^{local}=-\frac{8\left({m_{\phi}}^{2}\phi\, \left(
1+{\kappa}^{2}\xi\,{\phi}^{2 } \right)
-2\,{\kappa}^{2}{m_{\phi}}^{2}{\phi}^{3}\xi \right)
^{2}}{{\kappa}^{6}{m_{\phi}}^{6}{\phi}^{6}{\left( 1+{
\kappa}^{2}\xi\,{\phi}^{2} \right)}} \, \Bigg(
{-{\kappa}^{2}{m_{\phi}}^{2} \left(
1+{\kappa}^{2}\xi{\phi}^{2}+6\,{\kappa}^{2}{\xi}^{2}{\phi}^{2}
\right)}\hspace{1.5cm}\\\nonumber -2\,{{\kappa}^{2}{m_{\phi}}^{2}\phi\,
\left( 2\,{\kappa} ^{2}\xi\,\phi+12\,{\kappa}^{2}{\xi}^{2}\phi
\right)}+5{{{\kappa}^{4}{m_{\phi}}^{2}{\phi}^{2} \left( 1+{
\kappa}^{2}\xi\,{\phi}^{2}+6\,{\kappa}^{2}{\xi}^{2}{\phi}^{2}
\right) \xi}{\left( 1+{\kappa}^{2}\xi\,{\phi}^{2}
\right)^{-1}}}+\\\nonumber {\frac {2{\kappa}^{2}{m_{\phi}}^{2}\phi\,
 \left( 1+{\kappa}^{2}\xi\,{\phi}^{2}+6\,{\kappa}^{2}{\xi}^{2}{\phi}^{
2} \right)  \left( {m_{\phi}}^{2} \left( 1+{\kappa}^{2}\xi\,{\phi}^{2}
 \right) -4\,{m_{\phi}}^{2}{\phi}^{2}{\kappa}^{2}\xi \right) }{\left( {m_{\phi}}^{2}\phi\, \left(
1+{\kappa}^{2}\xi\,{\phi}^{2} \right)
-2{\kappa}^{2}{m_{\phi}}^{2}{\phi}^{3} \xi \right)}}-
{\frac{1}{2}{{m_{\phi}}^{2}{\phi}^{2}{\kappa}^{2} \left( 2
\,{\kappa}^{2}\xi+12\,{\kappa}^{2}{\xi}^{2} \right)
}}+\\\nonumber{\frac {2{\kappa}^{4}{m_{\phi}}^{2}{\phi}^{3} \left( 2\,{
\kappa}^{2}\xi\,\phi+12\,{\kappa}^{2}{\xi}^{2}\phi \right) \xi}{
 \left( 1+{\kappa}^{2}\xi\,{\phi}^{2} \right)}}+
 {\frac{{m_{\phi}}^{2}{\phi}^{2}{\kappa}^{2}
 \left( 2\,{\kappa}^{2}\xi\,\phi+12\,{\kappa}^{2}{\xi}^{2}\phi
 \right)  \left( {m_{\phi}}^{2} \left( 1+{\kappa}^{2}\xi\,{\phi}^{2} \right)
-4\,{m_{\phi}}^{2}{\phi}^{2}{\kappa}^{2}\xi \right) }{ \left(
{m_{\phi}}^{2}\phi\, \left( 1+{\kappa}^{2}\xi \,{\phi}^{2} \right)
-2\,{\kappa}^{2}{m_{\phi}}^{2}{\phi}^{3}\xi \right)}}-\\\nonumber {\frac
{4\,{\kappa}^{6}{m_{\phi}}^{2}{\phi}^{4} \left( 1+{\kappa}^{2}\xi\,
{\phi}^{2}+6\,{\kappa}^{2}{\xi}^{2}{\phi}^{2} \right) {\xi}^{2}}{
 \left( 1+{\kappa}^{2}\xi\,{\phi}^{2} \right) ^{2}}}-3{\frac {{\kappa}^{4}{m_{\phi}}^{4}{\phi}^{3}\left(
1+{\kappa}^{2}\xi{\phi}^{2}+6{\kappa}^{2}{\xi}^{2}{\phi}^{2}
\right)\xi}{\left({m_{\phi}}^{2}\phi\, \left(1+{
\kappa}^{2}\xi\,{\phi}^{2} \right)
-2\,{\kappa}^{2}{m_{\phi}}^{2}{\phi}^{3} \xi \right)}}-\\\nonumber {\frac
{2\,{\kappa}^{4}{m_{\phi}}^{2}{\phi}^{3}
 \left( 1+{\kappa}^{2}\xi\,{\phi}^{2}+6\,{\kappa}^{2}{\xi}^{2}{\phi}^{
2} \right) \xi\, \left( {m_{\phi}}^{2} \left(
1+{\kappa}^{2}\xi\,{\phi}^{2}
 \right) -4\,{m_{\phi}}^{2}{\phi}^{2}{\kappa}^{2}\xi \right) }{ \left( 1+{
\kappa}^{2}\xi\,{\phi}^{2} \right)\left( {m_{\phi}}^{2}\phi\, \left( 1+
{\kappa}^{2}\xi\,{\phi}^{2} \right)
-2\,{\kappa}^{2}{m_{\phi}}^{2}{\phi}^{3} \xi \right)}}-\\\nonumber
{\frac{{m_{\phi}}^{2}{\phi}^{2}{\kappa}^{2} \left( 1+{
\kappa}^{2}\xi\,{\phi}^{2}+6\,{\kappa}^{2}{\xi}^{2}{\phi}^{2}
\right)
 \left( {m_{\phi}}^{2} \left( 1+{\kappa}^{2}\xi\,{\phi}^{2} \right) -4\,{m_{\phi}}^{
2}{\phi}^{2}{\kappa}^{2}\xi \right) ^{2}}{\left( {m_{\phi}}^{2}\phi\,
\left( 1+{\kappa}^{2}\xi\,{ \phi}^{2} \right)
-2\,{\kappa}^{2}{m_{\phi}}^{2}{\phi}^{3}\xi \right) ^{2}}}\Bigg)
\frac{\left( 1+{\kappa}^{2}\xi\,{\phi}^{2}
 \right)^{3}}{\left(1+{\kappa}^{2}\xi\,{\phi}^{2}+6\,{\kappa} ^{2}{\xi}^{2}{\phi}^{2}
\right) ^{3}} \,.
\end{eqnarray}
\end{widetext}
Now we can substitute the field value in Einstein frame, that is, equation
(\ref{ein-phi(N)}) in Eqs. (\ref{ein-fNL}), (\ref{ein-TNL}) and
(\ref{ein-gNL}) to obtain the intended non-linear parameters as a
function of $N$. In the next section we perform some numerical
analysis on the model's parameter space in the Einstein. We study numerically
$\hat{f}_{NL}^{local}$, $\hat{\tau}_{NL}^{local}$ and $\hat{g}_{NL}^{local}$ to see the behaviors of
these quantities in Einstein frame and compare the results with their Jordan frame's counterparts.

\section{Confrontation with Observation}

\begin{figure*}[ht!]
\begin{center}
\scalebox{0.50}[0.50]{\includegraphics{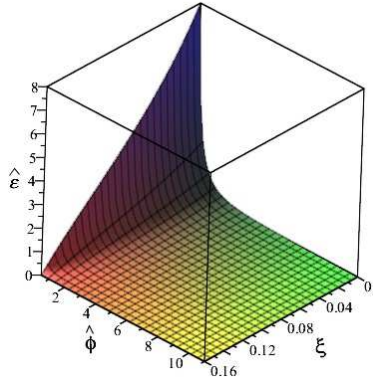}}
\caption{The behavior of $\hat{\epsilon}$ versus
$\hat{\phi}$ and $\xi$ for non-minimal inflation in Einstein frame.}
\label{fig:ein-epsilon}
\end{center}
\end{figure*}

\begin{figure*}[htp]
\flushleft\leftskip0em{
\includegraphics[width=.40\textwidth,origin=c,angle=0]{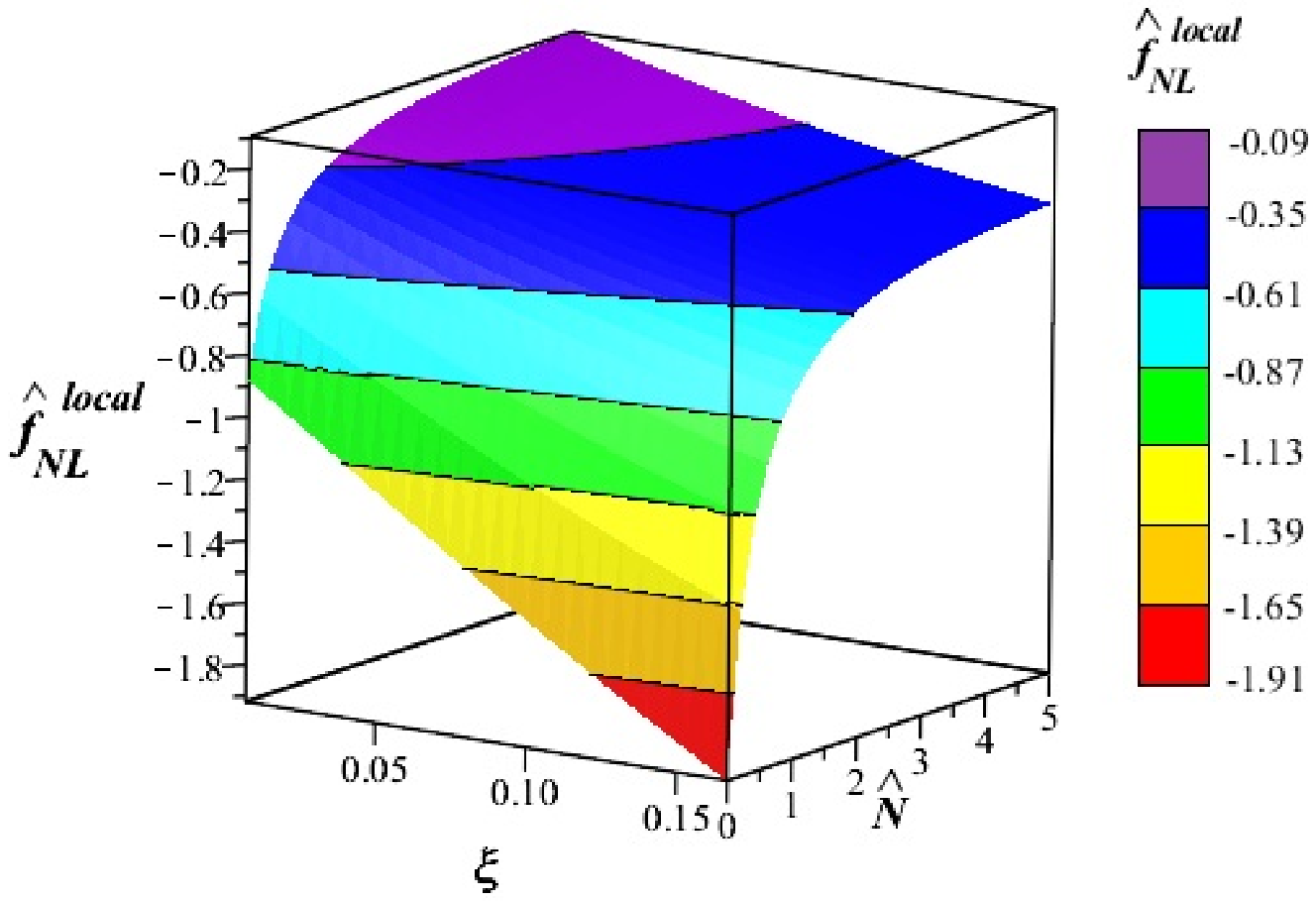}
\hspace{1cm}
\includegraphics[width=.42\textwidth,origin=c,angle=0]{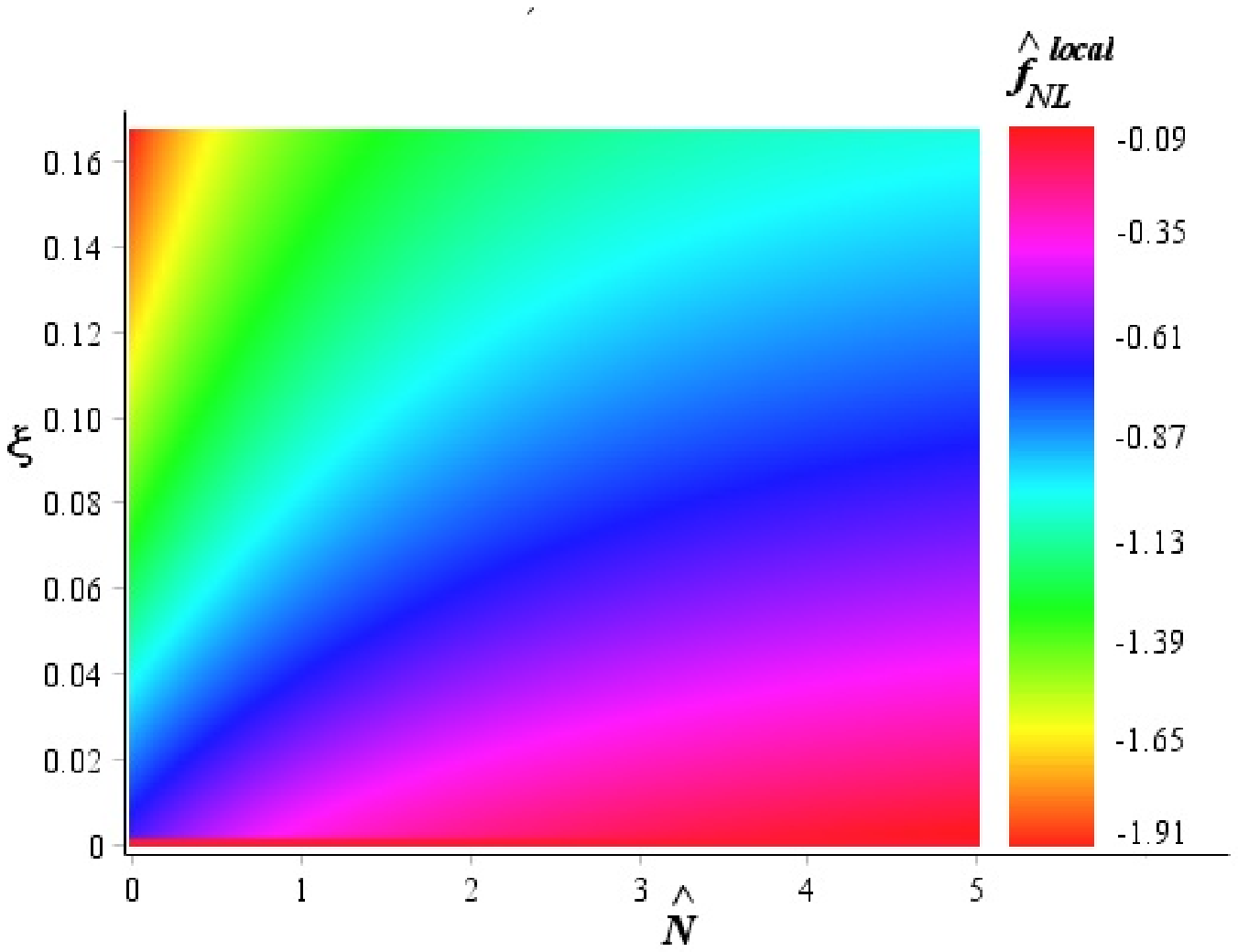}}
\caption{\label{fig:ein-fNL3d} The non-linear parameter
$\hat{f}_{NL}^{local}$ as a function of the non-minimal coupling
parameter $\xi$ and the number of e-folds $\hat{N}$ for non-minimal
inflation in Einstein frame.}
\end{figure*}

\begin{figure*}[htp]
\flushleft\leftskip0em{
\includegraphics[width=.45\textwidth,origin=c,angle=0]{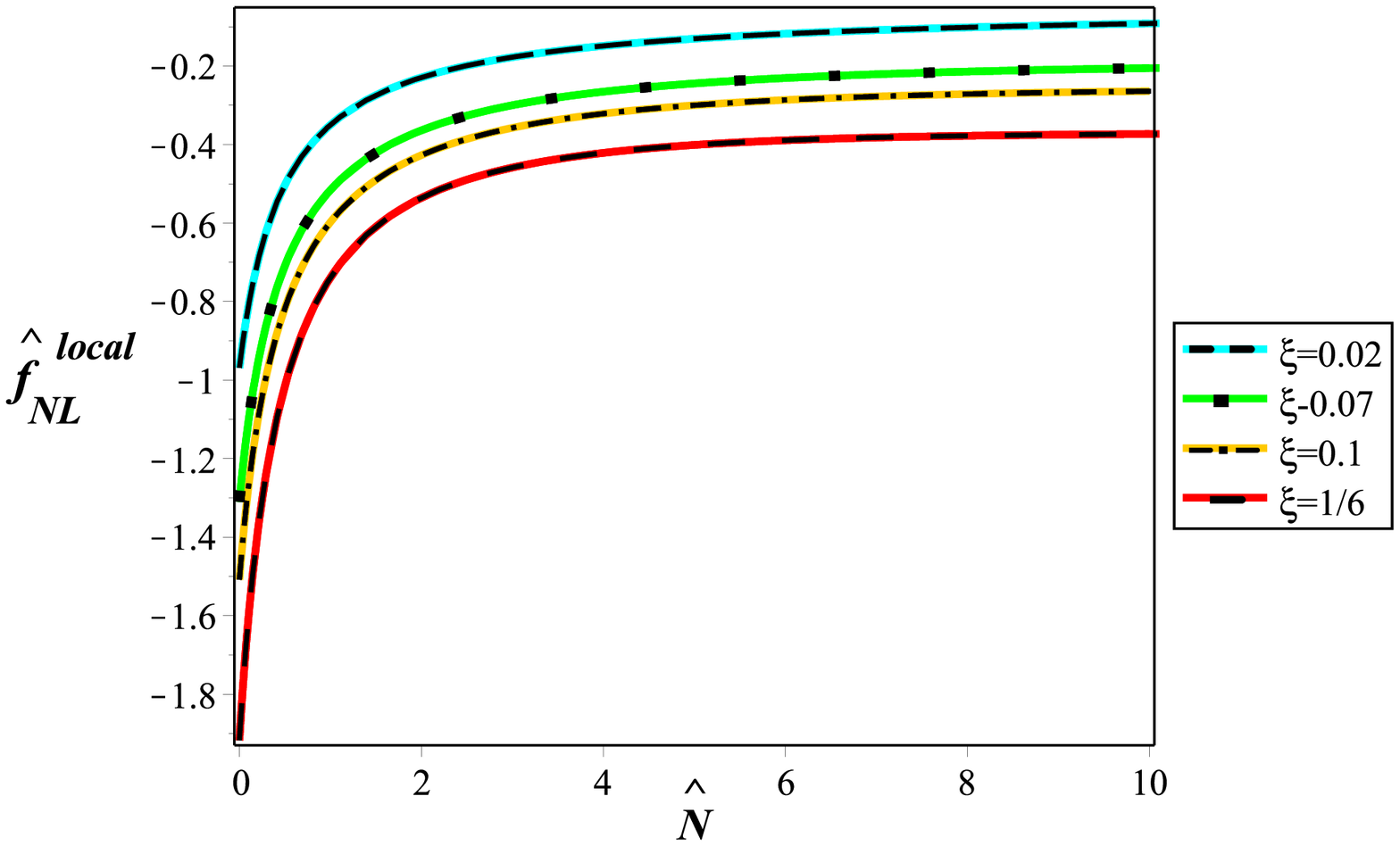}
\hspace{1cm}
\includegraphics[width=.40\textwidth,origin=c,angle=0]{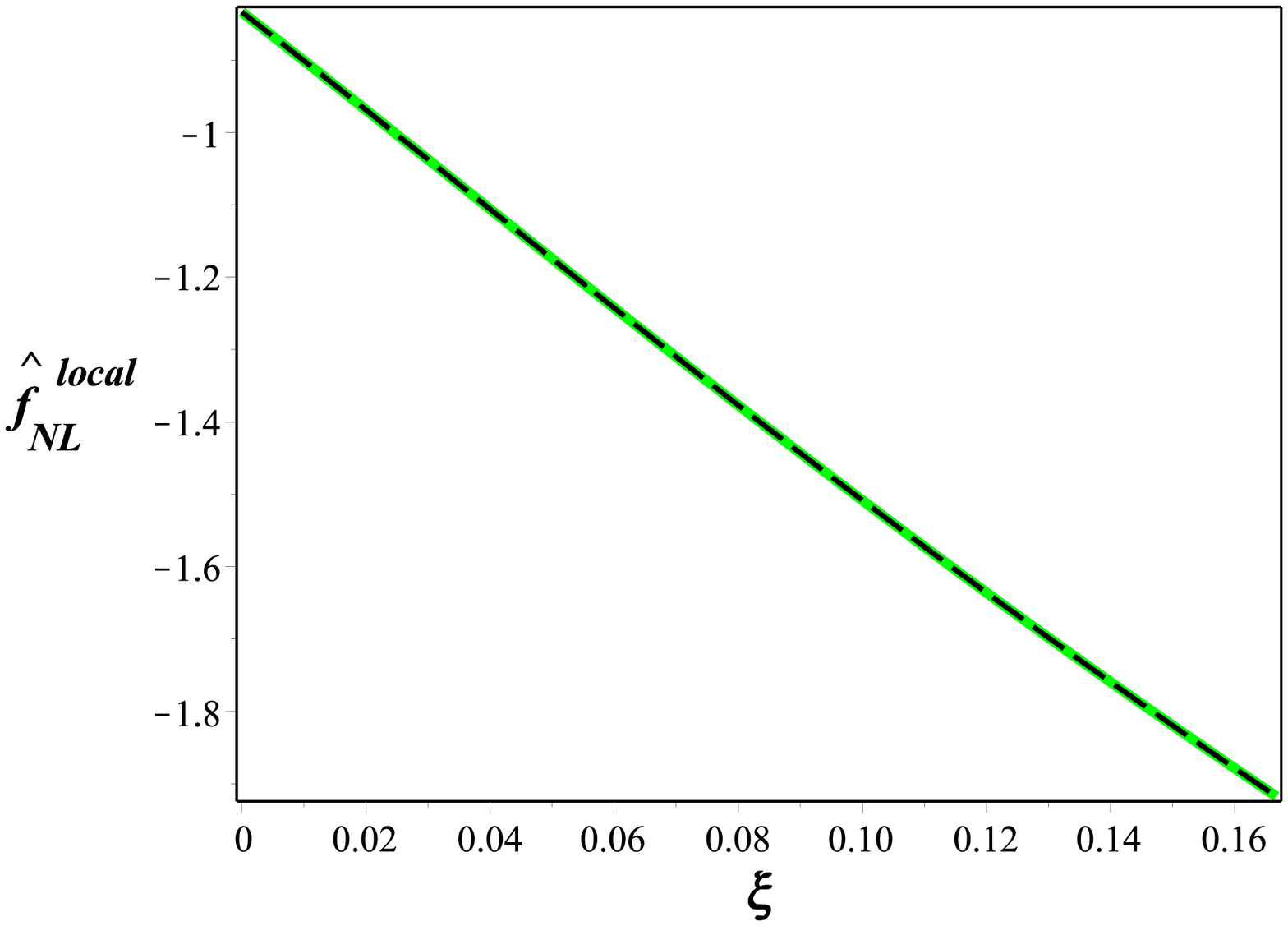}}
\caption{\label{fig:ein-fNL} Left Panel: The non-linear parameter
$\hat{f}_{NL}^{local}$ for the local type non-Gaussianity versus the
number of e-folds for various values of the non-minimal coupling
parameter in Einstein frame. Right Panel: The non-linear parameter
$\hat{f}_{NL}^{local}$ for the local type non-Gaussianity as a function
of the non-minimal coupling parameter in Einstein frame about the
time that cosmological scales exit the Hubble horizon.}
\end{figure*}

\begin{figure*}[htp]
\flushleft\leftskip0em{
\includegraphics[width=.40\textwidth,origin=c,angle=0]{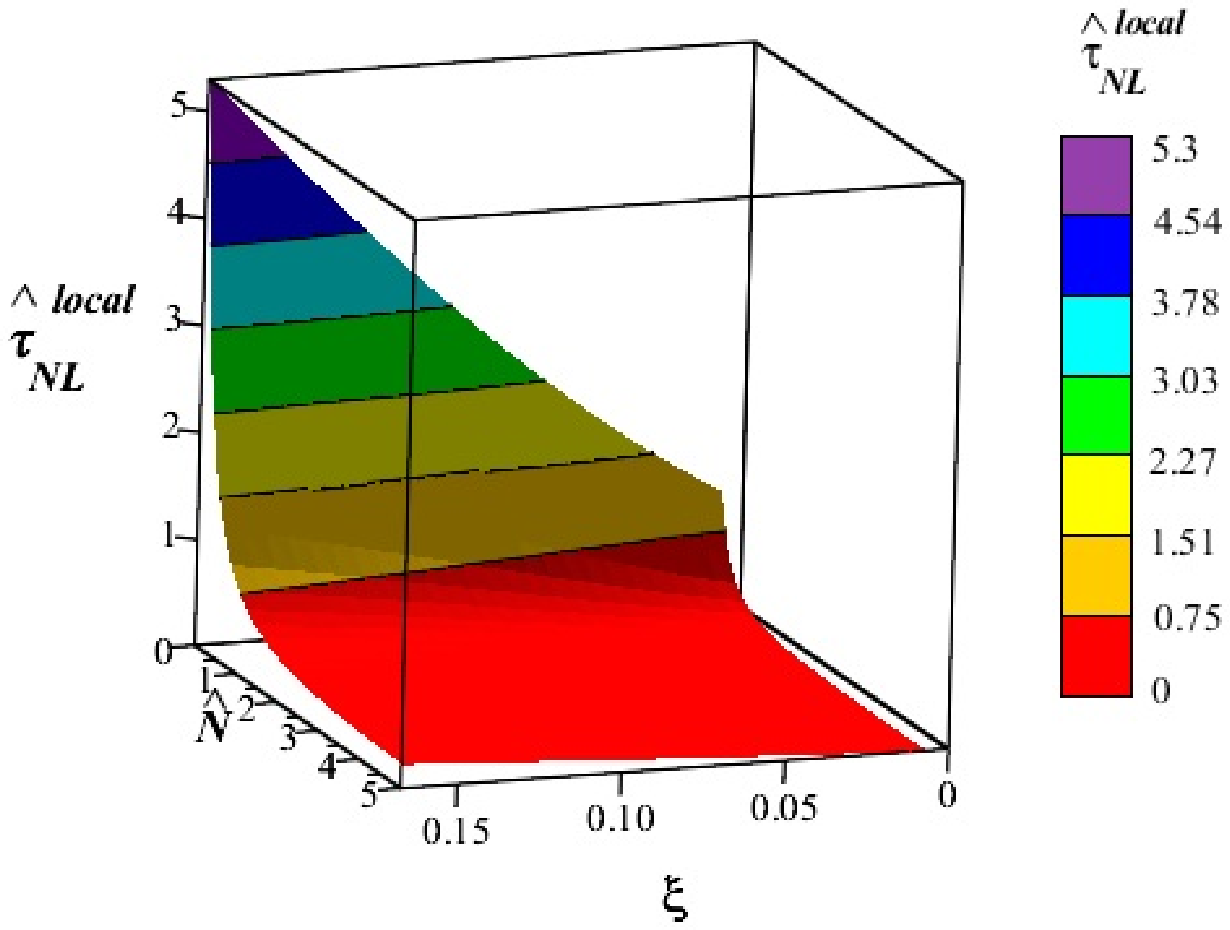}
\hspace{1cm}
\includegraphics[width=.44\textwidth,origin=c,angle=0]{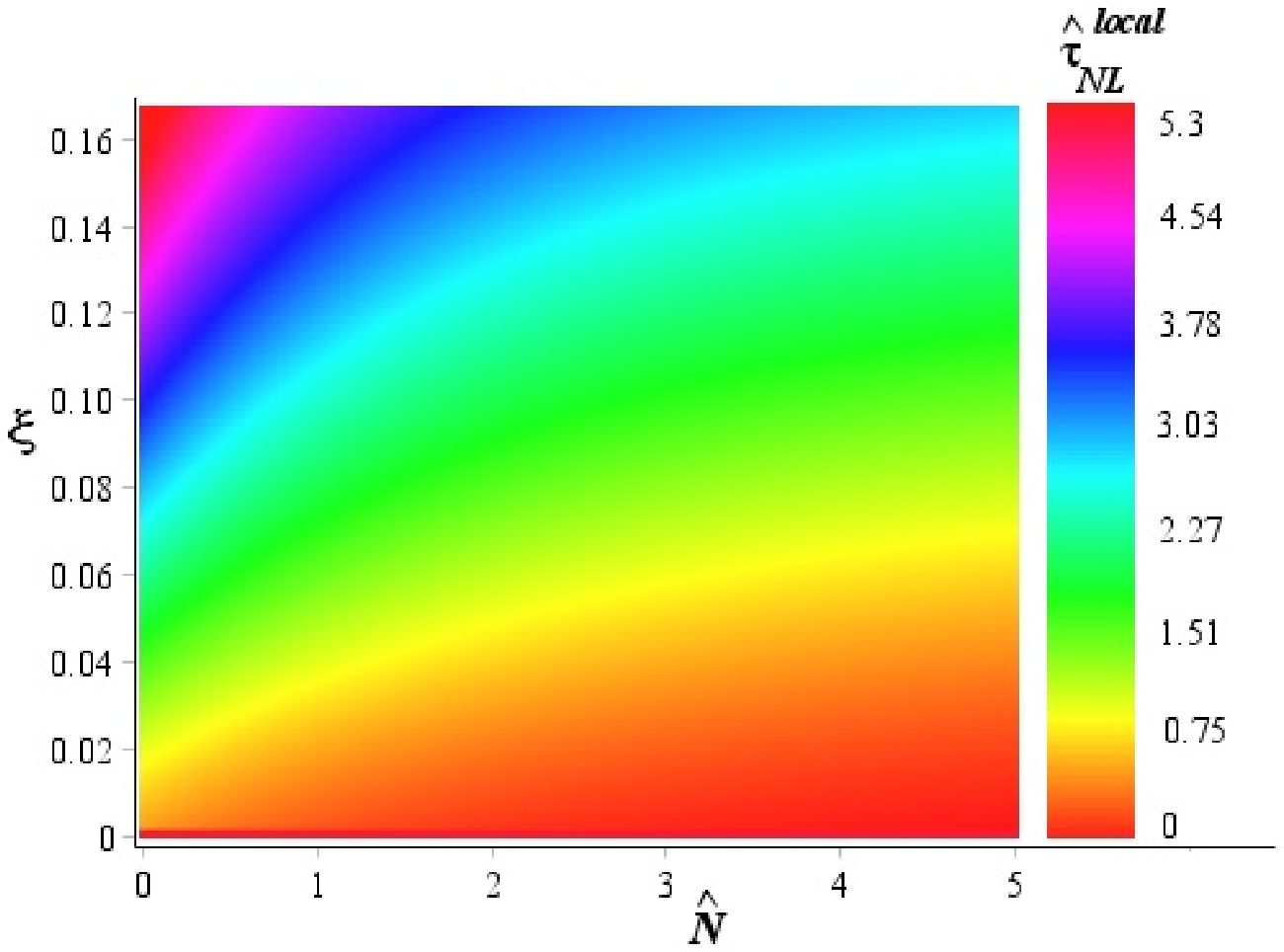}}
\caption{\label{fig:ein-TNL3d} The non-linear parameter
$\hat{\tau}_{NL}^{local}$ as a function of the non-minimal coupling
parameter $\xi$ and the number of e-folds $\hat{N}$ for non-minimal
inflation in Einstein frame.}
\end{figure*}

\begin{figure*}[htp]
\flushleft\leftskip0em{
\includegraphics[width=.42\textwidth,origin=c,angle=0]{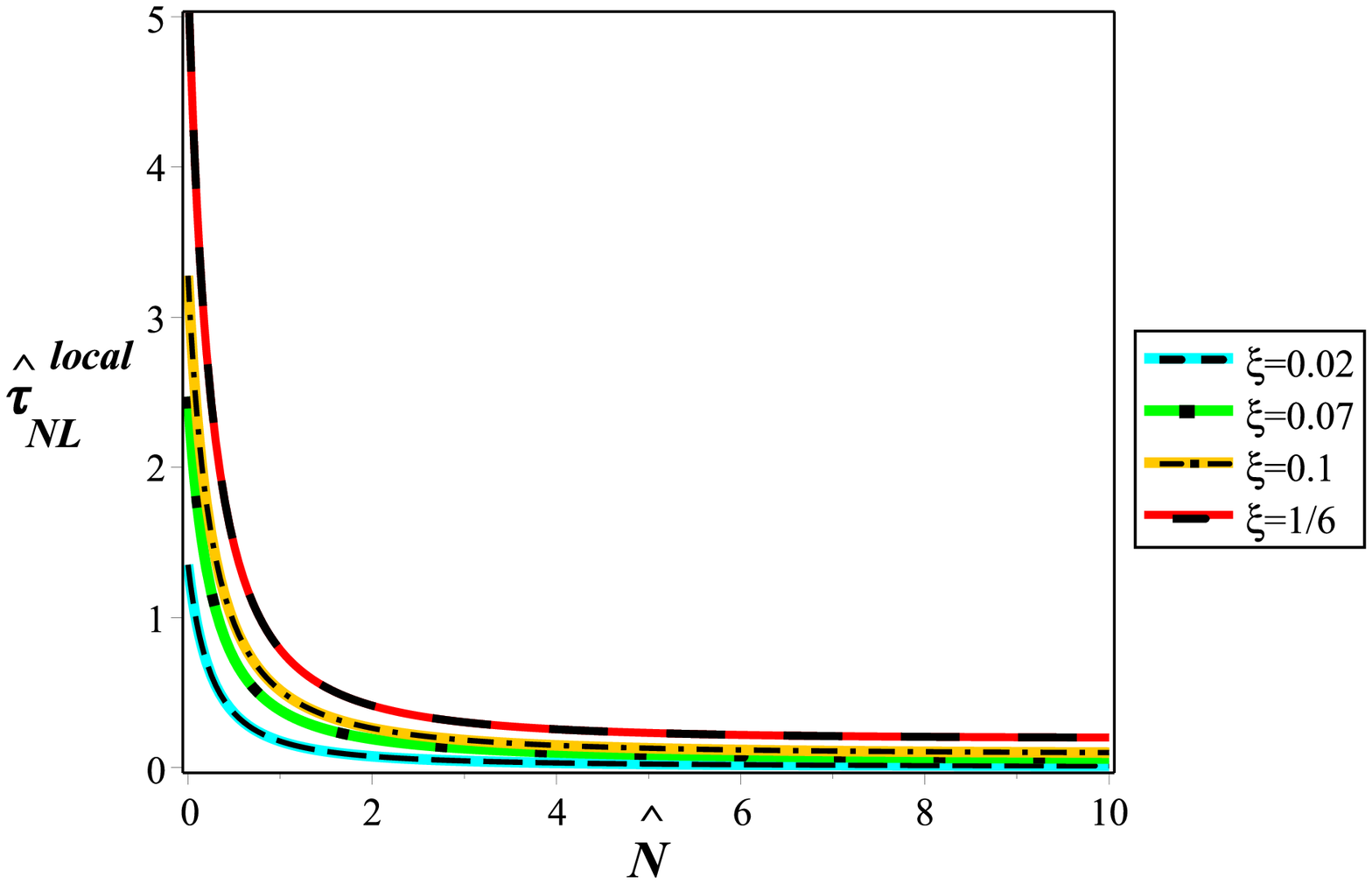}
\hspace{1.5cm}
\includegraphics[width=.37\textwidth,origin=c,angle=0]{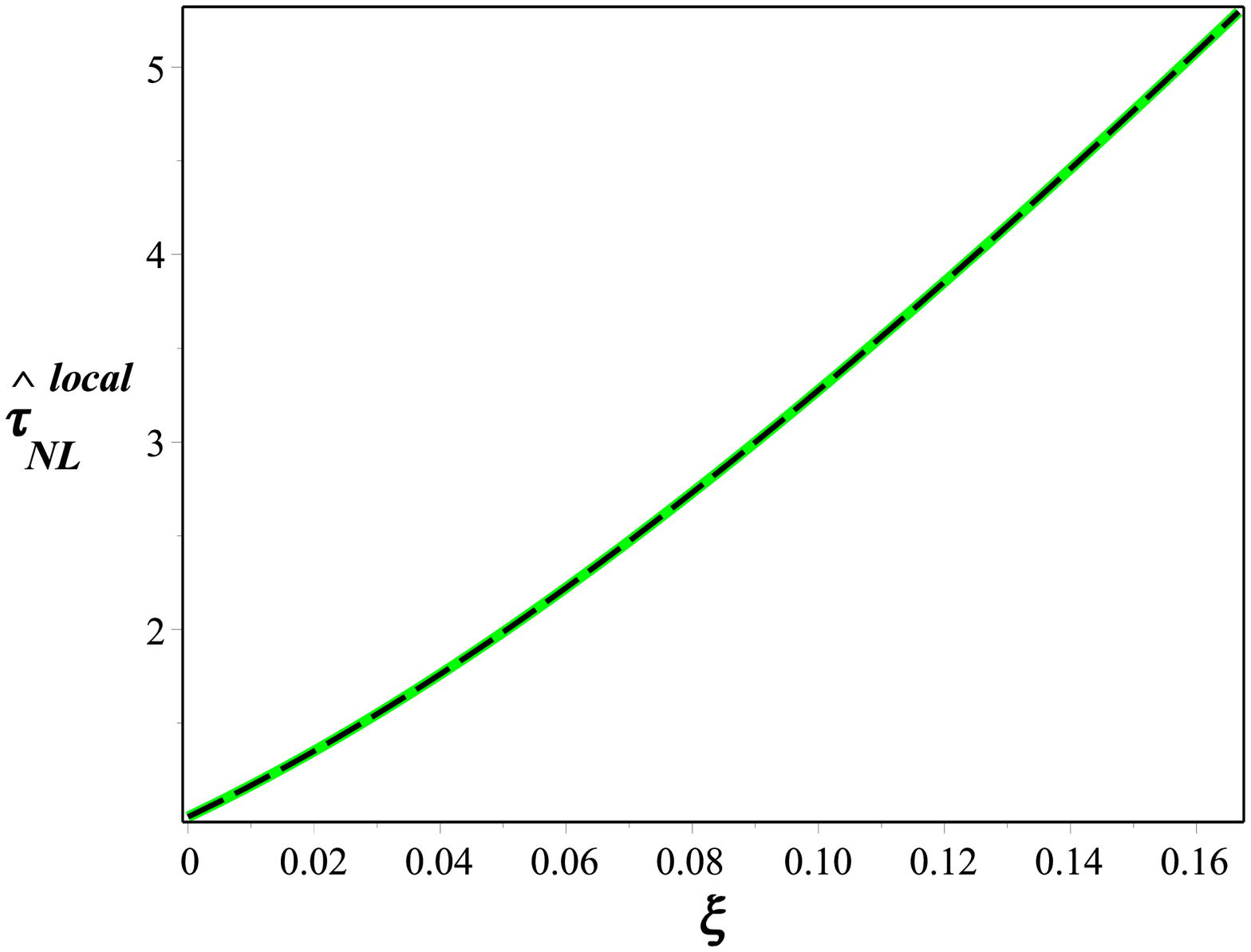}}
\caption{\label{fig:ein-TNL} Left Panel: The non-linear parameter
$\hat{\tau}_{NL}^{local}$ for the local type non-Gaussianity versus the
number of e-folds for various values of the non-minimal coupling
parameter for non-minimal inflation in Einstein frame. Right
Panel: The non-linear parameter $\hat{\tau}_{NL}^{local}$ for the local
type non-Gaussianity as a function of the non-minimal coupling
parameter about the time that cosmological scales exit the Hubble
horizon for non-minimal inflation in Einstein frame.}
\end{figure*}

\begin{figure*}[htb]
\flushleft\leftskip0em{
\includegraphics[width=.40\textwidth,origin=c,angle=0]{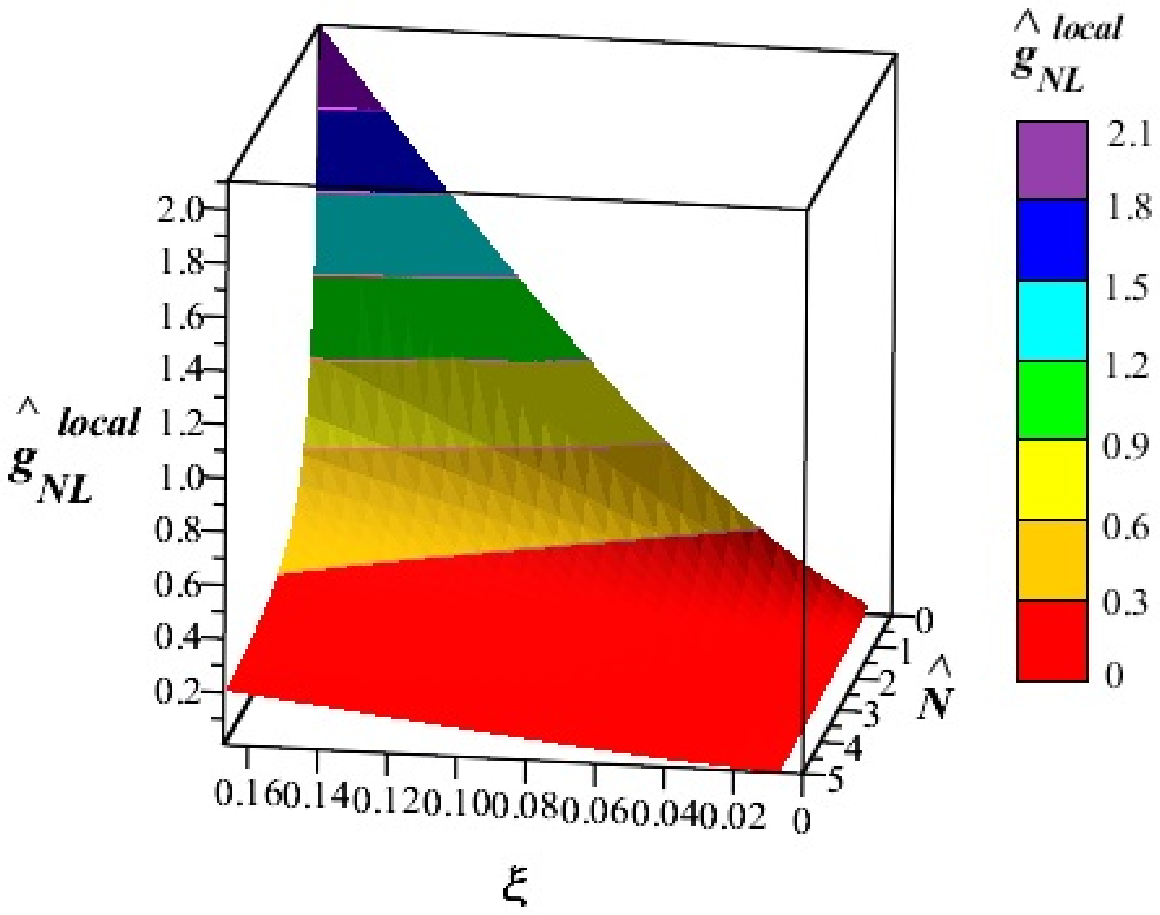}
\hspace{1cm}
\includegraphics[width=.44\textwidth,origin=c,angle=0]{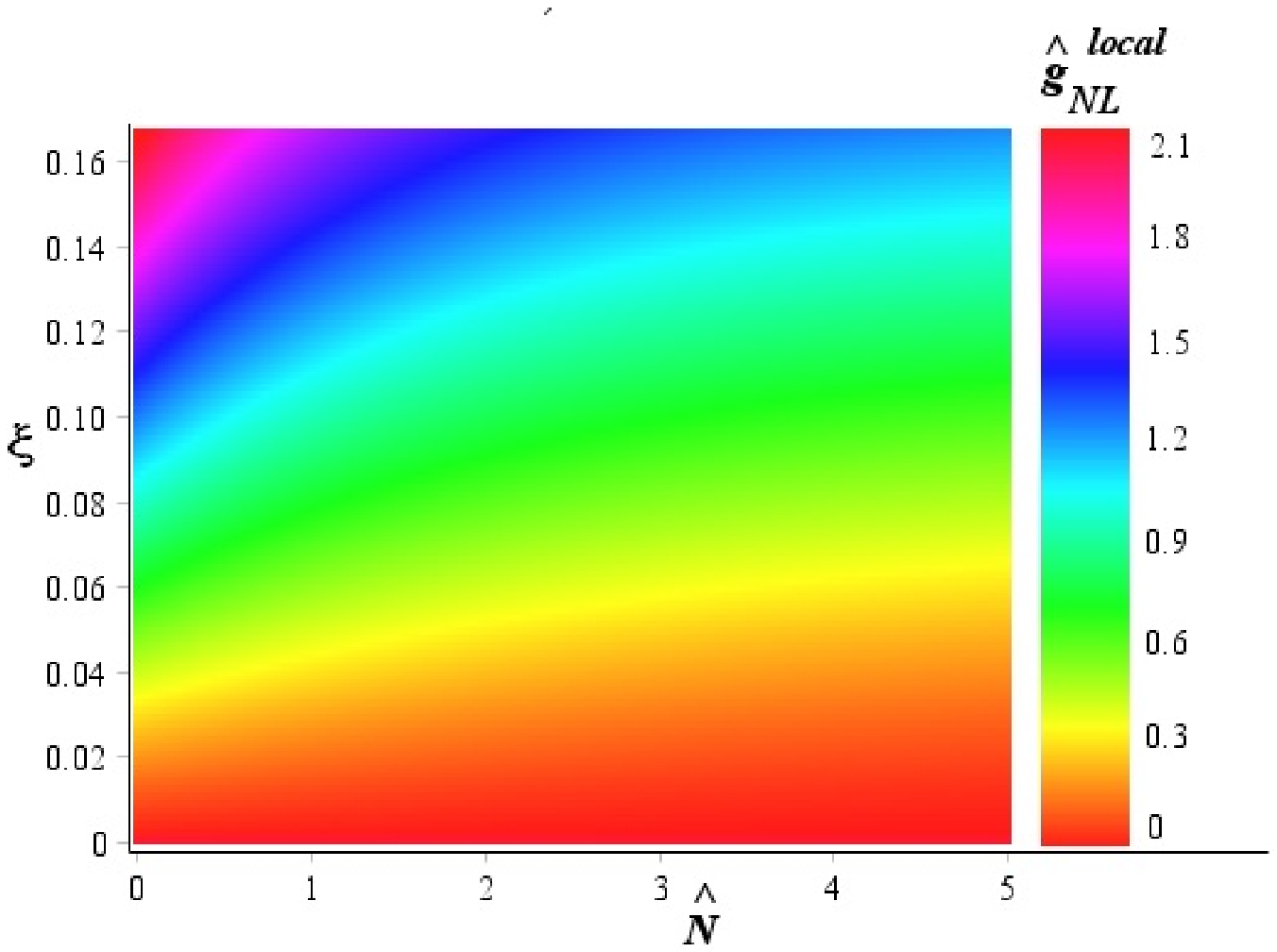}}
\caption{\label{fig:ein-gNL3d} The non-linear parameter
$\hat{g}_{NL}^{local}$ as a function of the non-minimal coupling
parameter $\xi$ and the number of e-folds $\hat{N}$ for non-minimal
inflation in Einstein frame.}
\end{figure*}

\begin{figure*}
\flushleft\leftskip0em{
\includegraphics[width=.44\textwidth,origin=c,angle=0]{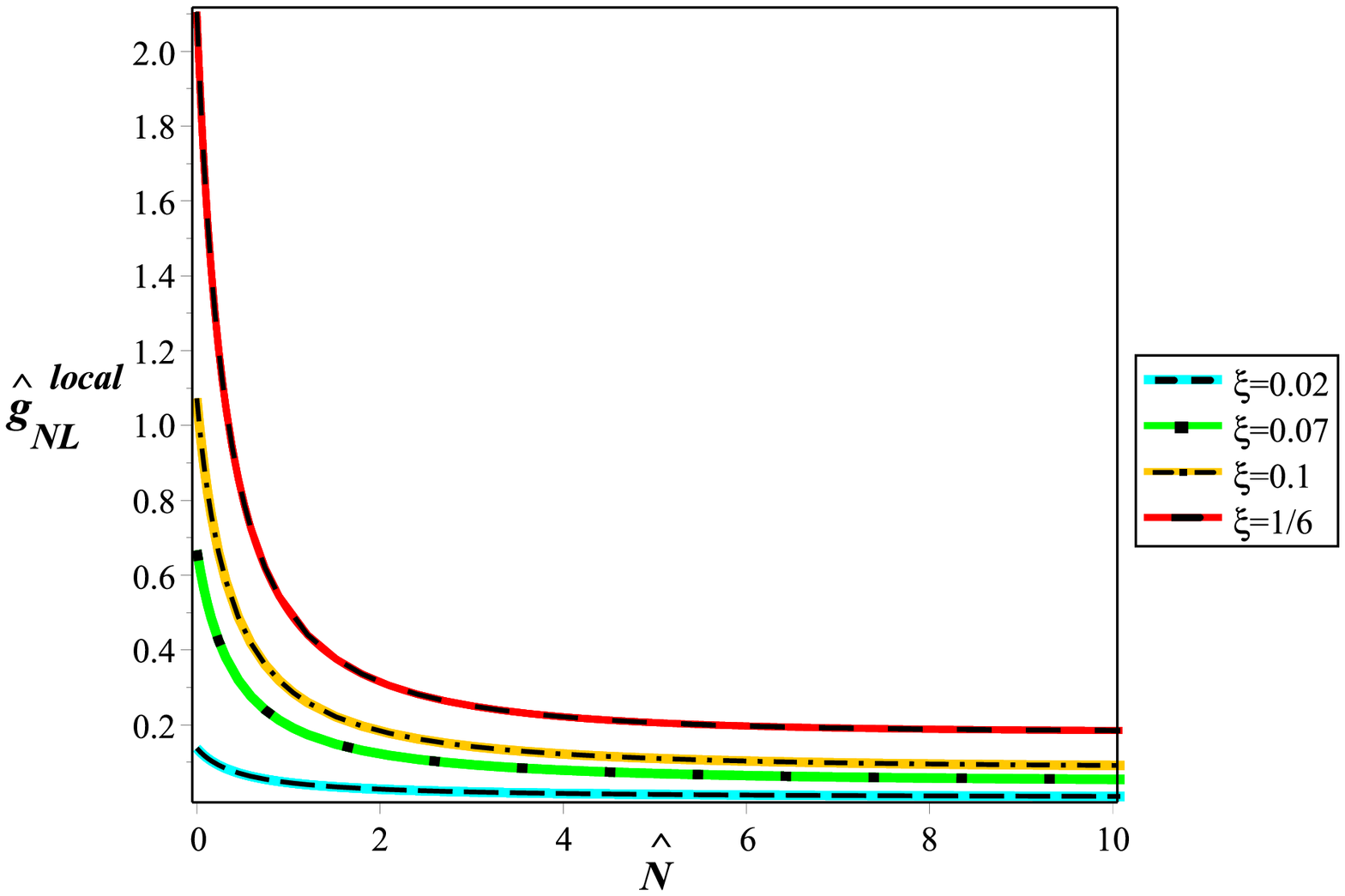}
\hspace{1cm}
\includegraphics[width=.39\textwidth,origin=c,angle=0]{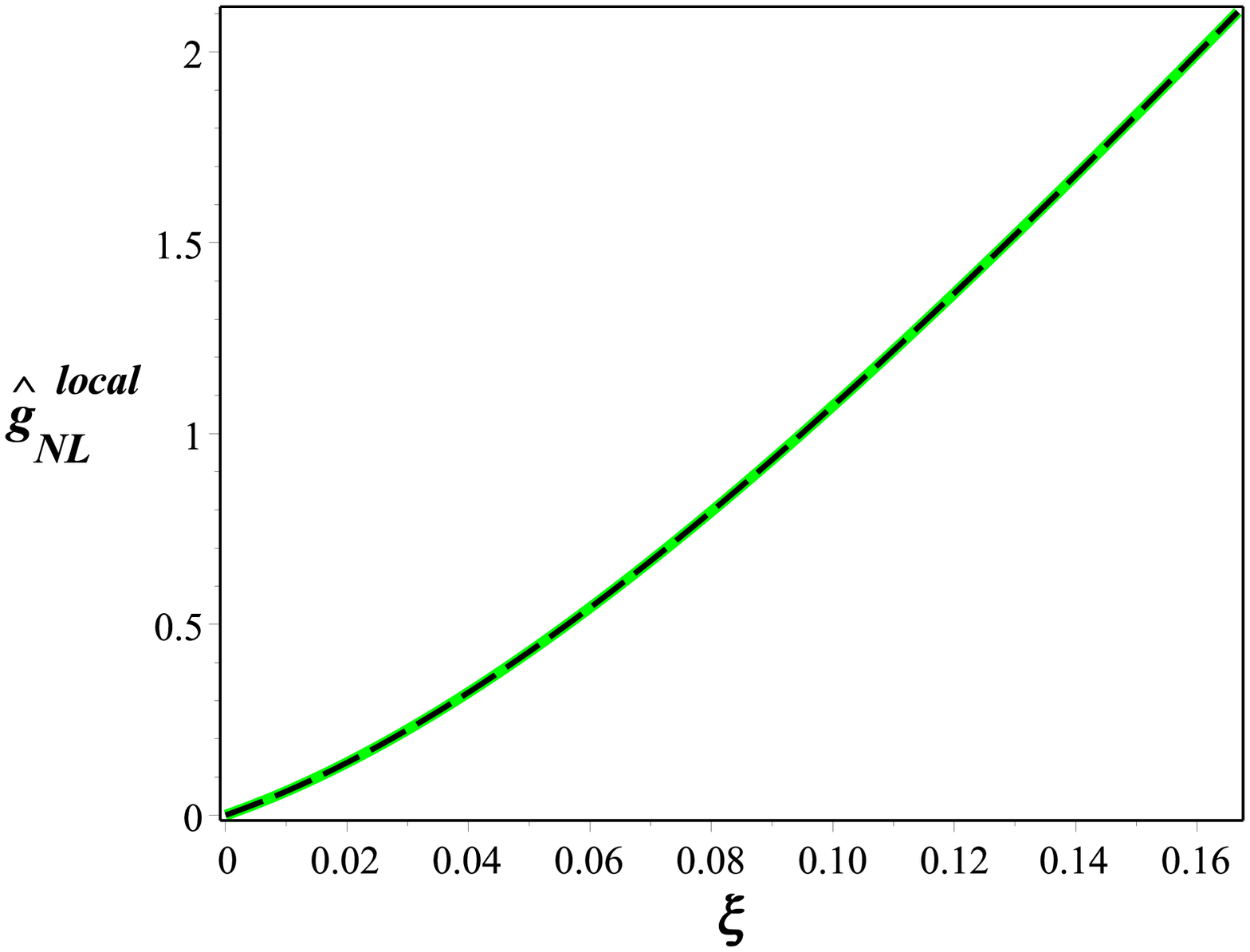}}
\caption{\label{fig:ein-gNL} Left Panel: The non-linear parameter
$\hat{g}_{NL}^{local}$ for the local type non-Gaussianity versus the
number of e-folds for various values of the non-minimal coupling
parameter for non-minimal inflation in Einstein frame. Right
panel: The non-linear parameter $\hat{g}_{NL}^{local}$ for the local
type non-Gaussianity as a function of the non-minimal coupling
parameter about the time that cosmological scales exit the Hubble
horizon for non-minimal inflation in Einstein frame.}
\end{figure*}

As for Jordan frame, we should firstly obtain the value of the inflaton field at the end of
inflation, $\phi_{e}$. By studying the evolution of the slow-roll
parameters, one can find the value of the inflaton field at the end of inflation (that is, for $\hat{\epsilon}\rightarrow 1$ or $\hat{\eta}\rightarrow 1$).
Using the definition of the slow-roll parameters in Einstein frame as $\hat{\epsilon}=\frac{1}{2\kappa^2}\Big(\frac{\hat{V}_{,\hat{\phi}}}{\hat{V}}\Big)^2$
and $\hat{\eta}=\frac{1}{\kappa^2}\Big(\frac{\hat{V}_{,\hat{\phi}\hat{\phi}}}{\hat{V}}\Big)$, we obtain the following expressions
\begin{eqnarray}\label{ein-epsilon2}
\hat{\epsilon}=\frac{1}{2\kappa^2}\frac{\Big(V_{,\phi}(1+2\kappa^2f)-4
\kappa^2Vf_{,\phi}\Big)^2}{V^2\Big(1+2\kappa^2f+6\kappa^2f_{,\phi}^2\Big)}\,,
\end{eqnarray}
and
\begin{eqnarray}\label{ein-eta2}
\hat{\eta}=\frac{(1+2\kappa^2f)^2}{\kappa^2 V}\frac{d^2}{d\hat{\phi}^2}\Bigg(\frac{V}{(1+2\kappa^2f)^2}\Bigg)\,.
\end{eqnarray}
Fig.~\ref{fig:ein-epsilon} shows the behavior of $\hat{\epsilon}$ versus
$\hat{\phi}$ and $\xi$. It is obvious from this figure that the value of
$\hat{\phi}_{e}$ is directly dependant on the non-minimal coupling
parameter. The value of the field at the end of inflation in
Einstein frame is given by the following expression
\begin{eqnarray}\label{ein-phi_e}
\hat{\phi_{e}}=\hspace{7cm}\\\nonumber{\frac {\sqrt {\xi\, \left(4\,\xi+1 \right) \left( -4
\,\xi-1+\sqrt {48\,{\xi}^{2}+16\,\xi+1} \right)
}}{\sqrt{2}\xi\,\kappa \left( 4\,\xi +1 \right)}}\,.
\end{eqnarray}
After obtaining the field value at the end of inflation in
Einstein frame, now we are in the position to study the
non-Gaussianity of the model at hand through its non-linear
parameters. Fig.~\ref{fig:ein-fNL3d} demonstrates the relation
between $\hat{f}_{NL}^{local}$ and parameters $\xi$ and $\hat{N}$ in Einstein
frame. This figure confirms that for any value of $\xi$, the
absolute value of $\hat{f}_{NL}^{local}$ increases during the
inflation era in this frame (similar to the Jordan frame case). This
property is more clarified in the left panel of
Fig.~\ref{fig:ein-fNL} which shows the behavior of this
parameter in the number of e-folds space for various values of
$\xi$. The right panel of Fig.~\ref{fig:ein-fNL} also shows the
behavior in $\xi$ space in Einstein frame about the end of
inflation. As an important result we can see from the left panel
of Fig.~\ref{fig:ein-fNL} that the first order non-Gaussian effect
is almost constant and small at the initial stage of the
inflationary era and then grows gradually to values of order unity at
the time of horizon crossing (similar to the Jordan frame case).
The behaviors of trispectrum non-linear parameters, $\hat{\tau}_{NL}^{local}$
and $\hat{g}_{NL}^{local}$ in $\xi$ and $\hat{N}$ spaces in Einstein frame
are shown in Figs.~\ref{fig:ein-TNL3d} and
~\ref{fig:ein-gNL3d}. Moreover, the behavior of the trispectrum
non-linear parameters versus $\hat{N}$ are
depicted in the left panels of Figs.~\ref{fig:ein-TNL} and
~\ref{fig:ein-gNL}. In order to see the effect of the
non-minimal coupling parameter on the level of trispectrum in
Einstein frame, the right panels of Figs.~\ref{fig:ein-TNL} and
~\ref{fig:ein-gNL} have been plotted about the end of
inflation. We note that while the overall behaviors of non-linear parameters in two frames
are the same, the values of these parameters are not the same in two frames as usual.
Specially, the value of the non-linear parameter $\hat{g}_{NL}^{local}$ in Einstein
frame is not as large as the $\hat{g}_{NL}^{local}$ values in Jordan frame.
This is reasonable since in Einstein frame one essentially deals with a
minimally coupled redefined scalar field and it is natural to recover the
standard results for a minimally coupled single scalar field as has been
reported in \citep{NGu}. Once again we emphasize that the larger values of $\hat{g}_{NL}^{local}$ in
Jordan frame in comparison with Einstein frame has its origin on the
non-minimal coupling between the scalar field and curvature in Jordan frame.

\begin{figure*}
\flushleft\leftskip0em{
\includegraphics[width=.43\textwidth,origin=c,angle=0]{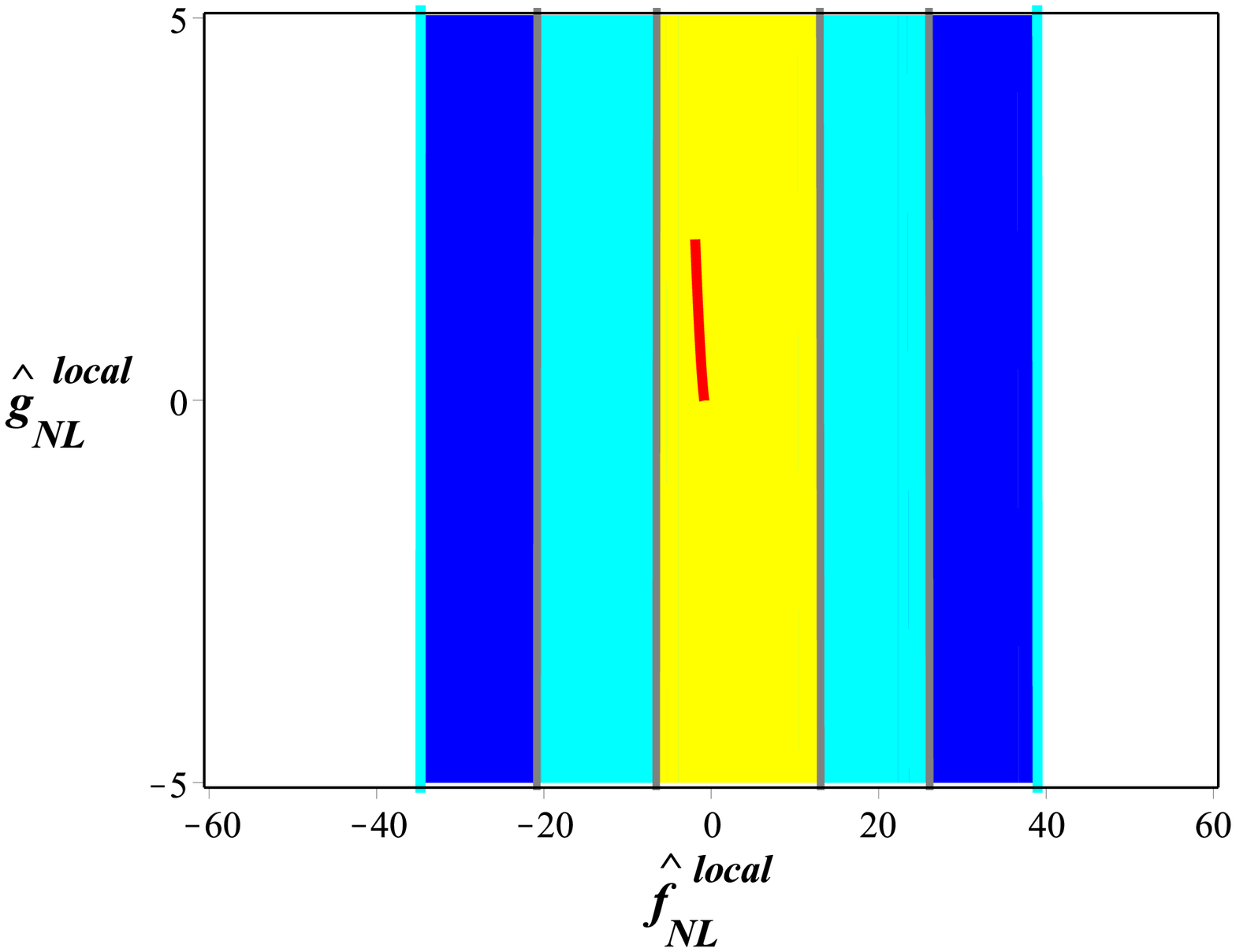}
\hspace{1cm}
\includegraphics[width=.42\textwidth,origin=c,angle=0]{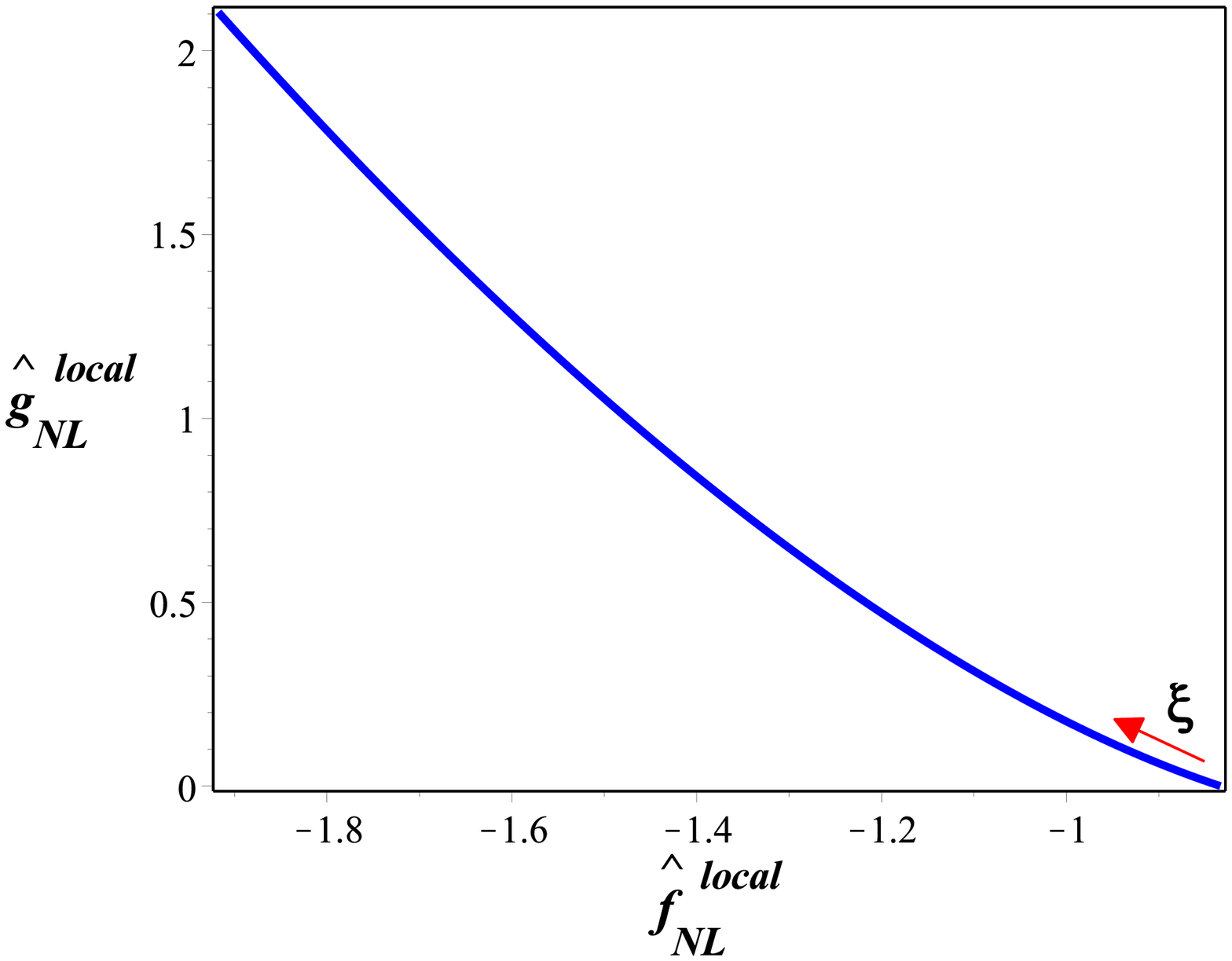}}
\caption{\label{fig:ein-gNL-fNL} The amplitude of $\hat{g}_{NL}^{local}$
versus $\hat{f}_{NL}^{local}$ for local configuration of non-Gaussianity
in Einstein frame in the background of Planck2015 data. These
figures are plotted about the end of inflation for a quadratic
potential and the non-minimal coupling function as
$f(\phi)\sim\xi\phi^2$. We note that the red
line in this figure shows the position of our result in the background of the observational data.
The right panel highlights more details of
$\hat{g}_{NL}^{local}$ versus $\hat{f}_{NL}^{local}$ in terms of $\xi$.}
\end{figure*}

\begin{figure*}
\flushleft\leftskip0em{
\includegraphics[width=.44\textwidth,origin=c,angle=0]{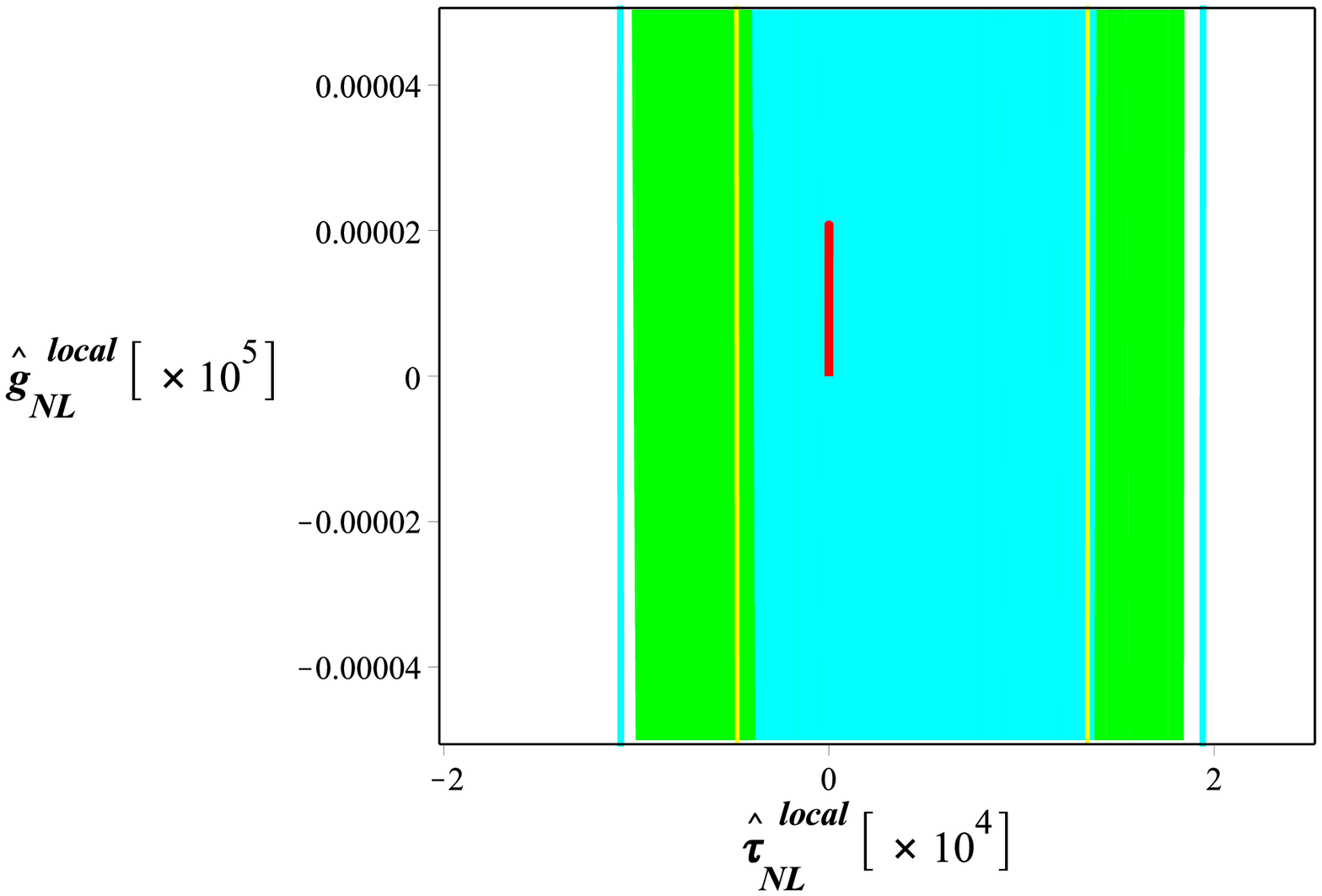}
\hspace{1cm}
\includegraphics[width=.41\textwidth,origin=c,angle=0]{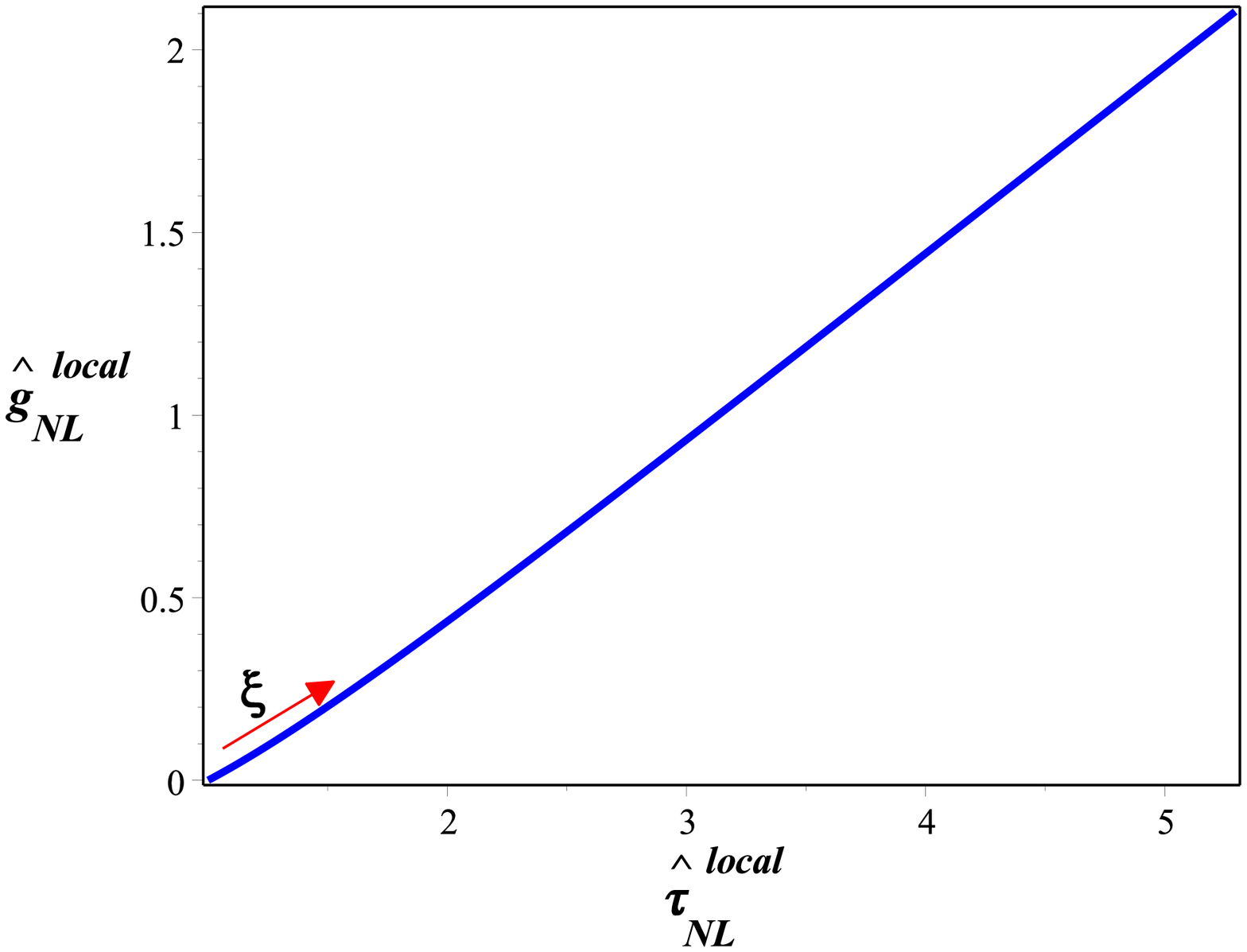}}
\caption{\label{fig:ein-gNL-TNL} The amplitude of $\hat{g}_{NL}^{local}$
versus $\hat{\tau}_{NL}^{local}$ for local configuration of
non-Gaussianity in Einstein frame in the background of Planck2013
data. These figures are plotted about the end of inflation for a
quadratic potential and the non-minimal coupling function as
$f(\phi)\sim\xi\phi^2$. We note that the red
line in this figure shows the position of our result in the background of
the observational data. The right panel highlights more details of
$\hat{g}_{NL}^{local}$ versus $\hat{\tau}_{NL}^{local}$ in terms of $\xi$.}
\end{figure*}

After illustrating the behavior of bispectrum and trispectrum
separately, now we are in the position to constrain the model in
Einstein frame with observations. Similar to our numerical
analysis in Jordan frame, we explore the behavior of
$\hat{f}_{NL}^{local}$ versus $\hat{g}_{NL}^{local}$ with local configuration
in the background of the Planck2015 data to see the viability of
the inflation in Einstein frame in confrontation with
the recent observations. This behavior is shown in
Fig.~\ref{fig:ein-gNL-fNL}. Similar to our result in Jordan
frame, this figure confirms that although our non-Gaussian results
in Einstein frame are somehow small, however
they are well inside the region allowed by the Planck observations.
This figure confirms also that our model is consistent with observation
for all values of $\xi$, that is $0<\xi\leq\frac{1}{6}$, in
Einstein frame. Furthermore, Fig.~\ref{fig:ein-gNL-TNL} shows
the amplitude of $\hat{\tau}_{NL}^{local}$ versus $\hat{g}_{NL}^{local}$ for
local configuration of non-Gaussianity in Einstein frame with
quadratic functions for $V(\phi)$ and $f(\phi)$ in the background
of Planck2013 data. It is again clear that a non-minimal inflation
in Einstein frame is consistent with observation for all values of $\xi$.
We note that the results obtained in this case are
compatible with the results obtained for $\hat{f}_{NL}^{local}$ versus
$\hat{g}_{NL}^{local}$. Moreover, comparing these results, which are
obtained in Einstein frame, with previous results obtained in
Jordan frame overall confirms that both cases match together as
well.

\section{Summary and Conclusion}

In this paper we have studied the dynamics of an inflationary
model driven by an scalar field that is coupled non-minimally with
gravity. At first, we have obtained the main equations of the
model. Then, imposing slow-roll approximation, we have studied
bispectrum and trispectrum of the curvature perturbations. We
focused on the local shape of non-Gaussianity which has a peak at
the squeezed limit (where two wave numbers are much larger than
the third one, $k_1=k_2\gg k_3$). This shape of non-Gaussianity
illustrates large super-Hubble interactions. On super-horizon
scales we can only use the evolution of unperturbed separate
universes and neglect spatial gradients. To proceed further, we
applied a formalism which provides a powerful tool in evaluating
the evolution of the curvature perturbation on this scales, the
so-called $\delta N$ formalism. The main advantage of this
approach is that allows the primordial curvature perturbation to
be related to the difference of $N$ between the perturbed universe
and the homogeneous background one. We calculated these values
between an initially flat hypersurface (with $t_{*}$ corresponding
to the time of horizon crossing) and a final uniform energy
density hypersurface (with $t_{e}$ referring to the end of
inflation) for the model at hand. By choosing the flat slicing
gauge and considering perturbations, we have expanded the inflaton
field around a homogeneous background and local perturbation and
found $\zeta$ as an expansion of $\delta \phi$. Then we have
employed the $\delta N$ expansion and used the two-point
correlation function to find the dimensionless power spectrum in
terms of the derivatives of $N$ with respect to the inflaton
field. In order to obtain the lowest order of non-Gaussianity, the
bispectrum $B_{\zeta}$, we applied the three-point correlation
function. Since in this work we have been interested in local
non-Gaussianities, which are confirmed on super-horizon scales, we
only need the momentum independent term of the bispectrum that
accounts for the super-horizon contribution. Next we have obtained
the non-linear parameter associated to the first order of local
non-Gaussianity, $f_{NL}^{local}$, in terms of $N_{,\phi}$ and
$N_{,\phi\phi}$. Furthermore, to describe the four-point
correlation function, or the trispectrum, using $\delta N$
formalism, we have obtained the relevant non-linear parameters,
$\tau_{NL}^{local}$ and $g_{NL}^{local}$. Although
$\tau_{NL}^{local}$ depends only on the first and second
derivatives of $N$ (which makes it capable to be expressed in
terms of $f_{NL}^{local}$), $g_{NL}^{local}$ depends on the third
order of this derivative, $N_{,\phi\phi\phi}$ too. After
explaining the numerical method for calculating derivatives of the
unperturbed number of e-folds with respect to the unperturbed
inflaton field at horizon crossing and showing how these
derivatives are related to the field description, we obtained the
exact form of parameters in terms of $\xi$ and $\phi$. After
deriving the mentioned non-linear parameters as a product of $N$'s
derivatives, we have studied their evolution. To this end, we have
first adopted a quadratic form for both potential and non-minimal
coupling function as $V(\phi)=\frac{1}{2}m_{\phi}^2 \phi^2$ and
$f(\phi)=\frac{1}{2}\xi\phi^2$. Finally, after obtaining the
field's value during slow-roll inflation in terms of the number of
e-folds, we were able to depict the evolution of the non-linear
parameters in $\xi$ and $N$ space. One can see
Fig.~\ref{fig:1}-~\ref{fig:7} to follow how $f_{NL}^{local}$,
$\tau_{NL}^{local}$ and $g_{NL}^{local}$ evolve in $N$ and $\xi$
space, both simultaneously and separately.

Eventually we have explored the behavior of the $f_{NL}^{local}$
versus $g_{NL}^{local}$ with local configuration in the background
of the Planck2015 data to see the viability of this theoretical
model. Our analysis confirms that for all chosen values of $\xi$
the model is consistent with observation. We have also studied the
behavior of $g_{NL}^{local}$ versus $\tau_{NL}^{local}$ using the
results of \citep{Fen15}. Our result in this case is also consistent with observation for all values of the
non-minimal coupling parameter, $\xi$.

As an important achievement of this study, although we haven't
obtained large local non-Gaussianities of perturbations in this non-minimal setup, however it is in well agreement with
Planck observations. Since there is no observational difference in between Jordan and
Einstein frames at least in the single field case, we have moved to
Einstein frame to see the situation in this frame. We have checked the consistency conditions in this frame. The non-Gaussianities are small in this frame too.

\newpage
{\bf Appendix \textbf{A}: Derivatives of the number of e-folds
with respect to the scalar field in Einstein frame}\\

Using Eq. (\ref{ein-N-3}), the first and second derivatives of the
number of e-folds take the following forms
\begin{eqnarray}\label{ein-N,phi}
\hat{N}_{,\phi}=-\kappa^2\frac{V\left[(1+2\kappa^2f)+6\kappa^2f_{,\phi}^2\right]}
{(1+2\kappa^2f)\left[V_{,\phi}(1+2\kappa^2f)-4\kappa^2Vf_{,\phi}\right]}\,,\hspace{0.7cm}
\end{eqnarray}
and
\begin{widetext}
\begin{eqnarray}\label{ein-Nz}
\hat{N}_{,\phi\phi}=-\kappa^2\frac{V_{,\phi}\left[(1+2\kappa^2f)+6\kappa^2f_{,\phi}^2\right]}
{(1+2\kappa^2f)\left[V_{,\phi}(1+2\kappa^2f)-4\kappa^2Vf_{,\phi}\right]}-
\kappa^2\frac{V\left[2\kappa^2f_{,\phi}+12\kappa^2f_{,\phi}f_{,\phi\phi}\right]}
{(1+2\kappa^2f)\left[V_{,\phi}(1+2\kappa^2f)-4\kappa^2Vf_{,\phi}\right]}\hspace{1cm}\\\nonumber
+\kappa^2\frac{V\left[(1+2\kappa^2f)+6\kappa^2f_{,\phi}^2\right]
\left[V_{,\phi\phi}(1+2\kappa^2f)+2\kappa^2V_{,\phi}f_{,\phi}-4\kappa^2V_{,\phi}f_{,\phi}-4\kappa^2Vf_{,\phi\phi}\right]}
{(1+2\kappa^2f)\left[V_{,\phi}(1+2\kappa^2f)-4\kappa^2Vf_{,\phi}\right]^2}\\\nonumber
+\kappa^2\frac{2\kappa^2Vf_{,\phi}\left[(1+2\kappa^2f)+6\kappa^2f_{,\phi}^2\right]}
{(1+2\kappa^2f)^2\left[V_{,\phi}(1+2\kappa^2f)-4\kappa^2Vf_{,\phi}\right]}\,,
\end{eqnarray}

respectively. Also, the third derivative of $N$ gives the following expression

\begin{eqnarray}\label{ein-Nppp}
\hat{N}_{,\phi\phi\phi}=\Bigg\{-\kappa^2V_{,\phi\phi}\left[(1+2\kappa^2f)+
6\kappa^2f_{,\phi}^2\right]-2\kappa^2V_{,\phi}\left[2\kappa^2f_{,\phi}+
12\kappa^2f_{,\phi}f_{,\phi\phi}\right]-\hspace{1.5cm}\\\nonumber
\kappa^2V\Big[2\kappa^2f_{,\phi\phi}+
12\kappa^2f_{,\phi\phi}^2+12\kappa^2f_{,\phi}f_{,\phi\phi\phi}\Big]+
\big[1+2\kappa^2f\big]^{-1}\Bigg[2\kappa^4f_{,\phi}V_{,\phi}\left[(1+2\kappa^2f)+
6\kappa^2f_{,\phi}^2\right]+\\\nonumber 2\kappa^4Vf_{,\phi\phi}
\Big[(1+2\kappa^2f)+6\kappa^2f_{,\phi}^2\Big]+2\kappa^4Vf_{,\phi}
\big[2\kappa^2f_{,\phi}+12\kappa^2f_{,\phi}f_{,\phi\phi}\big]\Bigg]-4\kappa^6Vf_{,\phi}^2(1+2\kappa^2f)^{-2}\times\\\nonumber
\big[(1+2\kappa^2f)+
6\kappa^2f_{,\phi}^2\big]+\Bigg[\kappa^2V_{,\phi}\left[(1+2\kappa^2f)+
6\kappa^2f_{,\phi}^2\right]\Big[V_{,\phi\phi}(1+2\kappa^2f)-2V_{,\phi}\kappa^2f_{,\phi}-4\kappa^2Vf_{,\phi\phi}\Big]+\\\nonumber
\kappa^2V\Big[2\kappa^2f_{,\phi}+12\kappa^2f_{,\phi}f_{,\phi\phi}\Big]\Big[V_{,\phi\phi}(1+2\kappa^2f)+
2\kappa^2V_{,\phi}f_{,\phi}-4\kappa^2V_{,\phi}f_{,\phi}-4\kappa^2Vf_{,\phi\phi}\Big]+\\\nonumber
\kappa^2V\big[(1+2\kappa^2f)+ 6\kappa^2f_{,\phi}^2\big]
\Bigg[V_{,\phi\phi\phi}(1+2\kappa^2f)-6\kappa^2V_{,\phi}f_{,\phi\phi}-
4\kappa^2Vf_{,\phi\phi\phi}\Bigg]\Bigg]\times\\\nonumber
\Bigg[V_{,\phi}(1+2\kappa^2f)-4\kappa^2Vf_{,\phi}\Bigg]^{-1}-\kappa^2V\Big[(1+2\kappa^2f)+6\kappa^2f_{,\phi}^2\Big]
\Big[V_{,\phi}(1+2\kappa^2f)-4\kappa^2Vf_{,\phi}\Big]^{-2}\times\\\nonumber
\Big[V_{,\phi\phi}(1+2\kappa^2f)+2V_{,\phi}\kappa^2f_{,\phi}-
4\kappa^2V_{,\phi}f_{,\phi}-4\kappa^2Vf_{,\phi\phi}\Big]^2\Bigg\}\Bigg[(1+2\kappa^2f)
\Big[V_{,\phi}(1+2\kappa^2f)-\\\nonumber
4\kappa^2Vf_{,\phi}\Big]\Bigg]^{-1}+
\Bigg\{\kappa^2V_{,\phi}\Big[(1+2\kappa^2f)+
6\kappa^2f_{,\phi}^2\Big]+ \kappa^2V\Big[2\kappa^2f_{,\phi}+
12\kappa^2f_{,\phi}f_{,\phi\phi}\Big]-2\kappa^2f_{,\phi}(1+2\kappa^2f)^{-1}\\\nonumber
\Big[\kappa^2V(1+2\kappa^2f)+6\kappa^2f_{,\phi}^2\Big]- \kappa^2V
\Big[V_{,\phi\phi}(1+2\kappa^2f)+2V_{,\phi}\kappa^2f_{,\phi}-
4\kappa^2V_{,\phi}f_{,\phi}-4\kappa^2Vf_{,\phi\phi}\Big]\times\\\nonumber
\Big[(1+2\kappa^2f)+6\kappa^2f_{,\phi}^2\Big]\Big[V_{,\phi}(1+2\kappa^2f)-4\kappa^2Vf_{,\phi}\Big]^{-1}\Bigg\}
{A}{\Bigg[(1+2\kappa^2f)\left[V_{,\phi}(1+2\kappa^2f)-4\kappa^2Vf_{,\phi}\right]\Bigg]^{-2}}\,,
\end{eqnarray}

where
\begin{eqnarray}\label{ein-A}
A=(1+2\kappa^2f)\Bigg[V_{,\phi\phi}(1+2\kappa^2f)-2\kappa^2V_{,\phi}f_{,\phi}-4\kappa^2V_{,\phi}f_{,\phi\phi}\Bigg]+\\\nonumber
2\kappa^2f_{,\phi}\left[V_{,\phi}(1+2\kappa^2f)-4\kappa^2Vf_{,\phi}\right]\,.
\end{eqnarray}
\end{widetext}

\newpage
\begin{widetext}
{\bf Appendix \textbf{B}: The first, second and third non-linear
parameters for the local configuration in Einstein frame}\\

\begin{eqnarray}\label{ein-fNL-V}
\frac{6}{5}\hat{f}_{NL}^{local}=-\frac{1}{\kappa^2}\frac{V_{,\phi}(1+2\kappa^2f)\left[V_{,\phi}(1+2\kappa^2f)-4\kappa^2Vf_{,\phi}\right]}
{V^2\left[(1+2\kappa^2f)+6\kappa^2f_{,\phi}^2\right]}-\hspace{4cm}\\\nonumber
\frac{1}{\kappa^2}\frac{\left[2\kappa^2f_{,\phi}+12\kappa^2f_{,\phi}f_{,\phi\phi}\right]
(1+2\kappa^2f)\left[V_{,\phi}(1+2\kappa^2f)-4\kappa^2Vf_{,\phi}\right]}
{V\left[(1+2\kappa^2f)+6\kappa^2f_{,\phi}^2\right]^2}+\hspace{3cm}\\\nonumber
\frac{1}{\kappa^2}\frac{(1+2\kappa^2f)\left[V_{,\phi\phi}(1+2\kappa^2f)+2\kappa^2V_{,\phi}f_{,\phi}-4\kappa^2V_{,\phi}f_{,\phi}-4\kappa^2Vf_{,\phi\phi}\right]}
{V\left[(1+2\kappa^2f)+6\kappa^2f_{,\phi}^2\right]}+
\frac{1}{\kappa^2}\frac{2\kappa^2f_{,\phi}\left[V_{,\phi}(1+2\kappa^2f)-4\kappa^2Vf_{,\phi}\right]}
{V\left[(1+2\kappa^2f)+6\kappa^2f_{,\phi}^2\right]}\,.
\end{eqnarray}

\begin{eqnarray}\label{ein-TNL-V}
\hat{\tau}_{NL}^{local}=\Bigg(-\frac{1}{\kappa^2}\frac{V_{,\phi}(1+2\kappa^2f)\left[V_{,\phi}(1+2\kappa^2f)-4\kappa^2Vf_{,\phi}\right]}
{V^2\left[(1+2\kappa^2f)+6\kappa^2f_{,\phi}^2\right]}-\hspace{4cm}\\\nonumber
\frac{1}{\kappa^2}\frac{\left[2\kappa^2f_{,\phi}+12\kappa^2f_{,\phi}f_{,\phi\phi}\right]
(1+2\kappa^2f)\left[V_{,\phi}(1+2\kappa^2f)-4\kappa^2Vf_{,\phi}\right]}
{V\left[(1+2\kappa^2f)+6\kappa^2f_{,\phi}^2\right]^2}+\hspace{4cm}\\\nonumber
\frac{1}{\kappa^2}\frac{(1+2\kappa^2f)\left[V_{,\phi\phi}(1+2\kappa^2f)+2\kappa^2V_{,\phi}f_{,\phi}-4\kappa^2V_{,\phi}f_{,\phi}-4\kappa^2Vf_{,\phi\phi}\right]}
{V\left[(1+2\kappa^2f)+6\kappa^2f_{,\phi}^2\right]}+
\frac{1}{\kappa^2}\frac{2\kappa^2f_{,\phi}\left[V_{,\phi}(1+2\kappa^2f)-4\kappa^2Vf_{,\phi}\right]}
{V\left[(1+2\kappa^2f)+6\kappa^2f_{,\phi}^2\right]}\Bigg)^2\,.
\end{eqnarray}

\begin{eqnarray}\label{ein--gNL-V}
\hat{g}_{NL}^{local}=\Bigg(\Bigg\{-\kappa^2V_{,\phi\phi}\left[(1+2\kappa^2f)+
6\kappa^2f_{,\phi}^2\right]-2\kappa^2V_{,\phi}\left[2\kappa^2f_{,\phi}+
12\kappa^2f_{,\phi}f_{,\phi\phi}\right]-\hspace{1.5cm}\\\nonumber
\kappa^2V\Big[2\kappa^2f_{,\phi\phi}+
12\kappa^2f_{,\phi\phi}^2+12\kappa^2f_{,\phi}f_{,\phi\phi\phi}\Big]+
\big[1+2\kappa^2f\big]^{-1}\Bigg[2\kappa^4f_{,\phi}V_{,\phi}\left[(1+2\kappa^2f)+
6\kappa^2f_{,\phi}^2\right]+\\\nonumber 2\kappa^4Vf_{,\phi\phi}
\Big[(1+2\kappa^2f)+6\kappa^2f_{,\phi}^2\Big]+2\kappa^4Vf_{,\phi}
\big[2\kappa^2f_{,\phi}+12\kappa^2f_{,\phi}f_{,\phi\phi}\big]\Bigg]-
4\kappa^6Vf_{,\phi}^2(1+2\kappa^2f)^{-2}\times\\\nonumber
\big[(1+2\kappa^2f)+6\kappa^2f_{,\phi}^2\big]+\Bigg[\kappa^2V_{,\phi}\left[(1+2\kappa^2f)+
6\kappa^2f_{,\phi}^2\right]\Big[V_{,\phi\phi}(1+2\kappa^2f)-2V_{,\phi}
\kappa^2f_{,\phi}-4\kappa^2Vf_{,\phi\phi}\Big]+\\\nonumber
\kappa^2V\Big[2\kappa^2f_{,\phi}+12\kappa^2f_{,\phi}f_{,\phi\phi}\Big]
\Big[V_{,\phi\phi}(1+2\kappa^2f)+2\kappa^2V_{,\phi}f_{,\phi}-4\kappa^2V_{,\phi}f_{,\phi}-
4\kappa^2Vf_{,\phi\phi}\Big]+\\\nonumber
\kappa^2V\big[(1+2\kappa^2f)+ 6\kappa^2f_{,\phi}^2\big]
\Bigg[V_{,\phi\phi\phi}(1+2\kappa^2f)-6\kappa^2V_{,\phi}f_{,\phi\phi}-
4\kappa^2Vf_{,\phi\phi\phi}\Bigg]\Bigg]\times\\\nonumber
\Bigg[V_{,\phi}(1+2\kappa^2f)-4\kappa^2Vf_{,\phi}\Bigg]^{-1}-
\kappa^2V\Big[(1+2\kappa^2f)+6\kappa^2f_{,\phi}^2\Big]
\Big[V_{,\phi}(1+2\kappa^2f)-4\kappa^2Vf_{,\phi}\Big]^{-2}\times\\\nonumber
\Big[V_{,\phi\phi}(1+2\kappa^2f)+2V_{,\phi}\kappa^2f_{,\phi}-
4\kappa^2V_{,\phi}f_{,\phi}-4\kappa^2Vf_{,\phi\phi}\Big]^2\Bigg\}
\Bigg[(1+2\kappa^2f)\Big[V_{,\phi}(1+2\kappa^2f)-\\\nonumber
4\kappa^2Vf_{,\phi}\Big]\Bigg]^{-1}+\Bigg\{\kappa^2V_{,\phi}\Big[(1+2\kappa^2f)+
6\kappa^2f_{,\phi}^2\Big]+ \kappa^2V\Big[2\kappa^2f_{,\phi}+
12\kappa^2f_{,\phi}f_{,\phi\phi}\Big]-2\kappa^2f_{,\phi}(1+2\kappa^2f)^{-1}\\\nonumber
\Big[\kappa^2V(1+2\kappa^2f)+6\kappa^2f_{,\phi}^2\Big]- \kappa^2V
\Big[V_{,\phi\phi}(1+2\kappa^2f)+2V_{,\phi}\kappa^2f_{,\phi}-
4\kappa^2V_{,\phi}f_{,\phi}-4\kappa^2Vf_{,\phi\phi}\Big]\times\\\nonumber
\Big[(1+2\kappa^2f)+6\kappa^2f_{,\phi}^2\Big]\Big[V_{,\phi}(1+2\kappa^2f)-4\kappa^2Vf_{,\phi}\Big]^{-1}\Bigg\}
{\Bigg[(1+2\kappa^2f)\left[V_{,\phi}(1+2\kappa^2f)-4\kappa^2Vf_{,\phi}\right]\Bigg]^{-2}}\times\\\nonumber
{\Bigg((1+2\kappa^2f)\Bigg[V_{,\phi\phi}(1+2\kappa^2f)-2\kappa^2V_{,\phi}f_{,\phi}-4\kappa^2V_{,\phi}f_{,\phi\phi}\Bigg]+
2\kappa^2f_{,\phi}\left[V_{,\phi}(1+2\kappa^2f)-4\kappa^2Vf_{,\phi}\right]\Bigg)}\Bigg)\times\\\nonumber
\Bigg(-\frac{(1+2\kappa^2f)^3\left[V_{,\phi}(1+2\kappa^2f)-4\kappa^2Vf_{,\phi}\right]^3}
{\kappa^6V^3\left[(1+2\kappa^2f)+6\kappa^2f_{,\phi}^2\right]^3}
\Bigg)\,.
\end{eqnarray}
\end{widetext}




\begin{thebibliography}{99}

\expandafter\ifx\csname natexlab\endcsname\relax\def\natexlab#1{#1}\fi
\providecommand{\url}[1]{\href{#1}{#1}}


\bibitem[{Ade {et~al.}(2013)}]{planck2013} Ade, P. A. R., et al. 2013, [arXiv:1303.5084].

\bibitem[{Ade {et~al.}(2015)}]{planck2015}  Ade, P. A. R., et al., 2015, [arXiv:1502.01592].

\bibitem[{Alabidi \& Lyth(2006)}]{NGq} Alabidi, L. \& Lyth, D. 2006, J. Cosmol. Astropart. Phys. \textbf{0608}, 006 [arXiv:astro-ph/0604569].

\bibitem[{Albrecht \& Steinhard(1982)}]{infc} Albrecht, A., \& Steinhard, P. 1982, Phys. Rev. D \textbf{48} 1220.

\bibitem[{Alishahiha {et~al.}(2004)}]{NGj} Alishahiha, M., Silverstein, E., \& Tong, D. 2004, Phys. Rev. D \textbf{70},
123505 [arXiv:hep-th/0404084].

\bibitem[{Babich {et~al.}(2004)}]{Bab04} Babich, D., Creminelli, P., \& Zaldarriaga, M. 2004, J. Cosmol. Astropart. Phys. \textbf{0408}, 009.

\bibitem[{Barnaby \& Cline(2006)}]{NGs} Barnaby, N., \& Cline, J. M. 2006, Phys. Rev. D \textbf{73}, 106012 [arXiv:astro-ph/0601481].

\bibitem[{Bartolo {et~al.}(2004{\natexlab{a}})}]{NGb} Bartolo, N., Komatsu, E., Matarrese, S., \& Riotto, A. 2004, Phys. Rept. \textbf{402}, 103 [arXiv:astro- ph/0406398].

\bibitem[{Bartolo {et~al.}(2004{\natexlab{b}})}]{pow1} Bartolo, N., Matarrese, S., \&  Riotto, A. 2004, Phys. Rev. Lett. \textbf{93}, 231301.

\bibitem[{Bartolo {et~al.}(2005)}]{4p2} Bartolo, N., Matarrese, S., \& Riotto, A. 2005, J. Cosmol. Astropart. Phys. \textbf{0508}, 010.

\bibitem[{Bartolo (2010{\natexlab{a}})}]{Bar1} Bartolo, N., Fasiello, M.,  Matarrese, S., \&  Riotto, A. 2010, J. Cosmol. Astropart. Phys. 1008 008.

\bibitem[{Bartolo (2010{\natexlab{b}})}]{Bar2} Bartolo, N., Fasiello, M.,  Matarrese, S., \&  Riotto, A. 2010, J. Cosmol. Astropart. Phys. 1009 035.

\bibitem[{Bassett {et~al.}(2006)}]{lssd} Bassett, B., Tsujikawa, S., \& Wands, D. 2006, Rev. Mod. Phys. \textbf{78},
537.

\bibitem[{Bernardeau \& Uzan(2003)}]{NGe} Bernardeau, F., \& Uzan, J. P. 2003, Phys. Rev. D \textbf{67}, 121301 [arXiv:astro-ph/0209330].

\bibitem[{Byrnes {et~al.}(2006)}]{NGu} Byrnes, C. T., Sasaki, M., \& Wands, D. 2006, Phys. Rev. D \textbf{74}, 123519.

\bibitem[{Byrnes {et~al.}(2007)}]{NGw} Byrnes, C. T., Koyama, K., Sasaki, M., \& Wands, D. 2007, J. Cosmol. Astropart. Phys. \textbf{0711}, 027.

\bibitem[{Byrnes {et~al.}(2008)}]{NGz} Byrnes, C. T., Choi, K. Y., \&  Hall, L. M. H. 2008, J. Cosmol. Astropart. Phys. \textbf{0810}, 008.

\bibitem[{Byrnes \& Tasinato(2009)}]{NG31} Byrnes C. T. \& Tasinato, G. 2009, J. Cosmol. Astropart. Phys. \textbf{0908}, 016.

\bibitem[{Byrnes \& Choi(2010)}]{NG33} Byrnes, C. T., \& Choi, K. Y. 2010, [arXiv:1002.3110].

\bibitem[{Calzetta(1995)}]{NGd} Calzetta, E., \& Hu, B. L. 1995, Phys. Rev. D \textbf{52}, 6770.

\bibitem[{Chambers \& Rajantie(2008{\natexlab{a}})}]{NG28} Chambers A. \&  Rajantie, A. 2008, Phys. Rev. Lett. \textbf{100}, 041302 [Erratum-ibid. 101, 149903].

\bibitem[{Chambers \& Rajantie(2008{\natexlab{b}})}]{NG29} Chambers, A., \& Rajantie, A. 2008, J. Cosmol. Astropart. Phys. \textbf{0808}, 002.

\bibitem[{Chen {et~al.}(2007)}]{equ1} Chen, X., Huang, M. -x., Kachru, S., \&  Shiu, G. 2007, J. Cosmol. Astropart. Phys. \textbf{0701}, 002.

\bibitem[{Chen \& Wang(2010{\natexlab{a}})}]{int1} Chen, X., \& Wang, Y. 2010, Phys. Rev.
D \textbf{81}, 063511.

\bibitem[{Chen \& Wang(2010{\natexlab{b}})}]{int2} Chen, X., \& Wang, Y. 2010, J. Cosmol. Astropart. Phys. \textbf{1004}, 027.

\bibitem[{Cogollo {et~al.}(2008)}]{NG27} Cogollo, H.R.S., Rodriguez, Y., \&
Valenzuela-Toledo, C.A. 2008, J. Cosmol. Astropart. Phys. \textbf{0808}, 029.

\bibitem[{Creminelli(2003)}]{NGg} Creminelli, P. 2003, J. Cosmol. Astropart. Phys. \textbf{0310}, 003 [arXiv:astro-ph/0306122].

\bibitem[{Creminelli {et~al.}(2006)}]{orth1} Creminelli, P.,
Nicolis, A., Senatore, L., Tegmark, M., \&  Zaldarriaga, M. 2006,
J. Cosmol. Astropart. Phys. \textbf{0605}, 004.

\bibitem[{Desjacques \& Seljak(2010)}]{lss6} Desjacques, V., \& Seljak, U. 2010, Phys. Rev., D \textbf{81}, 023006.

\bibitem[{Dodelson(2008)}]{lssc} Dodelson, S. 2008, Academic Press.

\bibitem[{Dvali {et~al.}(2004)}]{NGh} Dvali, G., Gruzinov A., \& Zaldarriaga, M. 2004, Phys. Rev. D \textbf{69}, 023505 [arXiv:astro- ph/0303591].

\bibitem[{Enqvist {et~al.}(2005)}]{NGo} Enqvist, K., Jokinen, A., Mazumdar, A., Multamaki, T., \& Vaihkonen, A. 2005, Phys. Rev. Lett. \textbf{94}, 161301 [arXiv:astro-ph/0411394].

\bibitem[{Feng {et~al.}(2015)}]{Fen15} Feng, C., Cooray, A., Smidt, J., O'Bryan, J., Keating, B., \& Regan, D. 2015, [arXiv:1502.00585].

\bibitem[{Fergusson {et~al.}(2010)}]{wmap3} Fergusson, J. R., Regan D. M., \&  Shellard, E. P. S. 2010b, ArXiv e-prints, [arXiv:1012.6039].

\bibitem[{Ferreira{et~al.}(1998)}]{bis2} Ferreira, P., Magueijo, J., \& Gorski, K. 1998, Astrophys. J. Lett.\textbf{503}, L1.

\bibitem[{Gangui {et~al.}(1994)}]{NGc} Gangui, A., Lucchin, F., Matarrese, S., \&
Mollerach, S. 1994, Astrophys. J. \textbf{430},
447.

\bibitem[{Giannantonio \& Ross (2014)}]{lss7} Giannantonio, T., \& Ross, A. J. 2014, Phys. Rev., D \textbf{89}, 023511.

\bibitem[{Guth(1981)}]{infa} Guth, A. 1981, Phys. Rev. D \textbf{23} 347.

\bibitem[{Heavens(1998)}]{bis1} Heavens, A. F. 1998, Mon. Not. R. Astron. Soc. \textbf{299}, 805.

\bibitem[{Hikage \& Matsubara(2012)}]{wmap4} Hikage, C., \& Matsubara, T. 2012, MNRAS, \textbf{425}, 2187.

\bibitem[{Hu(2001)}]{Hu01} Hu, W. 2001, Phys. Rev. D \textbf{64}, 083005.

\bibitem[{Jokinen \& Mazumdar(2006)}]{NGr} Jokinen, A., \& Mazumdar, A. 2006, J. Cosmol. Astropart. Phys. \textbf{0604}, 003 [arXiv:astro-ph/0512368].

\bibitem[{Kaiser (2010)}]{ein-2} Kaiser, D. I. 2010, Phys. Rev. D \textbf{81}, 084044
[arXiv:1003.1159 [gr-qc]].

\bibitem[{Kaiser {et~al.} (2013)}]{ein-3} Kaiser, D. I., Mazenc, E. A. and Sfakianakis, E. I. 2013, Physical Review D \textbf{87} 064004.

\bibitem[{Kenton} \& {Mulryne}(2015)]{Ken15} Kenton, Z., \& Mulryne, D. J. 2015, J. Cosmol. Astropart. Phys. \textbf{1510}, 018.

\bibitem[{Kenton \& Mulryne(2016)}]{Ken16} Kenton, Z., \& Mulryne, D. J. 2016, [arXiv:1605.03435[astro-ph]].

\bibitem[{Kogo \& Komatsu(2006)}]{4p3} Kogo, N., \&  Komatsu, E. 2006, Phys. Rev. D \textbf{73}, 083007.

\bibitem[{Komatsu {et~al.}(2010)}]{si1} Komatsu,  E. et al. 2010, [arXiv:1001.4538].

\bibitem[{Komatsu \& Spergel(2001)}]{NGa} Komatsu, E., \&  Spergel, D. N. 2001, Phys. Rev. D \textbf{63}, 063002 [arXiv:astro-ph/0005036].

\bibitem[{Koyama(2010)}]{NG32} Koyama, K. 2010, [arXiv:1002.0600].

\bibitem[{Langlois {et~al.}(2000)}]{ein-1} Langlois, D., Maartens R. and Wands, D. 2000, Phys. Lett. B, \textbf{489}, 259.

\bibitem[{Leistedt {et~al.}(2015)}]{lss8} Leistedt, B., Peiris H. V., \& Roth, N. 2015, Phys. Rev. Lett. \textbf{113}, 221301.

\bibitem[{Liddle \& Lyth(2000)}]{lssa} Liddle, A., \& Lyth, D. 2000, Cambridge University Press.

\bibitem[{Lidsey(1997)}]{lsse} Lidsey, J. E., Liddle, A. R., Kolb, E.W., Copeland, E. J.,
Barreiro, T., \& Abney, M. 1997, Rev. Mod. Phys. \textbf{69}, 373.

\bibitem[{Linde(1982)}]{infb} Linde, A. D. 1982, Phys. Lett. B \textbf{108} 389.

\bibitem[{Linde(1990)}]{infd} Linde, A. D. 1990, Harwood Academic Publishers, Chur, Switzerland,
[arXiv:hep-th/0503203].

\bibitem[{Lyth {et~al.}(2005)}]{NGk} Lyth, D. H., Malik, K. A., \& Sasaki, M. 2005, J. Cosmol. Astropart. Phys. \textbf{0505}, 004 [arXiv:astro-ph/0411220].

\bibitem[{Lyth(2005)}]{NGl} Lyth, D. H. 2005, J. Cosmol. Astropart. Phys. \textbf{0511}, 006 [arXiv:astro-ph/0510443].

\bibitem[{Lyth \& Liddle(2009)}]{inff} Lyth, D. H., \& Liddle, A. R. 2009, Cambridge University Press, Cambridge, England.

\bibitem[{Lyth \& Rodriguez(2005)}]{NGn} Lyth D. H., \& Rodriguez, Y. 2005, Phys. Rev. Lett. \textbf{95}, 121302 [arXiv:astro-ph/0504045].

\bibitem[{Lyth \& Zaballa(2005)}]{delN5} Lyth, D. H., \& Zaballa, I. 2005, J. Cosmol. Astropart. Phys. \textbf{10}, 005.

\bibitem[{Maldacena(2003)}]{NGf} Maldacena, J. M., JHEP \textbf{0305}, 013
[arXiv:astro-ph/0210603].

\bibitem[{Malik \& Lyth(2006)}]{NGp} Malik, K. A., \& Lyth, D. H. 2006,J. Cosmol. Astropart. Phys. \textbf{0609}, 008 [arXiv:astro-ph/0604387].

\bibitem[{Malik \& wands(2004)}]{NGi} Malik, K. A., \& Wands, D. 2004, Class. Quant. Grav. \textbf{21}, L65 [arXiv:astro-ph/0307055].

\bibitem[{Mukhanov {et~al.}(1992)}]{Muk92} Mukhanov, V. F., Feldman, H. A., \& Brandenberger, R. H. 1992, Phys. Rept. \textbf{215}, 203, Part 2.

\bibitem[{Mulryne  {et~al.}(2009)}]{NG30} Mulryne, D., Seery, D., \& Wesley, D. 2009, [arXiv:0911.3550].

\bibitem[{Nozari \& Asadi(2016)}]{NG39} Nozari, K., \&  Asadi, K. 2016, Phys. Rev. D \textbf{350} 339.

\bibitem[{Nozari \& Rashidi(2012)}]{NG35} Nozari, K., \& Rashidi, N. 2012, Phys. Rev. D \textbf{86} (2012) 043505.

\bibitem[{Nozari \& Rashidi(2013{\natexlab{a}})}]{NG36} Nozari, K., \& Rashidi, N. 2013, Phys. Rev. D \textbf{88} (2013) 023519.

\bibitem[{Nozari \& Rashidi(2013{\natexlab{b}})}]{NG37} Nozari, K., \& Rashidi, N. 2013, Phys. Rev. D \textbf{88} 084040.

\bibitem[{Nozari \& Rashidi(2014)}]{NG38} Nozari, K., \& Rashidi, N. 2014, Astrophys. Space Sci. \textbf{350} 339.

\bibitem[{Nozari \& Rashidi(2016{\natexlab{a}})}]{NG40} Nozari, K., \& Rashidi, N. 2016, Phys. Rev. D \textbf{93} 124022.

\bibitem[{Nozari \& Rashidi(2016{\natexlab{b}})}]{NG41} Nozari, K., \& Rashidi, N. 2016, Advances in High Energy Physics, \textbf{2016} , 1252689 [arXiv:1509.06240].

\bibitem[{Okamoto \& Hu(2002)}]{tris1} Okamoto, T., \& Hu, W. 2002, Phys. Rev. D \textbf{66}, 063008.

\bibitem[{Regan {et~al.}}(2013)]{wmap6} Regan, D., Gosenca, M., \& Seery, D. 2013, [arXiv:1310.8617].

\bibitem[{Rigopoulos \& Shellard (2003)}]{delN6} Rigopoulos, G., \& Shellard, E. 2003, Phys. Rev. D \textbf{68}, 123518.

\bibitem[{Riotto(2002)}]{infe} Riotto, A. 2002, [arXiv:hep-ph/0210162].

\bibitem[{Rodriguez \& Valenzuela-Toledo(2010)}]{NG34} Rodriguez,  Y., \& Valenzuela-Toledo, C. A. 2010, Phys. Rev. D \textbf{81},
023531.

\bibitem[{Salem(2005)}]{NGm} Salem, M. P. 2005 Phys. Rev. D \textbf{72}, 123516
[arXiv:astro-ph/0511146].

\bibitem[{Sasaki(2006)}]{NGt} Sasaki, M., Valiviita, J., \& Wands, D. 2006, Phys. Rev. D \textbf{74}, 103003.

\bibitem[{Sasaki \& Stewart(1996)}]{delN2} Sasaki, M., \& Stewart, E. D. 1996, Prog. Theor. Phys. \textbf{95}, 71.

\bibitem[{Sasaki \& Tanaka(1998)}]{delN3}Sasaki, M., \& Tanaka, T. 1998, Prog. Theor. Phys. \textbf{99}, 763.

\bibitem[{Seery {et~al.}(2007)}]{NGx} Seery, D., Lidsey, J. E., \& Sloth, M. S. 2007, J. Cosmol. Astropart. Phys. \textbf{0701}, 027 (2007).

\bibitem[{Seery \& Lidsey(2007)}]{NGv} Seery, D., \& Lidsey, J. E. 2007, J. Cosmol. Astropart. Phys. \textbf{0701}, 008.

\bibitem[{Sekiguchi \& Sugiyama(2013)}]{wmap5} Sekiguchi, T., \& Sugiyama, N., J. 2013, Cosmol. Astropart. Phys., \textbf{1309}, 002.

\bibitem[{Smidt {et~al.}(2010)}]{wmap2} Smidt, J., Amblard, A., Byrnes, C. T., et al., 2010, Phys. Rev. D, \textbf{81}, 123007.

\bibitem[{Smith {et~al.}(2015)}]{wmap7} Smith, K. M., Senatore, L., \& Zaldarriaga, M. 2015, [arXiv:1502.00635].

\bibitem[{Starobinsky(1985)}]{delN1} Starobinsky, A. A. 1985, Soviet Journal of
Experimental and Theoretical Physics Letters \textbf{42}, 152.

\bibitem[{Tanaka{et~al.}(2010)}]{Tan10} Tanaka, T., Suyama, T., \& Yokoyama, S. 2010, Class. Quant. Grav. \textbf{27}, 124003.

\bibitem[{Verde {et~al.}(2000)}]{3p1} Verde, L., Wang, L. M., Heavens, A., \& Kamionkowski, M. 2000, Mon. Not. Roy. Astron. Soc. \textbf{313}, L141.

\bibitem[{Vernizzi \& Wands(2006)}]{Ver06} Vernizzi, F., \& Wands, D. 2006, J. Cosmol. Astropart. Phys. \textbf{0605}, 019.

\bibitem[{Vielva \& Sanz(2010)}]{wmap1} Vielva, P. \& Sanz, J. 2010, Mon. Not. Roy. Astron. Soc., \textbf{404}, 895.

\bibitem[{Wands {et~al.}(2000)}]{delN4} Wands, D., Malik, K. A., Lyth, D. H., \&
Liddle, A. R. 2000, Phys. Rev. D \textbf{62}, 043527.

\bibitem[{Wang \& Kamionkowski(2000)}]{3p2} Wang, L. M., \& Kamionkowski, M. 2000, Phys. Rev. D \textbf{61}, 063504.

\bibitem[{Weinberg(2008)}]{lssb} Weinberg, S. 2008, Oxford University Press, Oxford, 9 England.

\bibitem[{White {et~al.} (2013)}]{whi-13} White, J.,  Minamitsujia, M. and Sasaki, M. 2013, JCAP \textbf{09} 015.

\bibitem[{Yokoyama {et~al.}(2008)}]{NGy} Yokoyama, S., Suyama, T., \& Tanaka, T. 2008, Phys. Rev. D \textbf{77}, 083511.


\end{thebibliography}
\end{document}